\documentclass{iopart} 

\usepackage{subfigure}
\usepackage{multirow}
\usepackage{epsfig}
\usepackage{times}
\usepackage{color}
\usepackage{graphicx}

\definecolor{darkgreen}{rgb}{0,0.5,0}
\newcommand{\cg}[1]{#1}
\newcommand{\pj}[1]{#1}

\newcommand{\Label}[1]{\label{#1}}

\newcommand{\BHM}{Bose-Hubbard model}
\renewcommand{\etal}{{\em et al.}}  
\newcommand{\DRAFT}%
{\renewcommand{\cg}[1]{\begingroup\color{blue}##1\endgroup}%
\renewcommand{\pj}[1]{\begingroup\color{darkgreen}##1\endgroup}%
\renewcommand{\Label}[1]{\label{##1}{\hbox to 0cm{\textcolor{magenta}{\hss\em $##1$\quad}}}}}

\newcommand{\nbar}{\overline{n}}
\newcommand{\Uc}{\overline{U}_c}
\newcommand{\nbarint}{[{\overline n}]}
\newcommand{\var}{{\rm var}[N]}
\newcommand{\half}{\textstyle \frac{1}{2}}
\newcommand{\quarter}{\textstyle \frac{1}{4}}
\newcommand{\rr}[1]{\multicolumn{1}{c}{#1}}

\newcommand{\supersection}[1]{\begin{center}\textbf{#1}\end{center}}

\begin{document}
\title{A phase-space method for the Bose-Hubbard model}
\author{P. Jain and C. W. Gardiner}
\address{School of Chemical and Physical Sciences, Victoria University of 
Wellington, New Zealand}
\date{}
\begin{abstract} 
\pj{We present a phase-space method for the Bose-Hubbard model based on the Q-function 
representation. In particular, we consider two model Hamiltonians in the mean-field 
approximation; the first is the standard "one site" model where quantum 
tunneling is approximated entirely using mean-field terms; the second "two site" 
model explicitly includes tunneling between two adjacent sites while treating 
tunneling with other neighbouring sites using the mean-field approximation. 
The ground state is determined by minimizing the classical energy functional 
subject to quantum mechanical constraints, which take the form of uncertainty 
relations. For each model Hamiltonian we compare the ground state results from 
the Q-function method with the exact numerical solution. The results 
from the Q-function method, which are easy to compute, give a good qualitative 
description of the main features of the Bose-Hubbard model including the
superfluid to Mott insulator. We find the 
quantum mechanical constraints dominate the problem and show there are some 
limitations of the method particularly in the weak lattice regime.}

\end{abstract}


\section{Introduction}
The prediction \cite{Jaksch1998b} that a Bose-Hubbard Hamiltonian 
could be realized with cold Bosonic atoms trapped in an optical 
lattice, and the transition between superfluid and Mott-insulator 
phases observed, led to the development of an experiment in which 
these predictions were verified \cite{Greiner2002a}, and has lent 
urgency to the quest for a computationally simple description of the 
phenomena involved.
\cg{It is desirable to find a description which can cover both of 
the two regimes of qualitatively different behaviour:}
\begin{itemize}
\item[a)] \emph{Weak Lattice}, when the atoms are delocalized, and are 
thus highly mobile; the system is weakly correlated.  \item[b)] 
\emph{Strong Lattice}, when the atoms are localized; the system is 
highly correlated.  In this case there are two subcases
\begin{itemize}
\item[i)] \emph{Commensurate}, when the number of atoms per site is  
integral, and the state is a product of number states on each site.  
In this case there can be no mean-field.  \item[ii)] 
\emph{Incommensurate}, when the number of atoms per site is 
non-integral, and in this case there can be a mean-field.
\end{itemize}
\end{itemize}
In the case of an integral average site filling, as the strength of 
the lattice varies, a transition from the delocalized to the localized 
situation takes place, and one passes from state with a non-zero 
mean-field to a state with no mean-field.  Conventionally, if there is 
no mean-field, one speaks of a \emph{Mott insulator phase}, while if 
there is a non-vanishing mean-field, the terminology \emph{superfluid 
phase} is applied.

The dynamics of the Bose-Hubbard model in states for the strong 
lattice region presents formidable technical difficulties.  The 
so-called Gutzwiller approximation \cite{Jaksch1998b} has been used in 
several treatments, but because it represents the wavefunction as a 
product of wavefunctions at different sites, the description of 
intersite correlations is necessarily very approximate.  Progress has 
been made \cite{Rey2002a, Oosten2001} in regions near to the weak 
lattice regime by using a self-consistent Bogoliubov method, in which 
the statistics of the quantum fluctuations are treated by what amounts 
to a Gaussian ansatz.  However, this method cannot be used near the 
strong lattice regime because in this case the statistics are far from 
Gaussian.

\subsection{Overview}

In this paper we want to introduce approximate methods which should be 
applicable in both the superfluid and the Mott insulator regimes, and 
to show their efficacy in the very simplest method used for the 
Bose-Hubbard model, that of static mean-field theory. We propose a phase space 
method from quantum optics based on the Q-function representation 
which involves an approximate reparamterization of the Bose-Hubbard 
Hamiltonian in terms of Gaussian variables.  The advantage of this 
methodology lies in the simple description of a wide range of quantum 
states for the system.

We apply this method to two approximations to the Bose-Hubbard Hamiltonian 
which we denote as the {\em one site} and {\em two site} models.  
The one site 
model treats all intersite correlations (ie.  tunnelling to adjacent 
sites) by a mean-field term, which essentially decouples the total 
Hamiltonian into a sum of one site Hamiltonians.  This has already 
been introduced with the Hamiltonian given by (\ref{bh1a}).  The two 
site model extends this formulation by explicitly including intersite 
correlations between two adjacent sites, while treating the 
interactions with neighbouring sites using the mean-field 
approximation.

For the one site Q-function approach, we find rather good agreement 
over all regimes, with the advantage that our phase space method can 
be evaluated very easily---in a matter of seconds on any reasonable 
workstation.  The accuracy is usually about 5\%, with the qualitative 
behavior being accurately given.  The essence of the result is that 
the ground state is essentially that wavefunction which minimizes the 
classical energy functional subject to the inequality constraints 
given by the uncertainty principle in two different forms.

These results are compared with an exact numerical solution to the one 
site model using arbitrary one site states.  In this case, the 
solution is found by reparameterizing the energy in terms of the 
number variance and finding the maximum mean-field for which the 
energy is minimized.  The formulation is equivalent to the application 
of the standard Gutzwiller ansatz.

The Q-function formulation is easily extended to the two site model 
when lattice homogeniety is assumed.  The correct description of the 
two site quantum statistics requires the inclusion of additional 
constraints which are used to determine the ground state solution.  
The results here give a good qualitative description of the overall 
features of the Bose-Hubbard model.  However, a shortcoming of the 
parameterization on two sites is highlighted by the failure of the 
method to correctly predict the Mott insulator phase, as determined by 
a vanishing mean-field, when compared to the results from sections 
\ref{sec-qfunc1s}, \ref{sec-exact1s} and \ref{sec-exact2s}.

The two site model is also solved by exact numerical minimization with 
the assumption of lattice symmetry giving a reduced Hilbert space.  
Here, the ground state results show the pertinent features of 
the Bose-Hubbard model, with a vanishing mean-field throughout the 
Mott insulator phase, which is an improvement over the two site 
Q-function formulation.  The results are compared with other reports 
from the literature, including density matrix renormalization group 
and Quantum Monte-Carlo methods, which yield numerically exact results 
for one dimensional finite lattices.

\section{Bose-Hubbard Hamiltonian}
\cg{In order to set our notation and define terminology, we will 
summarize the basics of the Bose-Hubbard model---first introduced by 
Fisher \etal\ \cite{fisher1989} to describe the superfluid-insulator 
transition in liquid $^{4}$He, but applicable more generally to any 
system with interacting bosons on a lattice.} The simplest form of the 
model is given when the atoms are loaded adiabatically into the 
lattice, so that they remain in the lowest vibrational state.  As we 
will be formulating the model using the Q-function representation, and 
noting that this distribution is adapted to antinormal ordering, we 
write the Bose-Hubbard Hamiltonian in the antinormally ordered form
\begin{eqnarray}\Label{bh1}
H &=& -\sum_{i,j}t_{ij}a_i a^\dagger_j + u\sum_i a_ia_ia_i^\dagger a_i^\dagger.
\end{eqnarray}
where $a_i^\dagger$ and $a_i$ respectively are the creation and 
annihilation operators for a boson on the $i$th lattice site.  The 
first term of the Hamiltonian describes the quantum tunelling 
(hopping) between neighboring sites with an amplitude $t_{ij}$ which 
is only non-zero when $ i$ and $ j$ are adjacent.  The second term 
represents the on-site interaction energy which is taken as repulsive 
with $u \geq 0$.  We note that this Hamiltonian is easily related to 
its normally ordered counterpart using the commutation relation $[a_i, 
a_j^{\dagger}] = \delta_{ij}$.

\subsection{The mean-field approximation}
In the mean-field approximation, one assumes that the effect of the the nearest 
neighbours on the site $ i$ is given by a c-number mean-field $ {\cal E}= 
\langle a_i \rangle$ 
which, by a choice of phase, we take as real, and
which is independent of $ i$ for a homogeneous system. Thus we can write
\begin{eqnarray}\Label{bh1a}
H&\to  & H_{\rm mf} = 
\sum_{i}\left\{
-{Z\over 2}{\cal E}( a^\dagger_i + a_i) + u a_ia_ia_i^\dagger a_i^\dagger
\right\}
\end{eqnarray}
where $ Z$ is proportional to the nearest neighbor value of $ t_{ij}$ 
multiplied by the number of nearest neighbours.  As noted in Sachdev's book 
\cite{Sachdev1990a}, to solve the system one finds the lowest eigenvalue of 
$ H_{\rm mf}$ on a single site, for which the value of 
$ {\cal E}$ must match that of $\langle a_i\rangle$.

\begin{figure}[t]
    \begin{center}
		\mbox{
		\subfigure[One site model]{\includegraphics[width=0.35\textwidth]{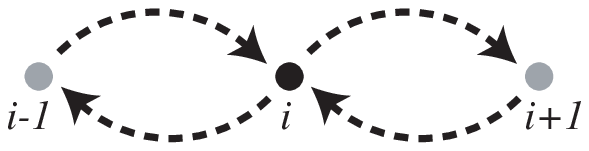}}
		\hspace{1cm}
		\subfigure[Two site model]{\includegraphics[width=0.50\textwidth]{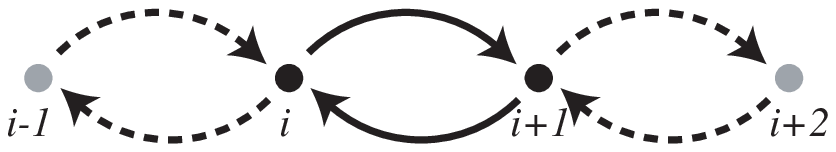}}
		}
		\caption{Treatment of hopping terms in the Bose-Hubbard model 
		for a lattice of dimension $d=1$.  The dashed arrows indicate 
		those hopping terms $t_{ij} a_i a_j^{\dagger}$ included by a 
		mean-field approximation, whereas the solid arrows indicate 
		those terms included explicitly in the Hamiltonian.  We 
		consider two formulations: (a) The hopping terms between the 
		$i$th site and it's $z=2d$ nearest neighbours are treated 
		solely by a mean-field approximation; (b) The hopping terms 
		between two adjacent sites $i$ and $i\!+\!1$ are included 
		explicitly, whereas the hopping terms between each of these 
		sites and and their (other) nearest $z^{\prime}=2d-1$ 
		neighbours are included using the mean-field approximation.}
		\label{fig:BH_diagrams}
		\end{center}
\end{figure}

Thus, one assumes a value for the mean-field,  computes the ground state eigenfunctions of the 
Hamiltonian, whose mean-field should equal that initially assumed.  Unlike the 
more conventional method, we do this at fixed mean occupation of the site, not 
fixed chemical potential, and compute the chemical potential from the energy. 
Since the solutions of this process are well known, it will provide an 
interesting testbed for the phase space method we wish to propose.

\begin{figure}[h]
\hbox{\includegraphics[width=6.5cm]{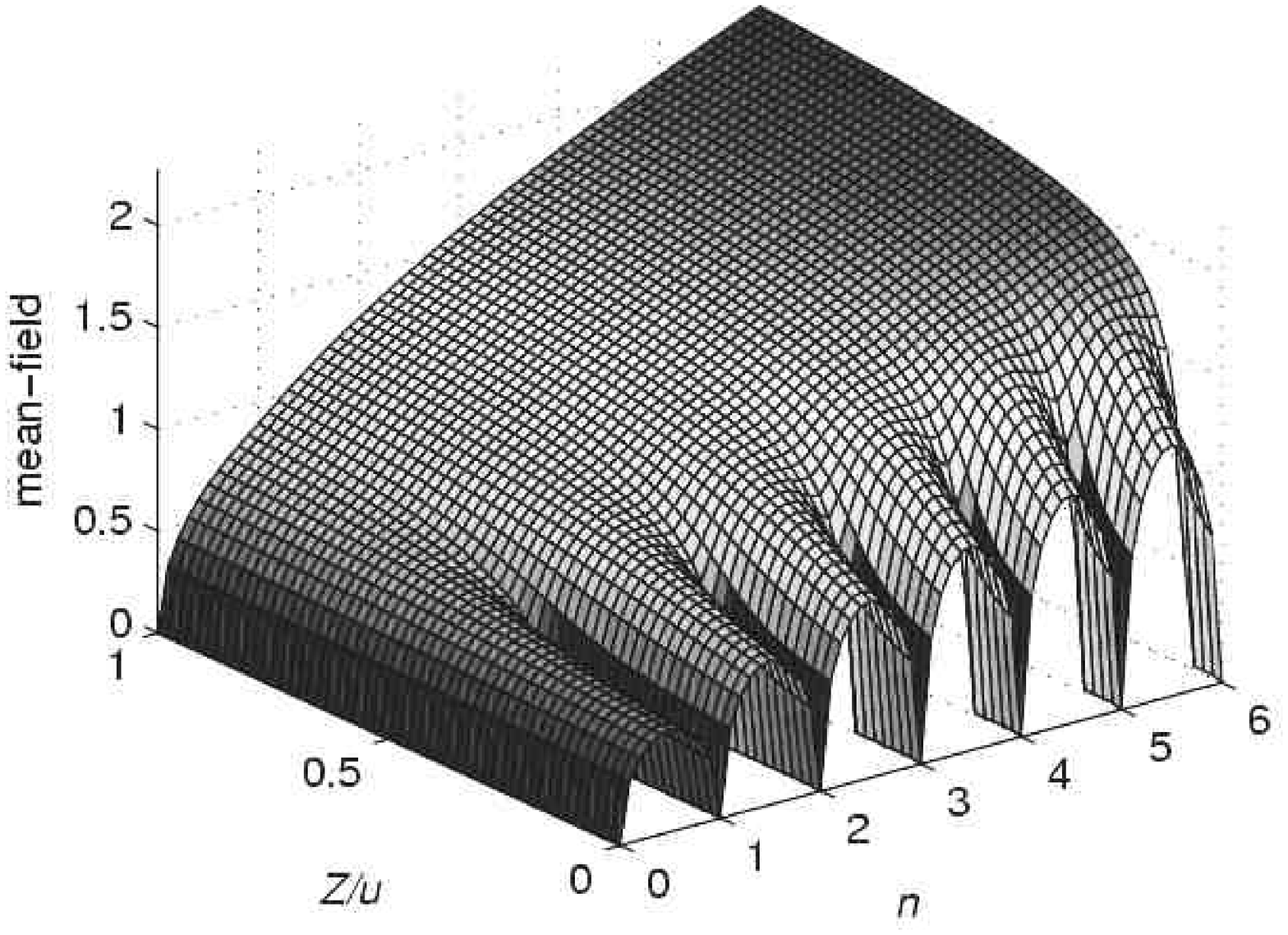}%
\includegraphics[width=6.5cm]{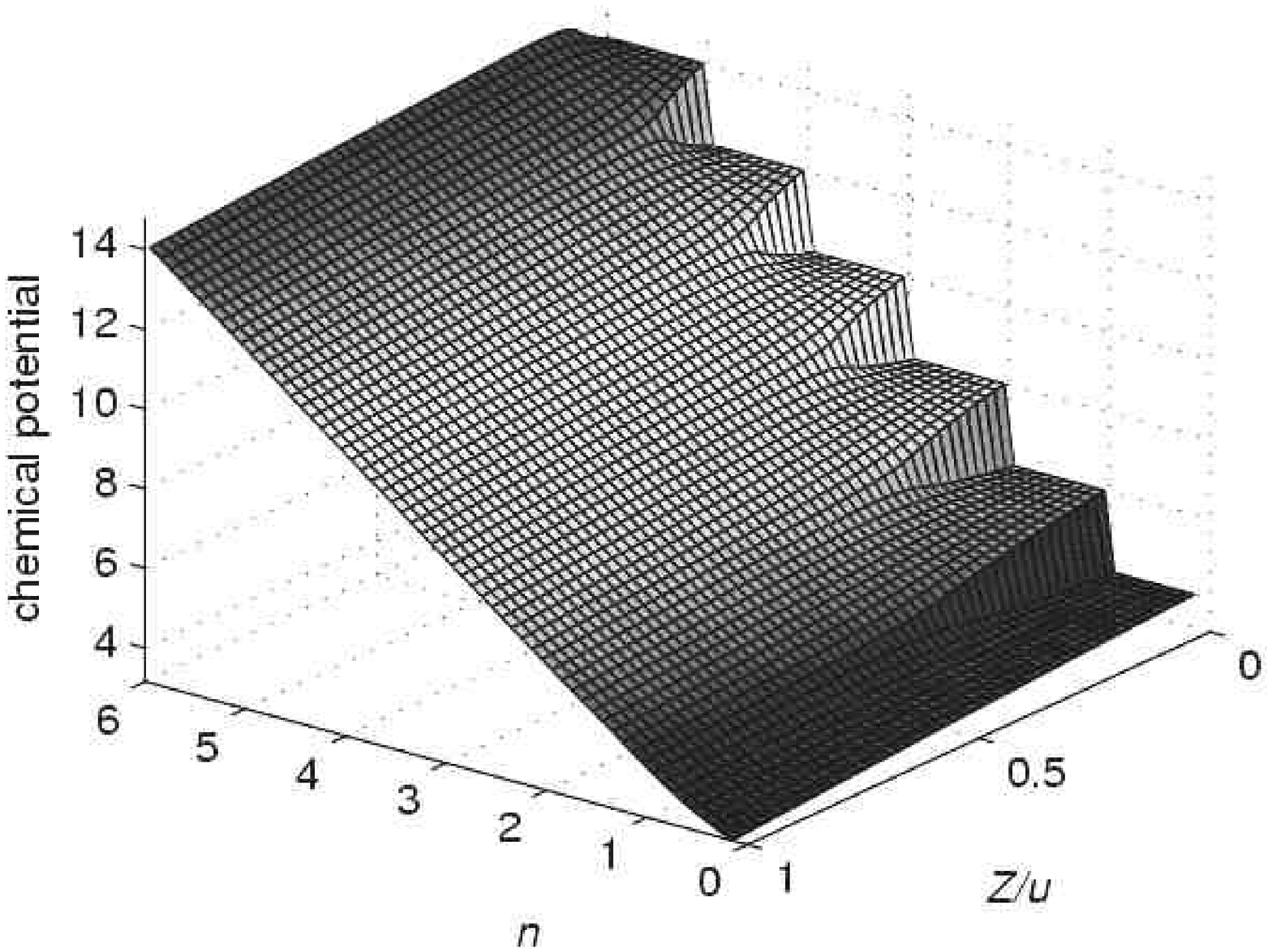}%
}
\caption{Ground state phase diagram for the Bose-Hubbard model using a 
self-consistent mean-field approach; the mean-field and chemical 
potential are both shown as a function of the relative interaction 
strength $Z/u$ and mean occupation number $n$ (normal ordered).}
\Label{fig:exact}
\end{figure}

\subsection{Self-consistent mean-field results}
The results obtained by this procedure are shown in 
Fig.\ref{fig:exact}.  The usual features of the superfluid to Mott 
insulator transition are present.  In particular, in the Mott 
insulator phase, the mean-field vanishes for commensurate occupations 
when the relative interaction strength ($Z/u$) is below a critical 
value.  There is a corresponding energy gap evident in the chemical 
potential which represents the energy required to add or remove a 
particle to the system; hence the Mott phase is {\em incompressible}.  
When the density is incommensurate or the tunnelling dominates, there 
is a transition to the superfluid phase where the mean-field is 
non-zero.

\supersection{PART I: ONE SITE FORMULATION}

\section{Q-functions for one site states}\label{sec-qfunc1s}

We first introduce the Q-function parameterization and its application 
to the one site Hamiltonian (\ref{bh1a}) for the Bose-Hubbard model.

\subsection{Phase space methods}
\cg{The field of quantum optics leads to various quasi-classical 
distributions \cite{QNoise} which can be used to treat quantum processes using 
a  c-number formalism in terms of coherent states $| 
\alpha \rangle$.
For a one-mode system with density operator $\rho$, 
the most widely used of these are given by
\begin{itemize}
\item The Q-function:
\begin{eqnarray}\Label{bh1b}
Q(\alpha,\alpha^*) &=&  \langle\alpha |\rho |\alpha\rangle/\pi .
\end{eqnarray} 
The Q-function is a quasiprobablity for \emph{antinormally} ordered operator 
averages, that is
\begin{eqnarray}\Label{bh1c}
\langle a^n(a^\dagger)^m\rangle &=&
 \int d^2\alpha\,\alpha^n({\alpha^*})^mQ(\alpha,\alpha^*)
\end{eqnarray}
It always exists and is positive 
\item The P-function: $P(\alpha,\alpha^*)$ 
with $\rho = \int d^2 \alpha P(\alpha,\alpha^*) | \alpha \rangle 
\langle \alpha |$, associated with normally ordered moments.  However, it does 
not 
always exist as a positive well behaved function.
\item The Wigner function: 
$W(\alpha,\alpha^*)$, is associated with symmetric operator ordering,  and 
always exists, but is not always positive.
 \item The positive P-function: 
$P(\alpha,\beta)$, a P-function defined in a doubled phase-space, 
always exists and is positive, but does present some technical difficulties 
\cite{QNoise}.
\end{itemize}
\noindent
We develop a treatment in terms of the Q-function, since 
this is always positive and well-defined, and is thus a genuine probability 
density to which we can apply 
probabilistic approximations.
It can be used to describe a wide range of 
states for the Bose-Hubbard model which interpolate between the extremes:
\begin{itemize}
\item[i)] Weak lattice: in the case that the hopping dominates ($ 
t_{ij}$ very large) the ground state is the product of coherent states 
at each site. 

\item[ii)] Strong lattice: in the other extreme, 
when the hopping is negligible, the ground state is  the 
product of eigenstates at each site.  If the mean occupation per site 
is $ n$ and $ [n]$ denotes the integer part of $ n$, then the state at 
each site is a superposition of number states of the form 
$\sqrt\lambda |[n]\rangle + \sqrt{1-\lambda}|[n]+1\rangle $, where $ 
\lambda$ is chosen to give the correct mean occupation.  These states 
are not Gaussian.
\end{itemize}
The Gaussian or non-Gaussian nature of the statistics in the \BHM\ is very 
important, and one of the virtues of the Q-function is that it is Gaussian when 
the quantum statistics is also Gaussian \cite{QNoise}. 

The Q-functions for 
the three principal kinds of states are:
\begin{itemize}
\item [i)] \emph{Number state:}
 The Q-function for a number state is 
\begin{eqnarray}\Label{bh2}
Q_n(\alpha,\alpha^*) &=& {1\over\pi}{e^{-|\alpha |^2}|\alpha |^{2n}\over n!}.
\end{eqnarray}
Examples of this distribution appear in Fig.\ref{Qfuns} (a) and (e).
The distribution has a characteristic ring shape, which is clearly 
non-Gaussian.

\item[ii)] \emph{Superposition of number states}
The Q-function for the superposition of number states with  $ n-1 $ and $ n$ 
atoms in proportions $ 1-\lambda : \lambda$ is
\begin{eqnarray}\Label{bh3}
Q_{{\rm sup}\lambda}(\alpha,\alpha^*) &=& {1\over\pi}
\left|\sqrt{1-\lambda} + \alpha \sqrt{\lambda\over n}\right|^2
{e^{-|\alpha |^2}|\alpha |^{2(n-1)}\over (n-1)!}.
\nonumber \\
\end{eqnarray}
This is the kind of state expected with an incommensurate filling in the limit 
of no hopping.  The Q-function then varies between a ring shaped distribution 
for $ \lambda =0,1$ to a rather distorted Gaussian distribution at 
$ \lambda = 1/2$; see Fig.\ref{Qfuns} (a)--(e).

\item[iii)] \emph{Coherent state}
In the superfluid case with mean filling per site $ n$, where the hopping is 
dominant, we expect an approximately coherent state with parameter 
$ \beta = \exp(i\theta)\sqrt{n}$, for some real $ \theta$, and then the 
Q-function has the Gaussian form
\begin{eqnarray}\Label{bh4}
Q_{{\rm coh},\beta}& = & {1\over\pi}\exp\left(-|\alpha - \beta |^2\right).
\end{eqnarray}
This is plotted in Fig.\ref{Qfuns} (f).
\end{itemize}


\begin{figure}[h]
    \begin{center}
		\mbox{
		\subfigure[$n=7$, $\lambda = 0$]{\includegraphics[width=0.33\textwidth]{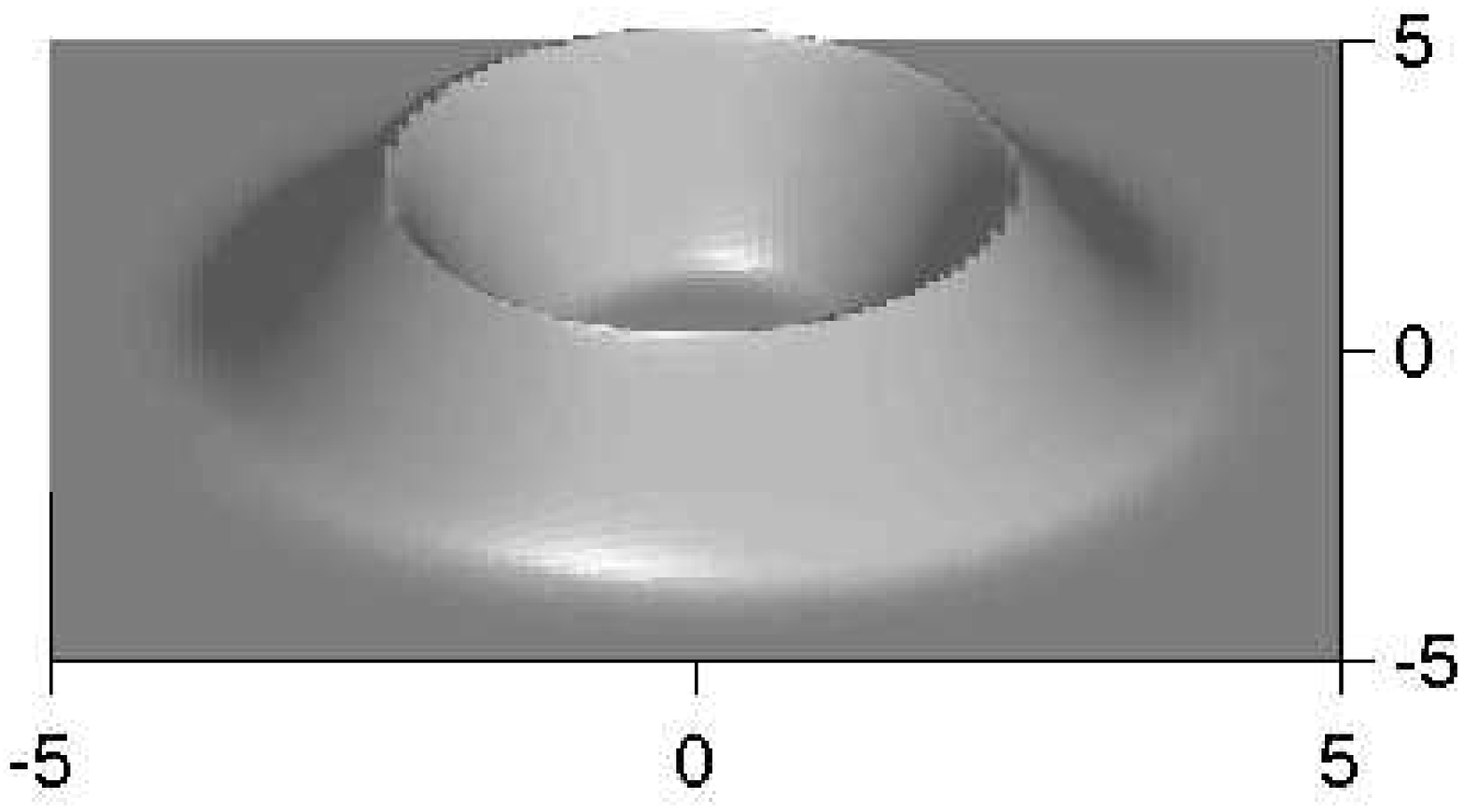}}
		\hspace{0mm}
		\subfigure[$n=7$, $\lambda = 0.15$]{\includegraphics[width=0.33\textwidth]{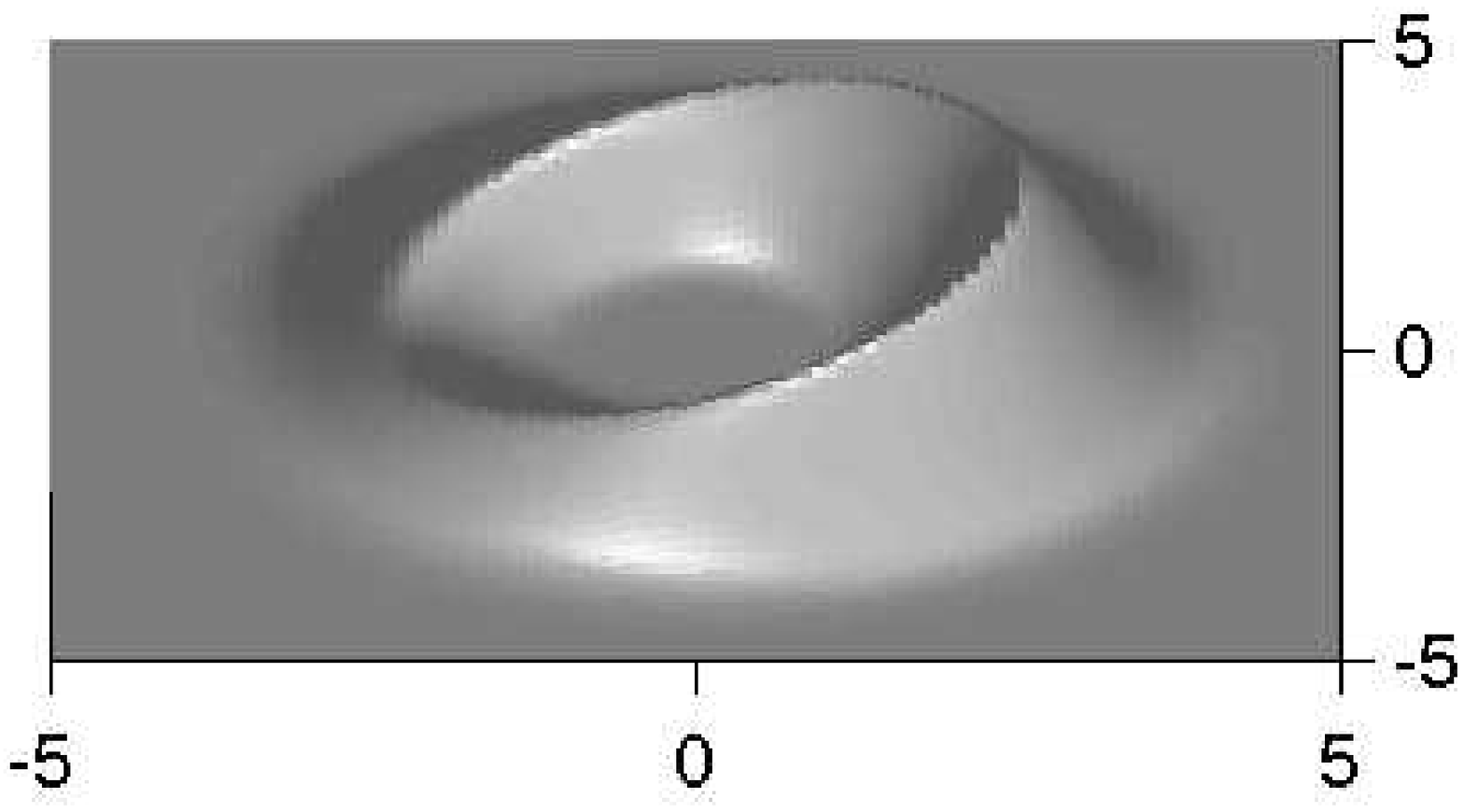}}
		\hspace{0mm}		
		\subfigure[$n=7$, $\lambda = 0.5$]{\includegraphics[width=0.33\textwidth]{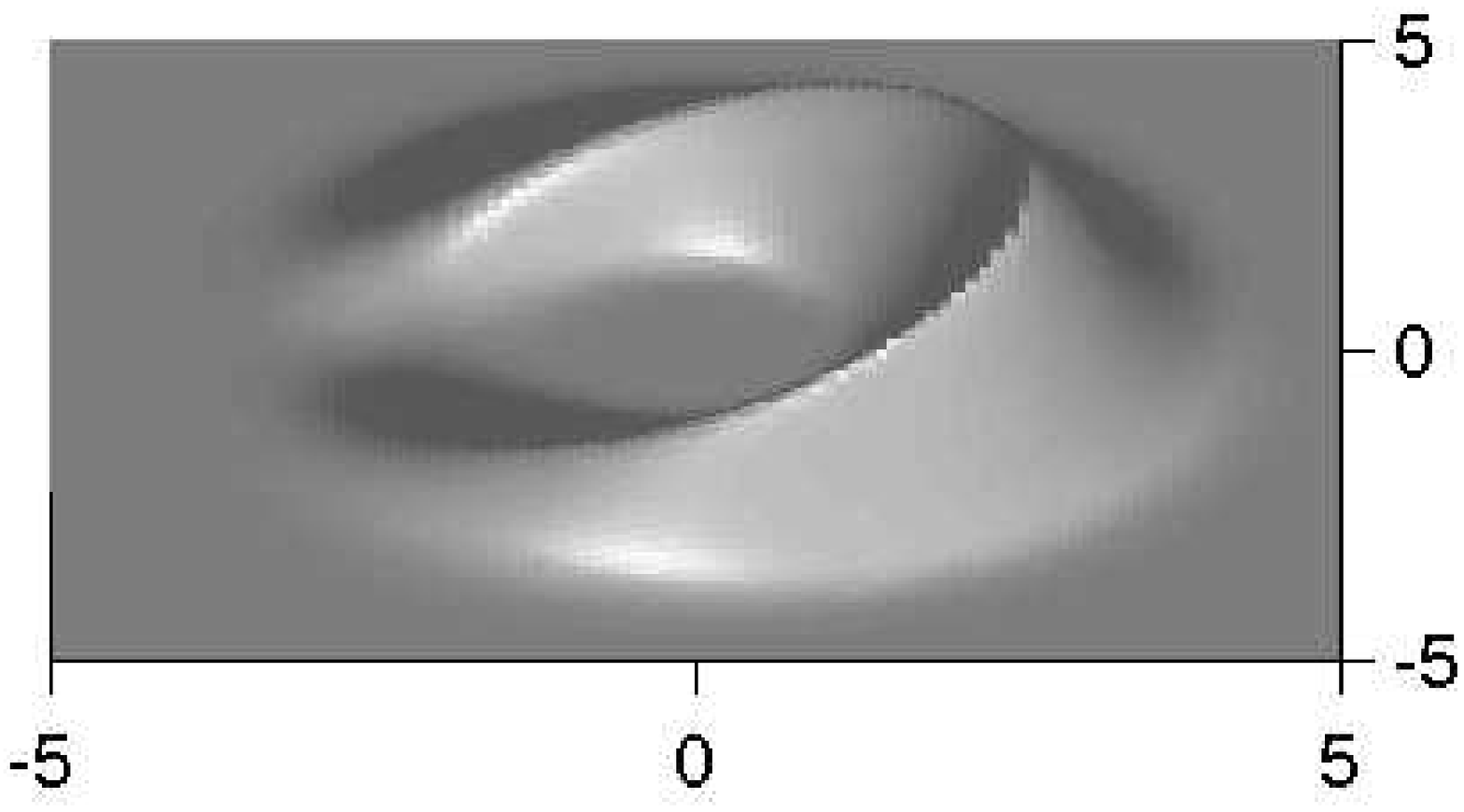}}
		}
		\mbox{
		\subfigure[$n=7$, $\lambda = 0.85$]{\includegraphics[width=0.33\textwidth]{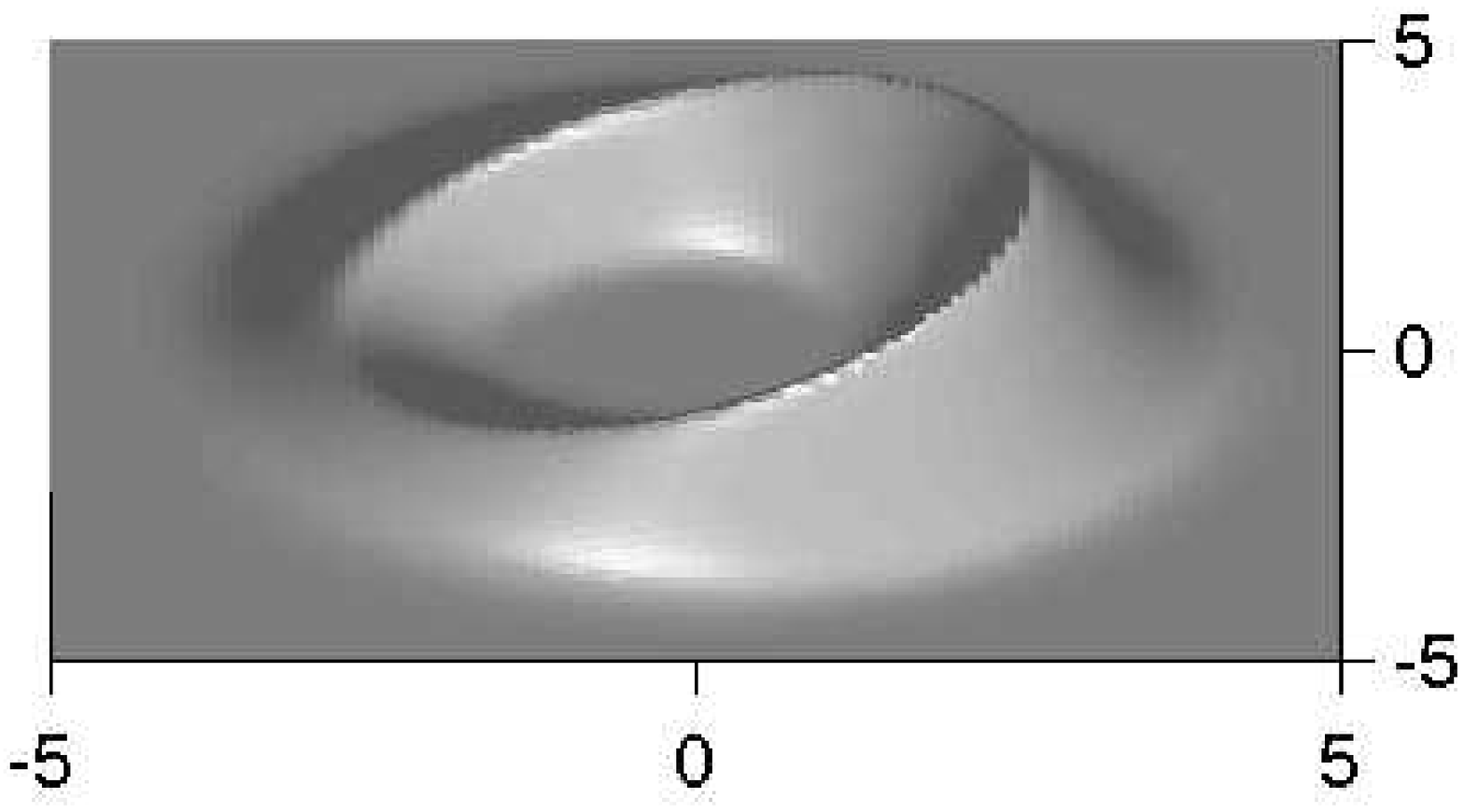}}
		\hspace{0mm}
		\subfigure[$n=7$, $\lambda = 1$]{\includegraphics[width=0.33\textwidth]{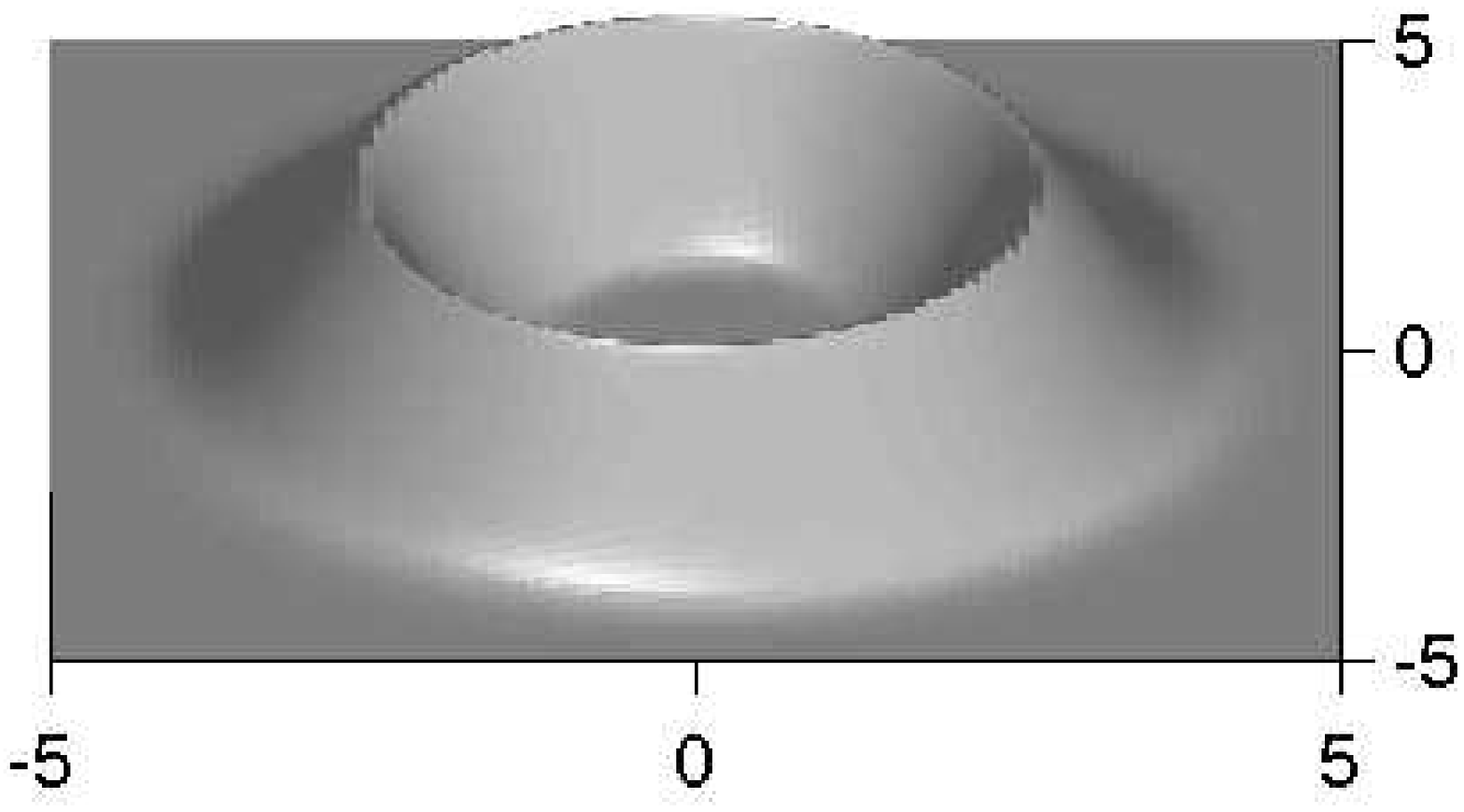}}
		\hspace{0mm}		
		\subfigure[coherent state, $\beta = \sqrt{6}$, $\lambda =
		 0$]{\includegraphics[width=0.33\textwidth]{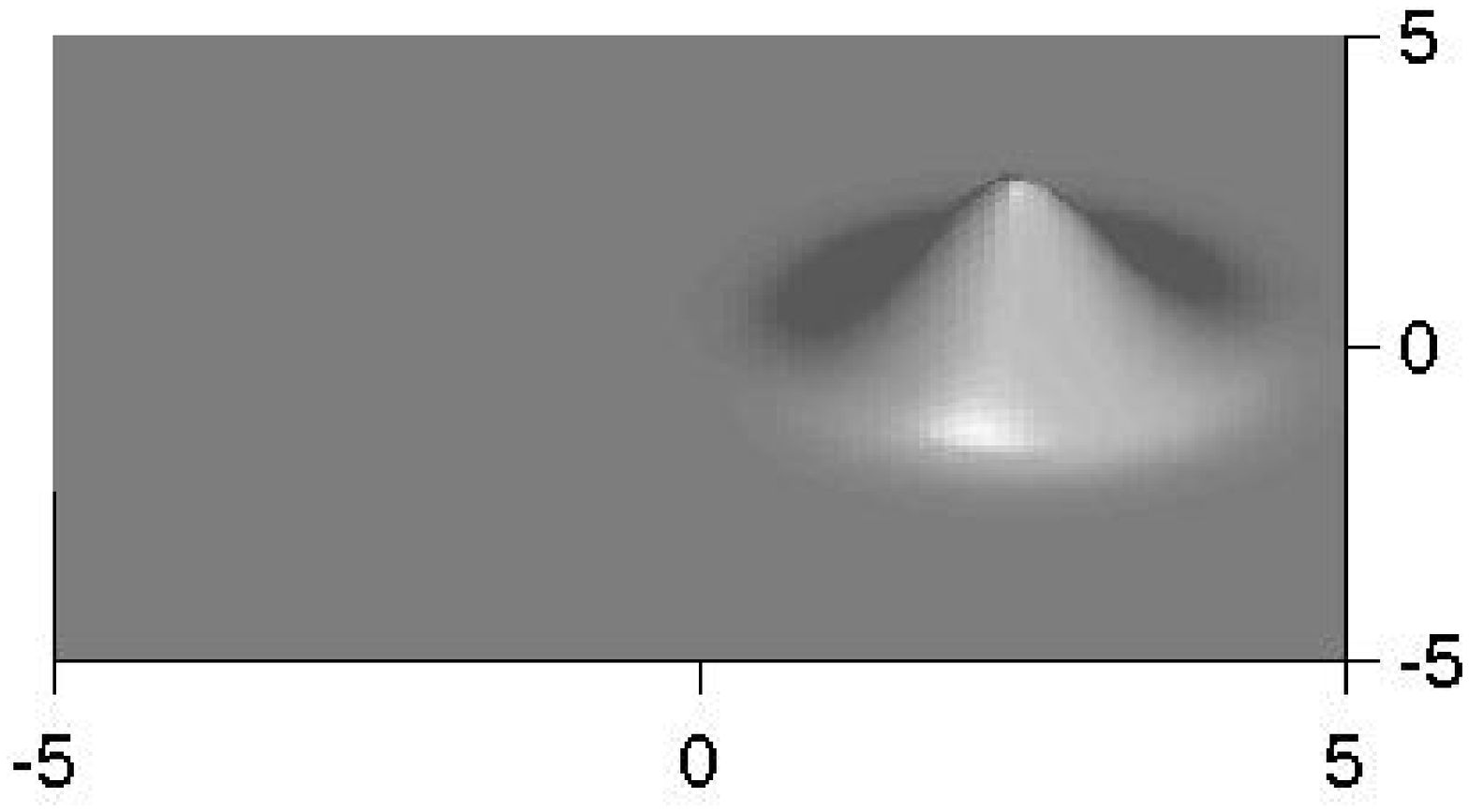}}
		}
		\caption{\label{Qfuns} Q-function for the superposition of $n=6$ and
$n=7$ states in proportion 1:0, 0.85:0.15, 0.5:0.5, 0.15:0.85, 0:1; and for a 
coherent state with $ \beta = \sqrt{6}$}.  
		\end{center}
\end{figure}

\subsection{Parametrization of the Q-function}
The probability distributions of 
Fig.\ref{Qfuns} can be approximately parametrized by writing the 
random variable $ {a}\to \alpha $ with}
\begin{eqnarray}\Label{par1}
\alpha &=& (v+ \delta)e^{i\theta}
\end{eqnarray}
where
\begin{itemize}
\item[a)]
$ v$ is a nonrandom positive real quantity;
\item[b)] 
$ \delta$ is a Gaussian random variable with zero mean and variance $ \sigma$;
\item[c)] $ \theta$ is a Gaussian random variable with a mean which is in 
principle nonzero, but which by choice of phase definition can be chosen to be 
zero, and with variance $ \Delta$.  In this case, the averages of powers of 
$ \exp(i\theta)$ are given by
\begin{eqnarray}\Label{par2}
\langle e^{pi\theta}\rangle = \exp\left({-p^2\Delta/2}\right) \equiv c^{p^2}.
\end{eqnarray}
\end{itemize}
Using properties of Gaussian variables this approximation leads to the values 
of 
the antinormally ordered moments:
\begin{eqnarray}\Label{par3}
\langle {a} \rangle &=& vc
\\ \Label{par4}
\langle {a}{a}^\dagger\rangle &=& (v^2 +\sigma)
\\ \Label{par5}
\langle {a}{a} \rangle &=& (v^2 + \sigma)c^4
\\ \Label{par6}
\langle {a}{a}{a}^\dagger\rangle &=& (v^3 + 3 \sigma v)c
\\ \Label{par7}
\langle {a} {a}{a}^\dagger{a}^\dagger\rangle 
&=& (v^4 + 6 v^2 \sigma + 3\sigma^2)
\end{eqnarray}
Note that the first of these moments (\ref{par3}) represents the 
mean-field.  A straightforward extension of this this parameterization 
to $L$ sites can be given by the substitution $a_k \rightarrow (v_k + 
\delta_k)e^{i\theta_k}$ for $k = 1, 2, ..., L$; we make use of this 
for the two site problem presented in section \ref{sec-qfunc2s}.

\subsubsection*{Validity of parameterization: values of the parameters 
in particular cases} To fit the kinds of distribution in 
Fig.\ref{Qfuns} we determine the parameters $ v$, {$ \sigma$} and $ c 
$ by fitting the moments $ \langle {a} \rangle$, $ \langle 
{a}{a}^\dagger\rangle$ and $ \langle {a} 
{a}{a}^\dagger{a}^\dagger\rangle $.  These are given in detail in 
\ref{Params}, where we show that for a wide variety of states we get 
very tolerable approximations.  Thus we can expect a good qualitative 
description of the system for an arbitrary lattice strength.  However, 
we note that the parameterization is least accurate in the case of an 
equal superposition of number states which we will find reflected in 
the ground state phase diagram in the strong lattice regime.

\subsection{Determination of the ground state}
Using the Hamiltonian (\ref{bh1a}) and the moments 
(\ref{par3})--(\ref{par7}), the average energy in the Q-function 
representation is
\begin{eqnarray}\Label{gs1}
E(n,c,v) &=& -Zc^2v^2 + u(3n^2-2v^4).
\end{eqnarray}
To find the ground state solution at a given interaction strength 
$Z/u$ and mean occupation $n$, this quantity is minimized with respect 
to the free parameters $v$ and $c$, with the bounds $v \geq 0$ and $0 
\leq c \leq 1$ as permitted by the Q-function parameterization.  

\subsubsection{Constraints}
The functional (\ref{gs1}) is a classical distribution that has no local 
minima so that the minimum must occur on the boundary.  However the system should be 
governed by the quantum mechanical nature of the problem, which we 
reintroduce in the form of two constraints on the minimization 
procedure, one derived from a restriction on the number variance, the 
other from the uncertainty relation for the conjugate variables of 
phase and number.

The most significant of the constraints enforced by the quantum mechanical 
nature of the problem are as follows.

\paragraph{\textbf{Fractionality constraint}}
If the mean occupation per site is non-integral, the variance of the 
occupation must be nonzero, and this has a major bearing on the problem, which 
we shall formulate precisely.
We will be considering cases where the mean occupation per site is fixed, so 
that 
\begin{eqnarray}\Label{con1}
\langle {a}{a}^\dagger\rangle &=& n \ge 1
\end{eqnarray}
so that
\begin{eqnarray}\Label{con2}
n&=& v^2 + \sigma \ge 1.
\end{eqnarray}
Setting 
\begin{eqnarray}
\Label{con6}
N &=& {a}{a}^\dagger
\end{eqnarray}
and using (\ref{par3})-(\ref{par7}) the variance of the site occupations is 
\begin{eqnarray}\Label{con3}
{\rm var}[N] &\equiv &\langle {a}{a}^\dagger{a}{a}^\dagger\rangle -
 \langle{a}{a}^\dagger\rangle^2 \nonumber \\ 
&=&  2n^2 - n - 2v^4.
\end{eqnarray}
In $ n$ is non-integral, with fractional part $ \delta n$, then the minimum 
variance for a given $ n$ occurs when only the two occupation numbers which 
bracket the value $ n$ are represented, and then this gives a variance of 
$ \delta n(1-\delta n)$, leading to the constraint
\begin{eqnarray}\Label{con4}
{\rm var}[N] &\ge &  \delta n(1-\delta n).
\end{eqnarray}
and using (\ref{con3}) we can write this as
\begin{eqnarray}\Label{con5}
0 &\le& v^4 \le n^2 -{1\over 2}\bigg(n + \delta n(1- \delta n)\bigg)
\end{eqnarray}

\paragraph{\textbf{Phase-number uncertainty relations}}
The uncertainty principle enters when we consider that phase and 
number are conjugate variables.  We can write a rigorous uncertainty 
relationship using 
\begin{eqnarray} \Label{con7}
X = {{a} +{a}^\dagger\over 2}; \qquad
Y = {{a} -{a}^\dagger\over 2i}
\end{eqnarray}
so that we have the commutation relations
\begin{eqnarray}\Label{con8}
[N,X] &=& -i Y; \qquad [N,Y] = iX
\end{eqnarray}
from which follow an uncertainty relation
\begin{eqnarray}\Label{con9}
\langle \delta Y ^2\rangle {\rm var} [N]
&\ge& {1\over 4}\langle X\rangle^2
\end{eqnarray}
In our formulation we have chosen a zero mean phase so that 
$ \langle Y \rangle=0$  (see (\ref{par11})) and thus 
$\langle \delta Y ^2\rangle=\langle  Y ^2\rangle $.  
Using (\ref{par11}--\ref{par15}), and antinormally ordering all the products 
involved, we find
\begin{eqnarray}\Label{con10a}
\langle \delta Y ^2\rangle &=& {n\over2}(1-c^4) -{1\over4}
\end{eqnarray}
and thus
\begin{eqnarray}
\Label{con10b}
\left[{n\over2}(1-c^4) -{1\over4}\right]\left(2n^2-{n}-2v^4\right) &\ge&
{1\over 4}v^2c^2.
\end{eqnarray}

\begin{figure}[!htp]
	\begin{center}
		\mbox{ \subfigure[$n = 
		1$, $Z/u = 0.5$]{\includegraphics[width=0.33
\textwidth]{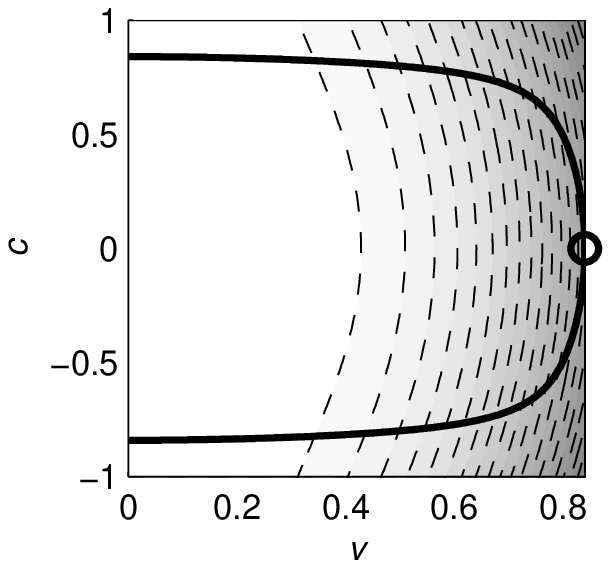}} 
		\hspace{0mm} \subfigure[$n = 
		1.5$, $Z/u = 0.5$]{\includegraphics[width=0.33
\textwidth]{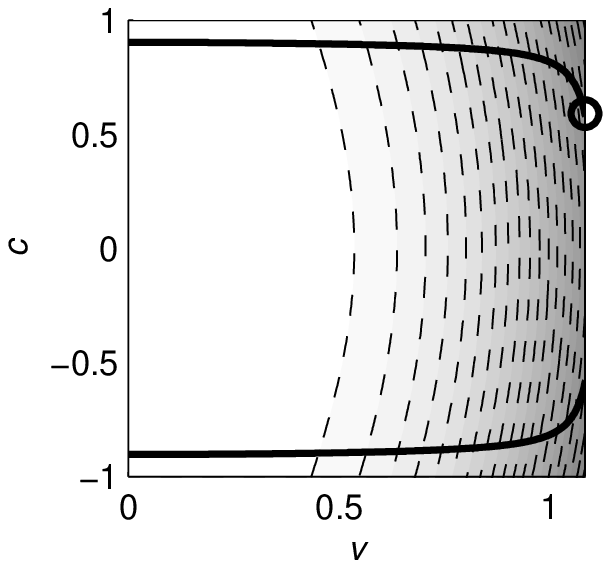}} 
    \hspace{0mm} \subfigure[$n = 1.5$, 
    $Z/u = 10$]{\includegraphics[width=0.325
\textwidth]{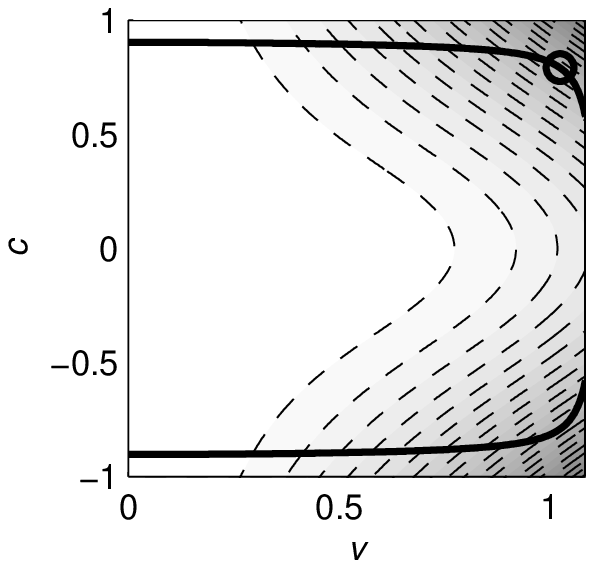}} 
		} \caption{\pj{Level curves of energy (\ref{gs1}) in the 
		Q-function representation for three different cases
	  are indicated by the dashed curves, with the 
		energy decreasing for darker regions; the permissable 
		parameter space for $v$ and $c$ is given by the bounds $v 
		\geq 0$ and $0 \leq |c| \leq 1$ (for purely illustrative 
		purposes we allow $c$ to be negative here because the problem 
		is symmetrical about $c=0$) where $v$ is further bounded by 
		(\ref{con5}).  Since (\ref{gs1}) has no local minima, the 
		minimum energy occurs within the shown bounded region and subject to the 
		constraint (\ref{con10b}) derived from the phase-number 
		uncertainty relation which is shown by the solid curve.  The solution in each
		case is indicated by a circle. In the first case ($n=1$, $Z/u=0.5$) this occurs
		for $c = 0$ so that the mean-field is zero; in the second case ($n=1.5$, $Z/u=0.5$) 
		this occurs for a non-zero $c$ so that there is a mean-field. In both these cases, 
		the solution saturates the upper bound for $v$. Conversely, in the third case 
		($n=1.5$, $Z/u=10$) where the lattice is weak, the solution occurs within the
		bounded region.}}
		\label{fig:ps-energycontours}
		\end{center}
\end{figure}

\subsubsection{Results}
Minimizing (\ref{gs1}) with the constraints (\ref{con5}) and 
(\ref{con10b}) leads to the ground state phase diagram shown in 
Fig.\ref{1site-qgs}.  These results compare well with the exact 
solution shown in Fig.\ref{fig:exact}, particularly with confirmation 
of the Mott insulator phase at commensurate occupations for a strong 
lattice (small $Z/u$), and of the superfluid phase for a weak lattice 
(large $Z/u$).

The results are dominated by the interplay between the constraints, 
which is made evident by considering Fig.\ref{fig:ps-energycontours}.  
\pj{In the strong lattice regime,} the ground state solution is given at the 
intersection of the fractionality constraint (\ref{con5}) which 
determines the upper bound for $v$, and the phase-number uncertainty 
constraint (\ref{con10b}), shown by the solid curve.  
Specifically for $Z/u = 0.5$, and for an 
integer $n=1$, this occurs at $c=0$ and the resulting vanishing 
mean-field is indicative of the Mott insulator phase; for 
a non-integer $n=1.5$ the solution occurs for $c \neq 0$ so that there 
is a non-zero mean-field corresponding to a superfluid component. 
\pj{Conversely for a weak lattice, with $n=1.5$ and $Z/u = 10$, the 
solution is determined solely by(\ref{con10b}).}

The fact that the quantum mechanical constraints dominate the problem 
is not unexpected as the phase transition between the Mott insulator 
and superfluid is driven by quantum fluctuations.

We note that for the Q-function results at half-integer values of $n$, 
the mean-field remains constant over some range of $Z/u$ in the strong 
lattice regime.  This contrasts with the numerically exact results 
obtained by a self-consistent mean-field approach as shown in 
Fig.\ref{fig:exact} where the mean-field is a monotonically increasing 
function of $Z/u$.  This difference arises from the weaker 
approximation of the Q-function parameterization in the case of an 
equal superposition of number states as shown in Fig.\ref{Qfuns} (c), 
and quantified in \ref{appendix-a}.

\begin{figure}[h]
\hbox{\includegraphics[width=6.5cm]{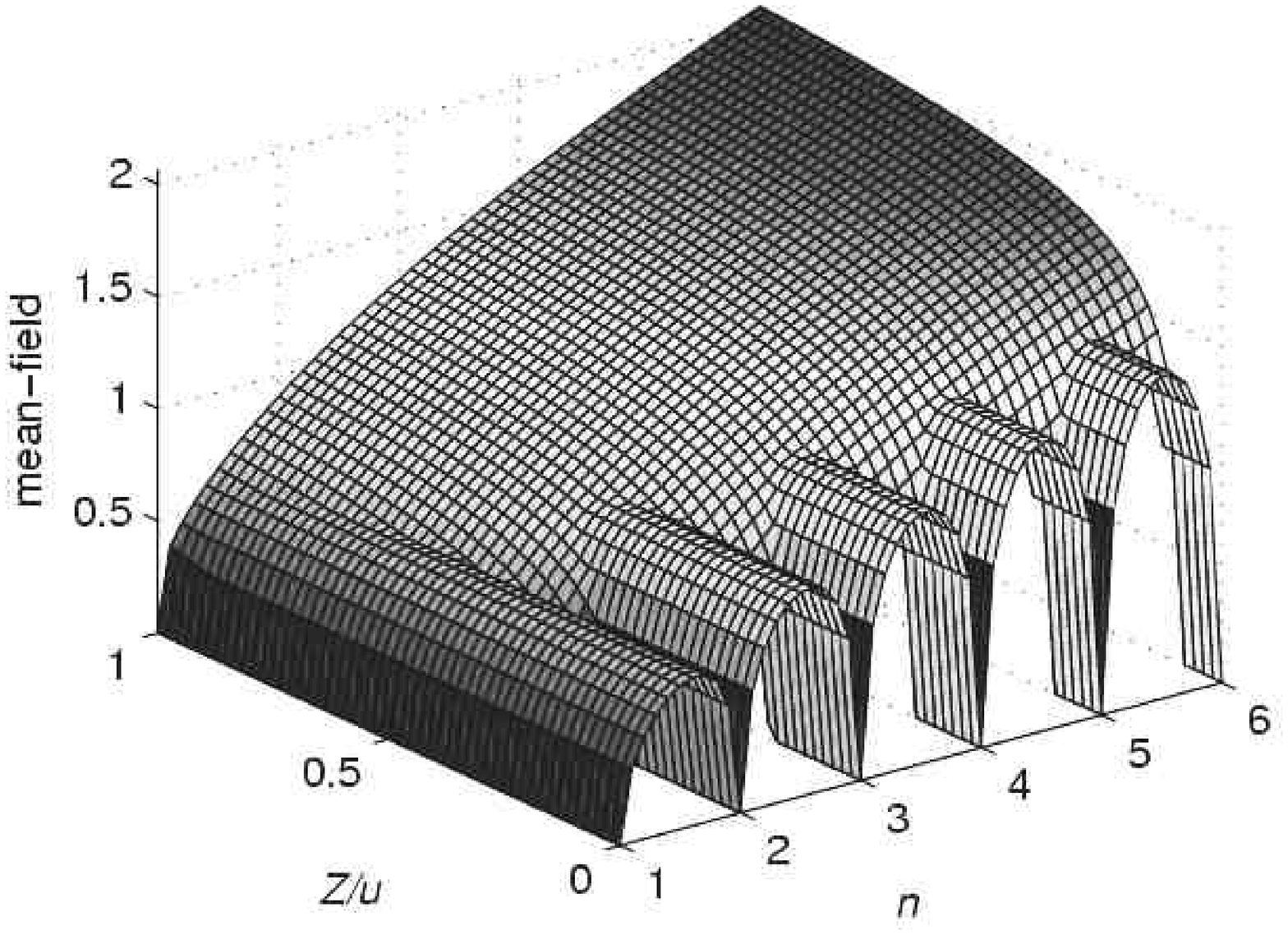}%
\includegraphics[width=6.5cm]{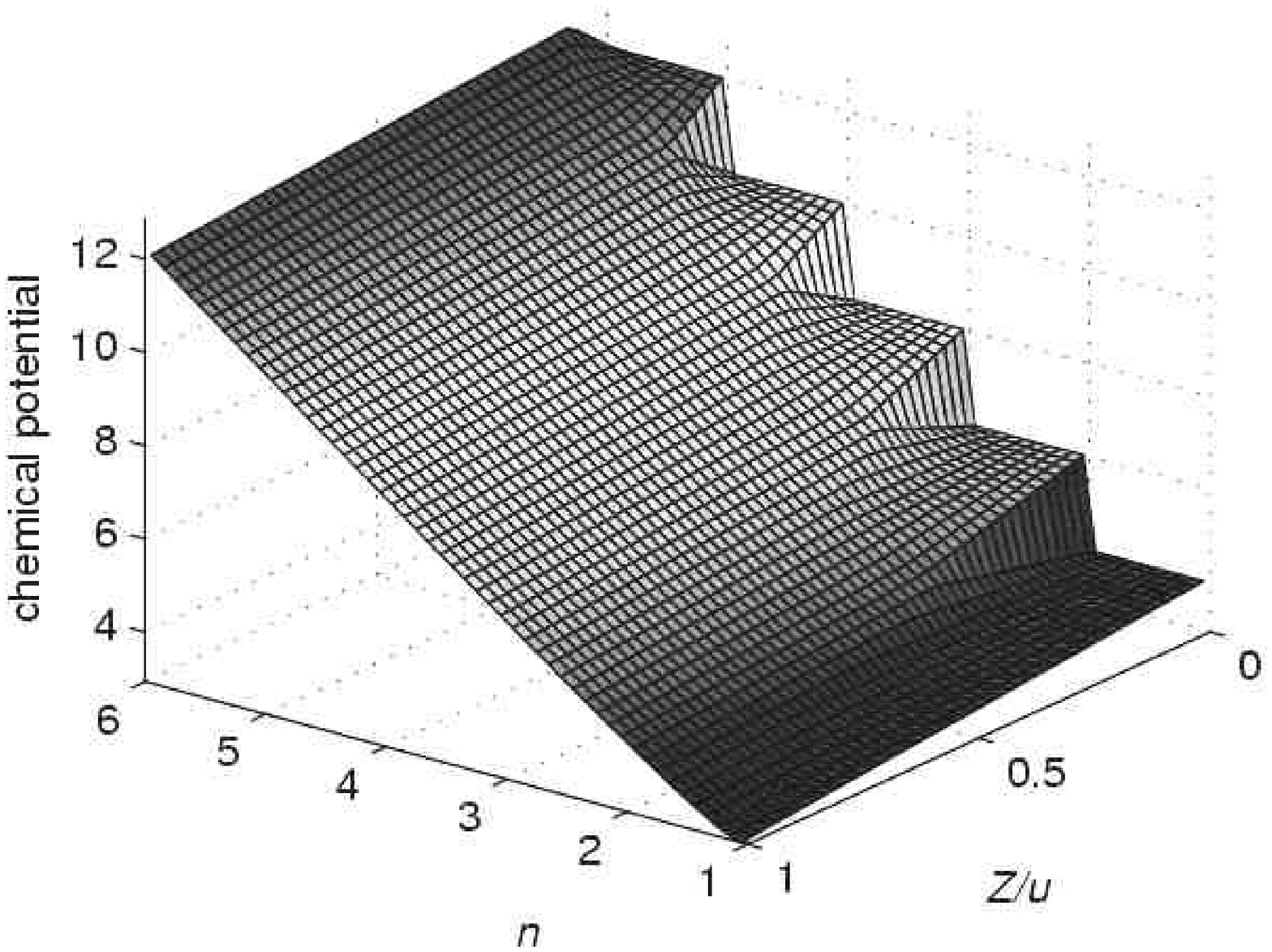}%
}
\caption{Ground state \pj{results} for Bose-Hubbard 
model with Q-function representation; \pj{the mean-field and chemical potential
are shown as a function of relative interaction strength $Z/u$ and (anti-normal 
ordered) occupation $n$.}}
\Label{1site-qgs}
\end{figure}

\section{Exact numerical minimization of one site problem}\label{sec-exact1s}

The inequalities (\ref{con10b}) and (\ref{con5}) restrict the mean-field to a certain 
maximum value, but this maximum may not be optimal.  Let us then pose the 
question: What is the largest value of the mean-field for a given mean number 
and variance. This value determines the minimum energy and permits the following basic strategy for finding the ground state phase diagram. Firstly, reparameterize the mean-field in terms of the mean number occupation and number variance; and secondly, minimize the energy with respect to the number variance at a fixed relative interaction strength $zt/u$ and mean occupation. The problem is then essentially reduced to a minimization of energy with respect to a single variable -- the number variance.

\subsection{Formulation using one site states}\label{sec-exact1sform}
\pj{Here we give an exact numerical procedure for determining the ground state of the Bose-Hubbard model in the one site mean-field approximation.} A one site quantum state can be written as a superposition of number states
\begin{eqnarray}\Label{1site-state}
| s \rangle = \sum_{n} c_{n} | n \rangle
\end{eqnarray} 
where the usual normalisation condition ($1 = \langle s | s \rangle$) applies. In this basis the mean-field is
\begin{equation}\Label{1site-mf}
\langle X \rangle = \langle s|\frac{a + a^{\dagger}}{2} |s\rangle = \Re\left\{ \sum \sqrt{n} \, c^*_{n-1} c_{n} \right\}
\end{equation}
The Hamiltonian (\ref{bh1a}) leads to the expression for the one site energy in the mean-field approximation
\begin{equation}\Label{1site-E}
\langle E \rangle = -Z \, {\langle X \rangle}^2 + u \big ( 2 + \var + \nbar^2 + 3\,\nbar \big )
\end{equation}
where $\nbar \equiv \langle a^{\dagger}\,a \rangle$ is the (normal ordered) mean number and $\var = \overline{n^{2}} - \nbar^2$ is the number variance of atoms on each site (note the normal and anti-normal ordered means are related by $\langle a^{\dagger} a \rangle = \langle a a^{\dagger} \rangle - 1$). Since the chemical potential term has been omitted in (\ref{bh1}), we treat the system using the canonical ensemble by taking a fixed number of atoms in the lattice, that is by explicitly using the constraint of fixed mean occupation. 

Clearly, the energy functional (\ref{1site-E}) is then minimised when the mean-field $\langle X \rangle$ is a maximum and subject to the constraints of normalisation and fixed number mean and variance. In terms of the one site state (\ref{1site-state}) these constraints are given by
\begin{equation} \Label{1site-norm}
\phi_1 = \sum_{n} |c_{n}|^2 - 1 = 0 \textrm{ (normalization)}
\end{equation}
\begin{equation} \Label{1site-mean}
\phi_2 = \sum_{n} |c_{n}|^2 \, n  - \nbar = 0 \textrm{ (fixed mean)}
\end{equation}
\begin{eqnarray} \Label{1site-var}
\phi_3 =  \sum_{n} |c_{n}|^2 \, n^{2} - \overline{n^{2}} = 0  \textrm{ (fixed variance)}
\end{eqnarray}
Moreover, since the constraints are independent of phase, the mean-field (\ref{1site-mf}) is maximized by choosing the set of $c_n$ to be real and positive at fixed $|c_n|$. To deal with the problem of constrained optimisation we apply the method of Lagrange multipliers; the mean-field is maximized when
\begin{equation} \Label{1site-lm}
0 = \frac{\partial \langle X \rangle}{\partial c_n} - \frac{1}{2} \lambda \, \frac{\partial \phi_1}{\partial c_n} + \frac{1}{2} \mu \, \frac{\partial \phi_2}{\partial c_n} + \frac{1}{2} \nu \, \frac{\partial \phi_3}{\partial c_n}
\end{equation}
Note the scalar factors that appear before the Lagrange multipliers ($\lambda$, $\mu$ and $\nu$) are included for convenience. Equation (\ref{1site-lm}) along with equations (\ref{1site-mf}) and (\ref{1site-norm})--(\ref{1site-var}) then lead to
\begin{equation}\Label{1site-eigproblem}
\sqrt{n} \, c_{n-1} + \sqrt{n\!+\!1} \, c_{n+1} + (\mu \, n + \nu \, n^2) \, c_n = \lambda \, c_n
\end{equation}
This is a Hermitian eigenvalue problem which has only one eigenvector solution with the set $c_n$ all positive, since the solutions form an orthonormal basis (solutions with some $c_n$ vanishing may cause ambiguities in principle however).

\begin{figure}[h]
	\begin{center}
		\includegraphics[width=0.50\textwidth]{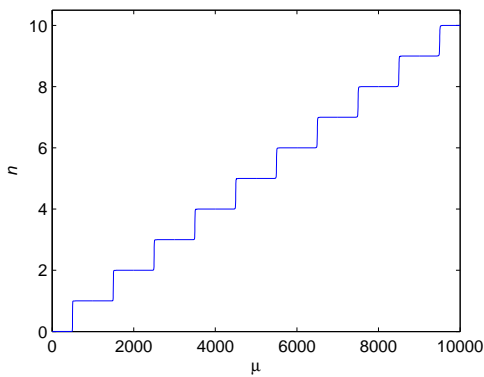}
	\end{center}
	\caption{Mean number occupation as a function of $\mu$ at $\nu = -500$ for the one site model. The Lagrange multiplier $\mu$ plays the role of \pj{(but is not equal to)} the chemical potential; the horizontal regions are indicative of the system near the incompressible Mott insulator phase where $\partial \nbar / \partial \mu = 0$.}
	\Label{fig:one_site_nbar_vs_mu}
\end{figure}

\subsection{Numerical Solutions}

Numerical solutions of (\ref{1site-eigproblem}) determine the maximum mean-field in terms of $\var$ and $\nbar$. By choosing a fixed $Z/u$ and $\nbar$ the calculation of the ground state energy then amounts to a direct minimization of (\ref{1site-E}) with respect to the variance $\var$. 

We outline the procedure as follows. First, for an appropriate range of the Lagrange multipliers $(\mu,\,\nu)$ the eigenvector solutions of (\ref{1site-eigproblem}) are used to calculate corresponding values of $\nbar$, $\var$ and the maximum mean-field $\langle X \rangle$. A fixed value of $\nu$ determines a $\nbar(\mu)$ curve (see for example Fig.\ref{fig:one_site_nbar_vs_mu}) which is inverted using linear interpolation - with an additional optimization step - to give $\mu$ on a uniform set of $\nbar$ points. That is, each value of $\nbar$ determines a curve in $\mu$ and $\nu$ from which the dependence of $\langle X \rangle$ on $\var$ is determined from solutions to (\ref{1site-eigproblem}). Since this relationship is determined on a finite set of points, it is necessary in practice to use cubic interpolation to determine $\langle X \rangle$ for arbitrary $\var$.

\begin{figure}[h]
\includegraphics[width=13cm]{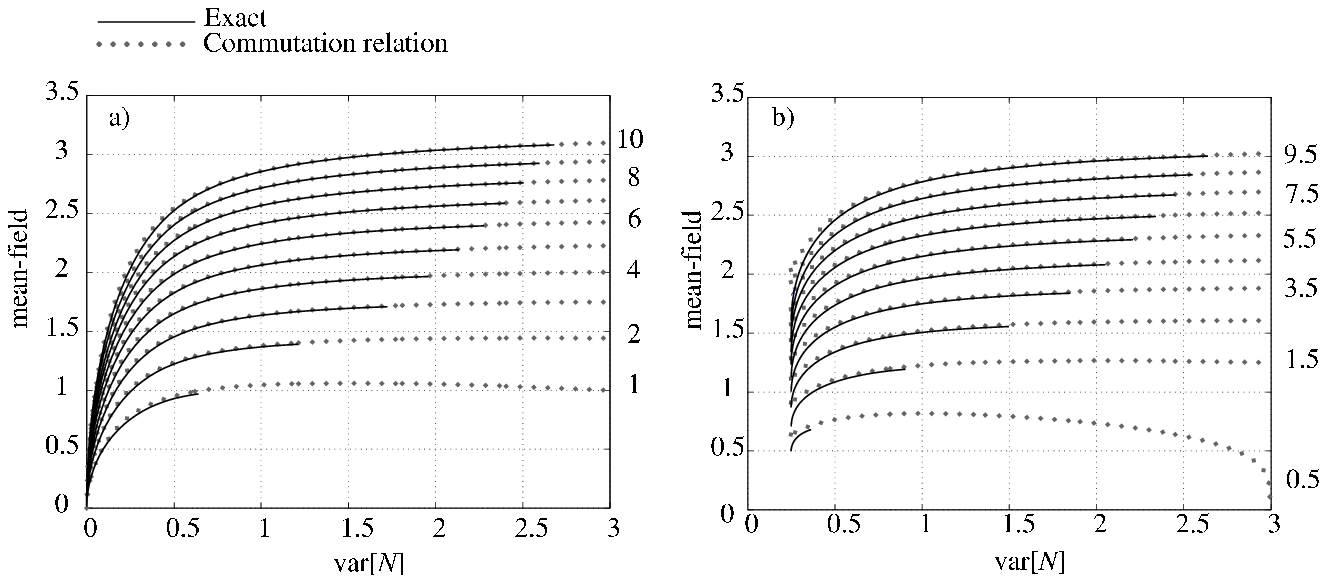}
\caption{Mean-field versus variance for commensurate (a) and incommensurate (b) mean site occupations 
for the one site case.}
\label{fig:X-max-compare}
\end{figure}

Calculations of $\langle X \rangle$ as a function of $\var$ are shown in Fig.\ref{fig:X-max-compare} for commensurate and incommensurate mean site occupations; the corresponding curves which arise from the uncertainty relation (\ref{con10b}) are shown to be very similar. Note that for commensurate mean occupations the mean-field and variance simultaneously approach zero, corresponding to the Mott insulator phase (a pure number state). In contrast, in the incommensurate case neither the mean-field nor variance approach zero; even in the strong lattice regime there is a non-zero superfluid component, corresponding to a superposition of two number states. Using these curves, the energy (\ref{1site-E}) can be minimized with respect to $\var$. \pj{This leads to ground state results that are the same numerically as those of the self-consistent mean-field approach shown in Fig.\ref{fig:exact}.}

\pj{Although the difference in the two sets of curves shown in Fig.\ref{fig:X-max-compare} appears slight, it is entirely responsible for the notable difference between the exact one site calculation (Fig.\ref{fig:exact}) and the approximate phase-space calculation (Fig.\ref{1site-qgs}). In the incommensurate case, the fact that the exact curves drop rapidly as $\var \rightarrow 1/4$ is responsible for the sloping behaviour of the mean-field arches shown in Fig.\ref{fig:exact} as opposed to the level behaviour shown in Fig.\ref{1site-qgs}.}

Numerical calculations were performed using the MATLAB software package. The state space was truncated with $n \leq 40$ as this provided a good trade-off between computational efficiency and accuracy (ie. this truncation introduced negligible error to solutions in the region of interest $\nbar \leq 10$). Note, to cover a sufficient $\nbar$ and $\var$ range, $(\mu, \nu)$ were chosen from the triangular region with $-800 \leq \nu \leq -0.01$ and $0 \leq \mu \leq -40\,\nu$. The minimization procedure used the MATLAB {\em fminbnd} (bounded minimization) function, with a lower bound for the number variance determined by the fractionality constraint (\ref{con4}).


\subsection{Calculating the phase transition boundary}
\pj{To check the validity of the one site formulation near the transition point, it is useful to calculate the 
the position of the phase boundary between the superfluid and Mott insulator phases. This can be done approximately 
by a perturbation method where the ground state energy is determined by states near the transition point.}
We consider a normalised state of commensurate occupation $\nbar$ near the transition point
\pj{
\begin{equation}
|s \rangle = \sqrt{1-2\lambda} \, |\nbar\rangle + \sqrt{\lambda} \, 
(e^{i \theta_1} |\nbar-1\rangle + e^{i \theta_2} |\nbar+1\rangle)
\end{equation}
where $\lambda$ is a real and small;} $\lambda=0$ corresponds to the Mott insulating phase with a commensurate occupation $\nbar$. States with $\lambda \geq 0$ characterize the superfluid phase where the mean-field is non-zero. The corresponding one site energy (\ref{1site-E}) is given by
\begin{eqnarray}
\overline{E} &=& -\lambda(1-2\lambda)\,(\sqrt{\nbar} \cos{\theta_1} + \sqrt{\nbar+1} \cos{\theta_2})^2 \nonumber \\
&& + \overline{u} (2+2\lambda+\nbar^2+3\nbar)
\end{eqnarray}
Note that in scaling by the factor $1/zt$ this form of the energy is dimensionless and we have defined the ratio $\overline{u} \equiv u/zt$ for the relative interaction strength. The ground state is determined by the minimum energy with respect to variations in the free parameters: $\lambda$, $\theta_1$ and $\theta_2$. \pj{Clearly this occurs for $\theta_1 = \theta_2 = 0$ (the result $\theta_1 = \theta_2 = \pi$ gives the same ground state energy and is equivalent
to a change in the sign of the mean-field) and when}
\pj{
\begin{equation}
\frac{\partial{\overline{E}}}{\partial{\lambda}} = 0 = -(1-4\lambda)(\sqrt{\nbar} \cos{\theta_1}
+\sqrt{\nbar+1} \cos{\theta_2})^2 + 2 \overline{u}
\end{equation}
}
The transition to a Mott insulator phase occurs in the limit $\lambda \rightarrow 0$ when the mean-field goes to zero. Applying this condition, the critial point for the transition is then given by
\begin{equation}\Label{1site-transition}
\overline{u}_c = \half (\sqrt{\nbar} + \sqrt{\nbar+1})^2
\end{equation}
For values of the relative interaction strength above the critical value $\overline{u} > \overline{u}_c$ the ground state solution saturates the bound at $\lambda = 0$ and the system remains in the Mott insulator phase.

\subsubsection*{Comparison to other work}

In order to compare our results with those reported elsewhere we define the parameter 
\begin{equation}\label{defn_criticalpt}
\overline{U}_c \equiv 2 (u/zt)_c
\end{equation}
\pj{to describe the relative interaction strength at the critical point. The factor of 2 is included because the on-site interaction term $u$ in the Hamiltonian (\ref{bh1}) is equivalent to the term $U/2$ which has been used elsewhere (see \cite{Oosten2001} and \cite{Zwerger2003} for example).}

Equation (\ref{1site-transition}) then becomes $\overline{U}_c = (\sqrt{\nbar} + \sqrt{\nbar+1})^2$, an expression also found elsewhere \cite{Oosten2001, Sheshadri93}; this yields the following values for the transition: $\Uc = 5.83$ for $\nbar = 1$, $\Uc = 9.90$ for $\nbar = 2$, $\Uc = 13.93$ for $\nbar = 3$.

\supersection{PART II: TWO SITE FORMULATION}


The one site formulation necessarily neglects correlations between neighbouring sites in the mean-field approximation. We can improve on this by using a Hamiltonian that explicity includes hopping between two adjacent sites while still treating interactions with other neighbouring sites with the mean-field approximation. In particular consider two adjacent sites labelled by $1$ and $2$ corresponding to sites $i$ and $i+1$ respectively as shown in Fig. \ref{fig:BH_diagrams}(b). Following the one site case (\ref{bh1}) we can write the Bose-Hubbard Hamiltonian \pj{in a two site mean-field approximation as}
\begin{eqnarray}\Label{2site-bh}
H_{\mbox{\scriptsize two site}} = && -t \left[ a_1 a_2^{\dagger} + a_2 a_1^{\dagger} + (2d\!-\!1)\left(\half {\cal E} (a_1 + a_1^{\dagger}) + \half {\cal E} (a_2 + a_2^{\dagger})\right) \right] \nonumber \\ && + u \, (a_1 a_1 a_1^{\dagger} a_1^{\dagger} + a_2 a_2 a_2^{\dagger} a_2^{\dagger})
\end{eqnarray}
for a homogenous lattice of dimension $d$ where ${\cal E} = \langle a_k \rangle$ (for $k \neq 1,\, 2$) is the mean-field representing the interaction with nearest neighbours, which as with the one site model, by a choice of phase is taken as real. 

\section{Q-function parameterization on two sites}\label{sec-qfunc2s}
In this case the Q-function parameterization becomes
\begin{equation}
a_1 \rightarrow \alpha_1 = (v_1 + \delta_1) e^{i\theta_1}
\end{equation}
\begin{equation}
a_2 \rightarrow \alpha_2 = (v_2 + \delta_2) e^{i\theta_2}
\end{equation}
We define the intersite correlation operators
\begin{eqnarray}
b \equiv a_1 a_2^{\dagger}; \qquad
b^{\dagger} \equiv a_2 a_1^{\dagger}
\end{eqnarray}
The corresponding moments are
\begin{equation}
\langle b \rangle = \langle b^{\dagger} \rangle = (v_1 v_2 + w)g
\end{equation}
where we have set
\begin{equation}
w \equiv \langle \delta_1 \delta_2 \rangle 
\end{equation}
\begin{equation}
g \equiv \langle e^{i (\theta_1 - \theta_2)} \rangle
\end{equation}
and assumed that the phase and number correlations are independent; although this assumption is possibly not strictly valid, without it the formalism becomes significantly more complicated. 

The two site Hamiltonian (\ref{2site-bh}) leads to the expression for the average energy
\begin{eqnarray}\label{ps_2s_energy}
E & = &  \langle H_{\mbox{\scriptsize two site}} \rangle \nonumber \\ 
& = & -\overline{t} \left [ 2(v_1 v_2 + w)g + (2d-1)(v_1^2 c_1^2 + v_2^2 c_2^2) \right ] \nonumber \\
& & + 3 n_1^2 - 2 v_1^4 + 3 n_2^2 - 2 v_2^4 
\end{eqnarray}
which has been normalised by dividing through by $u$ and defining the relative interaction strength $\overline{t} \equiv t/u$.

\subsection{Constraints}

Following the one site case, to find the ground state solution, the energy (\ref{ps_2s_energy}) is minimized subject to  constraints arising from any applicable uncertainty relations. In particular, in addition to the constraints (\ref{con5}) and (\ref{con10b}) that have already been introduced for the one site statistics, we can derive further constraints that account for the two site statistics by considering commutation relations between bilinear operators.

\subsubsection*{Variances and covariance}
In deriving the necessary constraints for the two site problem, we first consider the following operators
\begin{equation}
N \equiv \frac{N_1 + N_2}{2}
\end{equation}
\begin{equation}
M \equiv \frac{N_1 - N_2}{2}
\end{equation}
The corresponding variances for each of these operators are given by 
\begin{equation}\label{ps_2s_varN}
\textrm{var}[N] = 1/4(\textrm{var}[N_1] + \textrm{var}[N_2] + 2 \, \textrm{cov}(N_1, N_2))
\end{equation}
\begin{equation}\label{ps_2s_varM}
\textrm{var}[M] = 1/4(\textrm{var}[N_1] + \textrm{var}[N_2] - 2 \, \textrm{cov}(N_1, N_2))
\end{equation}
where the number covariance function is given by
\begin{eqnarray} 
\textrm{cov}(N_1, N_2) & \equiv & \langle N_1 N_2 \rangle - \langle N_1 \rangle \langle N_2 \rangle \nonumber \\
& = & 2(2 v_1 v_2 w + w^2)
\end{eqnarray}
with the one site variances, $\textrm{var}[N_1]$ and $ \textrm{var}[N_2]$, given analogously to (\ref{con3}).

\subsubsection*{Commutation relations}

We define the following operators
\begin{equation}\label{ps2s-corr}
V \equiv \frac{b + b^{\dagger}}{2}
\end{equation}
\begin{equation}\label{ps2s-anticorr}
W \equiv \frac{b - b^{\dagger}}{2 i}
\end{equation}
$\langle V \rangle$ is the {\em intersite correlation} and gives a measure of particle exchange, or tunnelling, between adjacent sites. Eqs. (\ref{ps2s-corr}) and (\ref{ps2s-anticorr}) have the following commutation relations
\begin{equation}\Label{ps_2s_comm1}
[V, M] = i W
\end{equation}
\begin{equation}\Label{ps_2s_comm2}
[M, W] = i V
\end{equation}
\begin{equation}\Label{ps_2s_comm3}
[W, V] = i M
\end{equation}

\subsubsection*{Quantum mechanical constraints}
Using (\ref{ps_2s_comm1})--(\ref{ps_2s_comm3}) and noting that $\langle W \rangle = 0$, we can write the corresponding uncertainty relations
\begin{equation}\label{ps_2s_constrA}
\left ( \langle V^2 \rangle - {\langle V \rangle}^2 \right ) \textrm{var}[M] \geq
 \quarter {\langle [\delta V, \delta M]_{+} \rangle}^2
\end{equation}
\begin{equation}\label{ps_2s_constrB}
\textrm{var}[M] \langle W^2 \rangle  \geq \quarter {\langle V \rangle}^2 
+ \quarter {\langle [\delta M, \delta W]_{+} \rangle}^2
\end{equation}
\begin{equation}\label{ps_2s_constrC}
\left ( \langle V^2 \rangle - {\langle V \rangle}^2 \right ) \langle W^2 \rangle \geq \quarter {\langle M \rangle}^2 
+ \quarter {\langle [\delta V, \delta W]_{+} \rangle}^2
\end{equation}
with
\begin{equation}
\langle V^2 \rangle = \frac{1 + g^4}{2} \left ( n_1 n_2 + \textrm{cov}(N_1, N_2) \right ) - \frac{n_1 + n_2}{4}
\end{equation}
\begin{equation}
\langle V \rangle = (v_1 v_2 + w)g\
\end{equation}
\begin{equation}
\langle W^2 \rangle = \frac{1 - g^4}{2} \left ( n_1 n_2 + \textrm{cov}(N_1, N_2) \right ) - \frac{n_1 + n_2}{4}
\end{equation}

\begin{figure}[!htp]
    \begin{center}
		\mbox{
		\subfigure[Mean-field, d=1]{\includegraphics[width=0.50\textwidth]{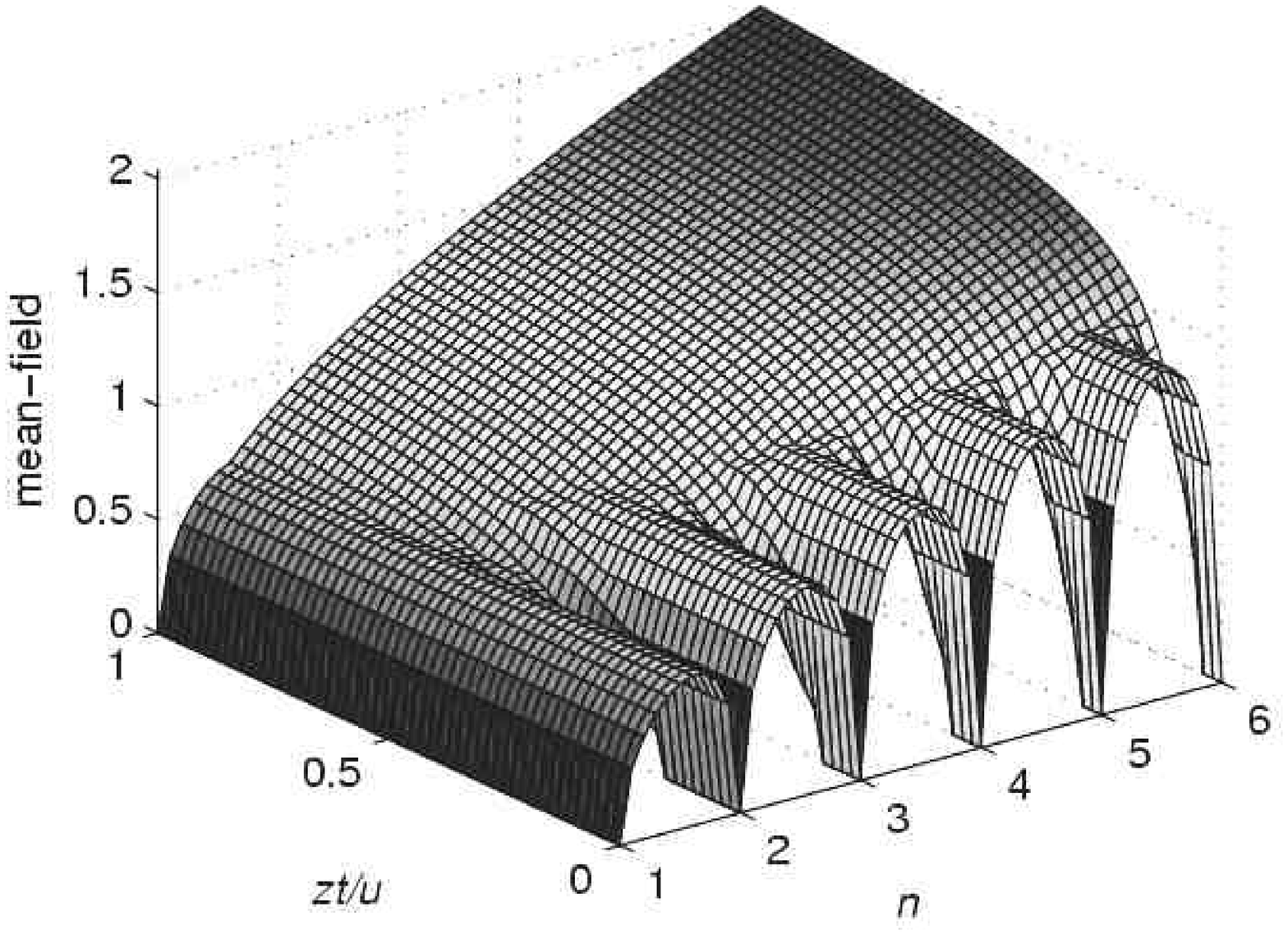}}
		\hspace{3mm}
		\subfigure[Intersite correlation, d=1]{\includegraphics[width=0.50\textwidth]{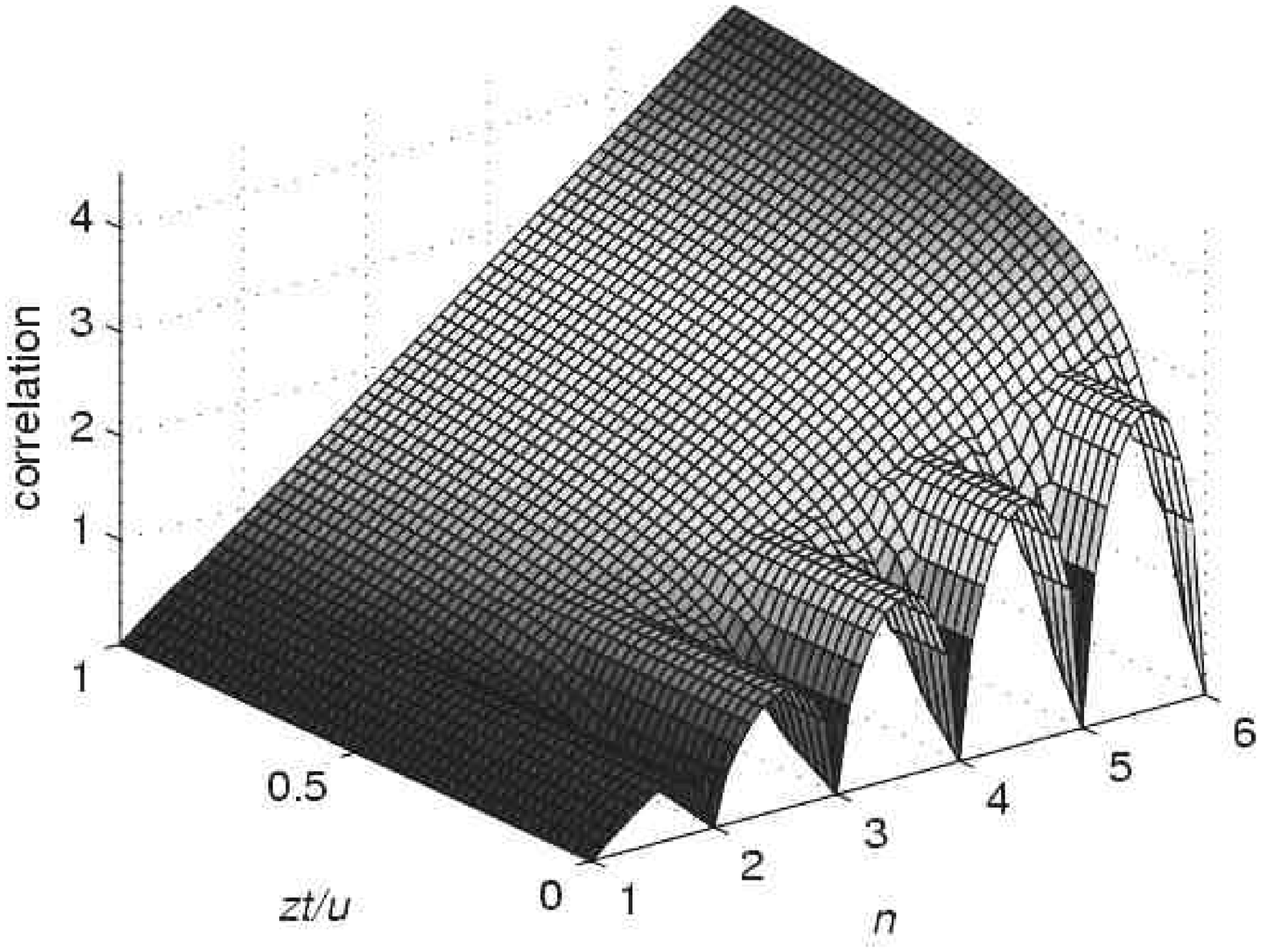}}
		}
		\mbox{
		\subfigure[Mean-field, d=2]{\includegraphics[width=0.50\textwidth]{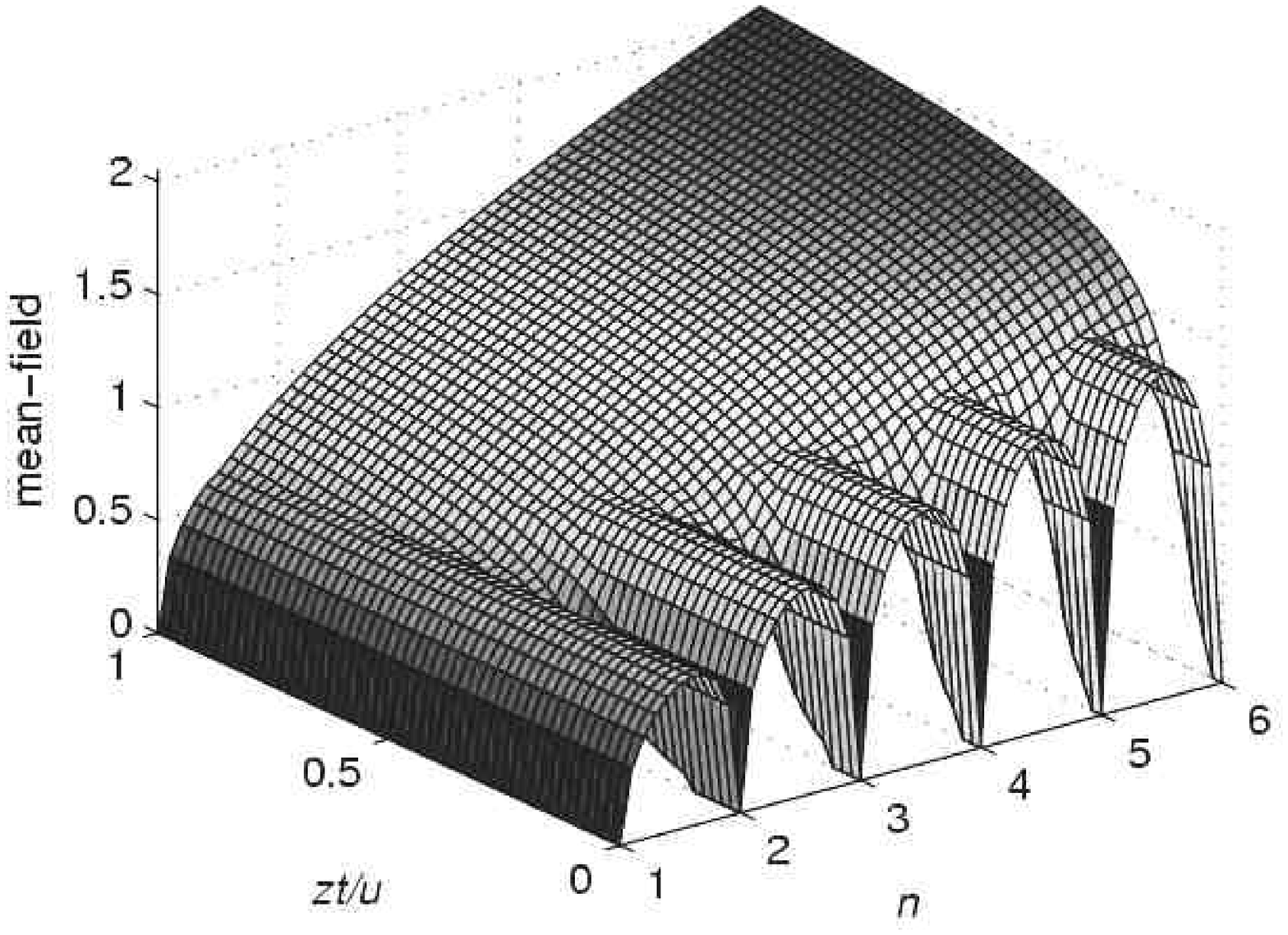}}
		\hspace{3mm}
		\subfigure[Intersite Correlation, d=2]{\includegraphics[width=0.50\textwidth]{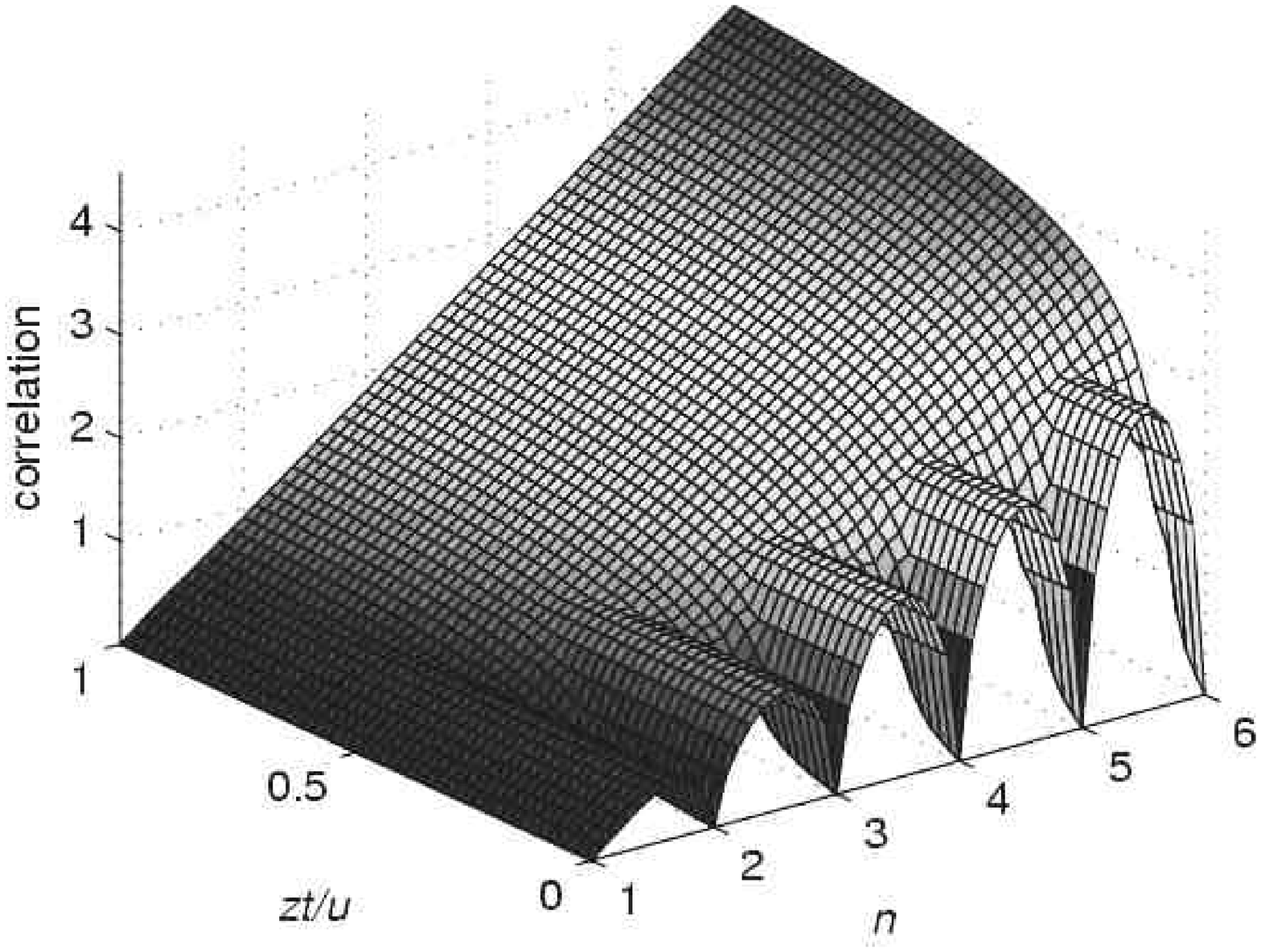}}
		}
		\mbox{
		\subfigure[Mean-field, d=3]{\includegraphics[width=0.50\textwidth]{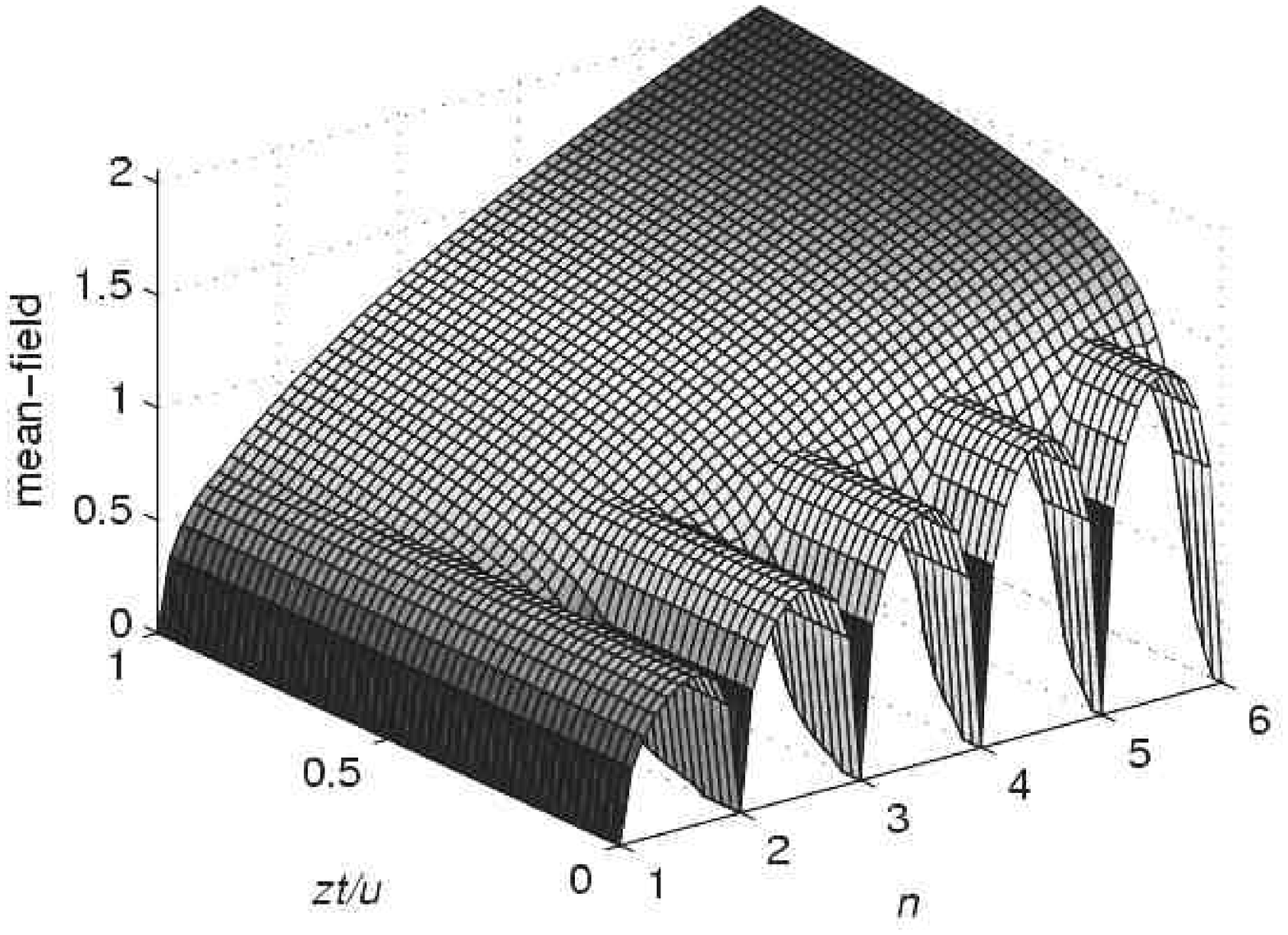}}
		\hspace{3mm}
		\subfigure[Intersite Correlation, d=3]{\includegraphics[width=0.50\textwidth]{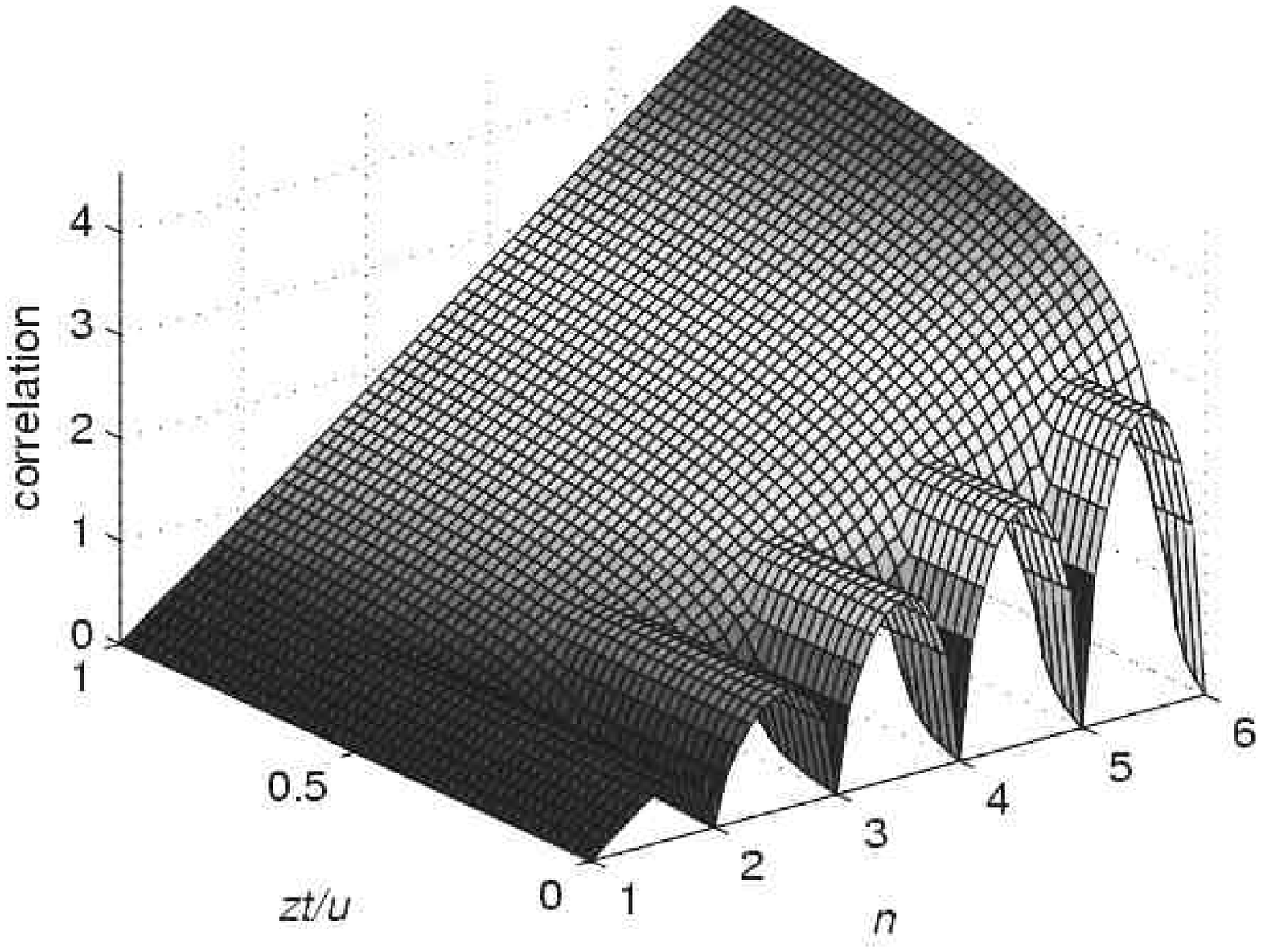}}
		}
		\caption{Ground state phase diagram for two site Bose-Hubbard model using the Q-function representation in the mean-field approximation. The mean-field and intersite correlations are shown as a function of the mean (anti-normal ordered) occupation $n$ and the relative interaction strength $zt/u$.} 
		\label{fig:ps-2s-phasediagram}
		\end{center}
\end{figure}

\subsubsection*{Applying homogeneity}
These constraints further simplify when homogeneity is assumed. In particular, if we assume the ground state is symmetric to the $1 \leftrightarrow 2$ interchange, we can set
\begin{equation}
v_1 = v_2 = v
\end{equation}
\begin{equation}
c_1 = c_2 = c
\end{equation}
\begin{equation}
n_1 = n_2 = n
\end{equation}
Clearly then 
\begin{equation}
\langle M \rangle = 0
\end{equation}
Moreover, under the above interchange the following moments are equal
\begin{equation} 
\langle b^{\dagger} N_2 \rangle = \langle b N_1 \rangle
\end{equation}
\begin{equation}
\langle b N_2 \rangle = \langle b^{\dagger} N_1 \rangle
\end{equation}
It follows that the anti-commutators in equations (\ref{ps_2s_constrA})--(\ref{ps_2s_constrC}) are given by
\begin{equation}
[\delta V, \delta M]_{+} = 0
\end{equation}
\begin{equation}
[\delta M, \delta W]_{+} = 0
\end{equation}
\begin{equation}
[\delta V, \delta W]_{+} = 0
\end{equation}

\subsubsection*{Two site phase-number uncertainty relation}
Similarly, the commutation relation ($i = 1$, $2$)
\begin{equation}
[N, Y_i] = \half i X_i
\end{equation}
leads to the constraint
\begin{equation}\label{ps_2s_constrD}
\textrm{var}[N] \langle Y_i^2 \rangle \geq \frac{1}{16} {\langle X_i \rangle}^2
\end{equation}

\subsubsection*{Two site fractionality constraint}
A lower limit on the two site variance $\textrm{var}[N]$ can also be given by considering two site states with minimum variance.
Such a state has the form
\begin{eqnarray}\label{2site-lowvarstate}
|s \rangle & = & \sqrt{\alpha} \, | [n], [n] \rangle + \sqrt{\beta/2} \, (| [n], [n]+1 \rangle + | [n]+1, [n] \rangle) 
\nonumber \\ && + \sqrt{\gamma} \, | [n]+1, [n]+1 \rangle
\end{eqnarray}
where $[n] = n - \delta n$ is the integer part of $n$, $\delta n$ being the fractional part. 
We then have
\begin{equation}
\textrm{var}[N_i] = \frac{\beta}{2} \left ( 1 - \frac{\beta}{2} \right)
\end{equation}
\begin{equation}
\textrm{cov}(N_1, N_2) = -{\beta}^2/4
\end{equation}
so that we can write the minimum variance
\begin{equation}
\textrm{var}[N]_{min} = \quarter \beta (1 - \beta)
\end{equation}
Using (\ref{ps_2s_varN}) and applying homogeneity this leads to the fractionality constraint for the two site variance
\begin{equation}\label{ps_2s_constr_frac}
n^2 - \half n - v^4 + 2 v^2 w + w^2 \geq \quarter \beta (1 - \beta)
\end{equation}
We note the following two cases
\begin{enumerate}
\item{$0 \leq \delta n \leq 0.5$ :}
In this case $\gamma = 0$ and $\beta = 2 \delta n$.
\item{$0.5 \leq \delta n \leq 1$ :}
In this case $\alpha = 0$ and $\beta = 2(1 - \delta n)$.
\end{enumerate}

\subsection{The ground state solution}
To find the ground state solution we assume homogeneity and minimize the two site energy (\ref{ps_2s_energy}) at a given occupation $n$ and relative interaction strength $\overline{t}$ with respect to the free parameters $v$, $c$, $w$ and $g$. This is done subject to the constraints (\ref{con5}), (\ref{con10b}), (\ref{ps_2s_constrA})--(\ref{ps_2s_constrC}), (\ref{ps_2s_constrD}) and (\ref{ps_2s_constr_frac}). The parametrization requires that
\begin{equation}
v \geq 0
\end{equation} 
Additionally, by noting the variables $c$ and $g$ both take the form $\langle e^{i \phi} \rangle$, the following bounds apply
\begin{equation}
0 \leq c \leq 1
\end{equation}
\begin{equation}
0 \leq g \leq 1
\end{equation}

\subsection{Results}
The results of this calculation are shown in Fig.\ref{fig:ps-2s-phasediagram}. The solutions reproduce the general features of the Bose-Hubbard model in the superfluid limit and for incommensurate occupations in the limit of zero tunneling. However, in the case of a commensurate mean occupation, the Q-function parametrization on two sites does not correctly demonstrate the existence of the Mott insulator phase. Specifically we find that for a three dimensional lattice, the mean-field only vanishes in the limit of $zt/u \rightarrow 0$. Moreover, in the one and two dimensional cases, the predicted Mott insulator phase only occurs for very small values of $zt/u$. See sections \ref{sec-qfunc1s}, \ref{sec-exact1s} and \ref{sec-exact2s} for comparison; in particular, the results from section \ref{sec-exact2s} show a well defined Mott phase with a vanishing mean-field.
\pj{Improved agreement can be seen between the intersite correlations calculated by the Q-function representation and the exact numerical results of section \ref{sec-exact2s}. In both cases the intersite correlations vanish in the limit of zero tunneling for commensurate occupations.}

\section{Exact numerical minimization of two site problem}\label{sec-exact2s}

\pj{Following the one site formulation discussed in section \ref{sec-exact1sform}, we apply a two site generalization of the exact numerical calculation using arbitrary two site states. The formulation uses the assumption that the lattice is translationally invariant so that we can apply symmetry for the two sites in the ground state.}

The Hamiltonian (\ref{2site-bh}) leads to an expression for the two site energy
\begin{eqnarray}\Label{2site-E}
\langle E \rangle_{\mbox{\scriptsize two site}} & = & -2t \left[ \langle V \rangle + (2d\!-\!1) \langle X \rangle^2 \right] \nonumber \\ && + 2u \left( 2 + \var + \nbar^2 + 3\,\nbar \right)
\end{eqnarray}
where $V$ is the intersite correlation operator given by (\ref{ps2s-corr}) and 
\begin{equation}\Label{2site-mf}
\hat{X} = \half(a_i + a_i^\dagger)
\end{equation}
is the mean-field operator, which by homogeneity, does not depend on the site index.


Note that in the limit that $\langle V \rangle \! \to \! {\langle X \rangle}^2$ the two site energy tends to twice the one site energy. This corresponds to the case where the two site state is factorizable $| \psi_{12} \rangle \rightarrow | \psi_1 \rangle \, | \psi_2 \rangle$ and the one site behaviour is returned. 

In general, for $n$ and $m$ atoms on sites $1$ and $2$ respectively, the two site quantum state can be written as
\begin{eqnarray}\Label{2site-state}
|s \rangle & = & \sum_{n,m} c_{n,m} |n, m \rangle \nonumber \\
& = & \sum_{N,\,p \neq 0} C_{N,\,p} \left ( \big | \half(N\!+\!p), \half(N\!-\!p) \big\rangle + \big | \half(N\!-\!p), \half(N\!+\!p) \big\rangle \right ) \nonumber \\
&& + \sum_{N,\,p = 0} \, C_{N,\,0} | \half N,\half N \rangle
\end{eqnarray}
where we have introduced the quantum numbers $N = n + m$ and $p = |n-m| \leq N$ to account for the lattice symmetry \begin{math} c_{n,m} = c_{m,n} \end{math} (we should note this symmetry argument can only be applied when the ground state exhibits homogeneity). The correlation function is
\begin{eqnarray}\Label{two_site_corr}
\langle V \rangle & = & \langle a_{i} a^\dagger_{j} \rangle \nonumber \\
& = & \sum_{n,\,m} c_{n,\,m} c_{n+1,\,m-1} \sqrt{m(n+1)} \nonumber \\
& = & \sum_{N,\,p \geq 0} C_{N,\,p} \, C_{N,\,p+2} \sqrt{(N\!-\!p)(N\!+\!p\!+\!2)} \nonumber \\
&& + \ \frac{1}{2} \, \delta_{p,\,1} \, \sum_{N} (C_{N,\,1})^2 (N\!+\!1)
\end{eqnarray}
and the mean-field is
\begin{eqnarray}\Label{two_site_mf}
\langle X \rangle &=& \langle a_{i} \rangle \nonumber \\
& = & \sum_{n,\,m} c_{n,\,m} c_{n+1,\,m} \sqrt{n+1} \nonumber \\
&=& \frac{1}{\sqrt{2}} \sum_{N,\,p>0} C_{N,\,p} \left [ C_{N+1,\,p+1} \sqrt{N\!+\!p\!+\!2} + C_{N+1,\,p-1}  \sqrt{N\!-\!p\!+\!2} \right ] \nonumber \\ 
&& + \frac{1}{\sqrt{2}} \, \delta_{p,\,0} \, \sum_{N} C_{N,\,0} \, C_{N+1,\,1} \sqrt{N\!+\!2}
\end{eqnarray}
From (\ref{2site-E}) it is evident that, for a given $\nbar$, $d$ and $t/u$, the ground state solution can be found by finding the maximum kinetic (hopping) energy 
\begin{equation}\Label{2site-hoppingenergy}
F = \langle V \rangle + (2d\!-\!1)\langle X_{i} \rangle^2
\end{equation} subject to the constraints
\begin{enumerate}
	\item normalisation, $1=\langle s | s \rangle$
	\item fixed mean occupation (per site), $\nbar = \half\langle a_1^{\dagger}a_1 + a_2^{\dagger}a_2 \rangle$
	\item fixed variance (per site), $\overline{n^2} = \half\langle a_1^{\dagger}a_1 a_1^{\dagger}a_1 + a_2^{\dagger}a_2 a_2^{\dagger}a_2 \rangle$
\end{enumerate}
In terms of the symmetric state (\ref{2site-state}) these constraints are respectively given by
\begin{equation}\Label{2site-norm}
\phi_1 = \left( 2 \! \sum_{N,p>0} C^2_{N,p} + \sum_{N} C^2_{N,0} \right)\!-1 = 0 \textrm{ (normalisation)}
\end{equation}
\begin{equation}\Label{2site-nbar}
\phi_2 = \left( \sum_{N,p>0} C^2_{N,p} N + \frac{1}{2} \sum_{N} C^2_{N,0} N \right)\! - \nbar = 0 \textrm{ (fixed mean)}
\end{equation}
\begin{eqnarray}\Label{2site-var}
\phi_3 = \left( \frac{1}{2} \sum_{N,p>0} C^2_{N,p} (N^2\!+\!p^2) + \frac{1}{4} \sum_{N} C^2_{N,0} N^2 \right)\!- \overline{n^2} = 0 \nonumber \\ \textrm{ (fixed variance)}
\end{eqnarray}
Using the method of Lagrange multipliers, $F$ is then maximized when
\begin{equation}\Label{2site-lm}
0 = \frac{\partial F}{\partial C_{N\!,\,p}} - \frac{1}{2} \lambda \, \frac{\partial \phi_1}{\partial C_{N\!,\,p}} + \mu \, \frac{\partial \phi_2}{\partial C_{N\!,\,p}} + \nu \, \frac{\partial \phi_3}{\partial C_{N\!,\,p}}
\end{equation}

\subsubsection{Vectorizing the problem}
We can write this set of simultaneous differential equations in matrix form when the set of coefficients $C_{N,\,p}$ is expressed as an ordered vector. To truncate the problem space in terms of a maximum two site occupation $N_{\mathrm{max}}$ (with a corresponding $p_{\mathrm{max}}$), it is convenient to use an ordering for the quantum numbers $p$ and $N$ with increasing $p$ (for each $N$ value) within increasing $N$. That is,
\begin{equation}
\hspace{-8mm}
\mathbf{C} = [C_{0,0}, C_{1,1}, C_{2,0}, C_{2,2}, C_{3,1}, C_{3,3}, C_{4,0}, C_{4,2}, C_{4,4} \,\dots\, C_{N_{\mathrm{max}},p_{\mathrm{max}}}]
\end{equation}

\begin{figure}[ht]
\begin{center}
\includegraphics[width=0.7\textwidth]{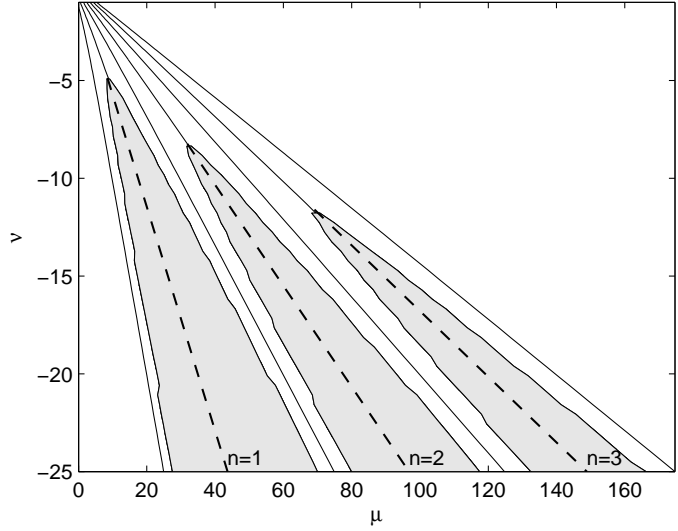}
\caption{A contour plot of the mean occupation $\nbar$ with respect to the Lagrange multipliers $\mu$ and $\nu$ reveals the Mott lobes for commensurate occupations, shown for $\nbar = 1$, $2$, $3$, where the solutions to (\ref{2site-eig}) are degenerate with respect to $\mu$, indicating the incompressibility of the phase. The Mott insulator phase does not exist for incommensurate occupations as shown by the remaining curves at $\nbar = 0.5$, $1.5$, $2.5$ and $3.5$ from left to right in the figure. The dashed lines represent possible (non-degenerate) trajectories for the Mott insulator phase as discussed in section \ref{sec-mott2s}.}
\Label{fig:2site-contours}
\end{center}
\end{figure}

The correlation and mean-field can then be written respectively in terms of the vector quadratic forms
\begin{eqnarray}
\langle V \rangle = \mathbf{C}^T A_V \mathbf{C} \label{2site-quadform-mf}\\
\langle X \rangle = \mathbf{C}^T A_X \mathbf{C} \label{2site-quadform-corr}
\end{eqnarray}
where the construction of matrices $A_V$ and $A_X$ follows directly from equations (\ref{ps2s-corr}) and (\ref{2site-mf}). In particular,

\begin{equation}
\mbox{
  \begin{small} 
  $\displaystyle
A_V = 
  \left(
    \begin{array}{c|c|cc|cc|ccc|c}
      0 & \rr{0} & 0 & \rr{0} & 0 & \rr{0} & 0 & 0 & \rr{0} & \\\cline{1-2}
      0 & 1 & 0 & \rr{0} & 0 & \rr{0} & 0 & 0 & \rr{0} & \\\cline{2-4}
      \rr{0} & 0 & 0 & 2\sqrt{2} & 0 & \rr{0} & 0 & 0 & \rr{0} & \\
      \rr{0} & 0 & 0 & 0 & 0 & \rr{0} & 0 & 0 & \rr{0} & \\\cline{3-6}
      \rr{0} & \rr{0} & 0 & 0 & 2 & 2\sqrt{3} & 0 & 0 & \rr{0} & \\
      \rr{0} & \rr{0} & 0 & 0 & 0 & 0 & 0 & 0 & \rr{0} & \\\cline{5-9}
      \rr{0} & \rr{0} & 0 & \rr{0} & 0 & 0 & 0 & 2\sqrt{6} & 0 & \\
      \rr{0} & \rr{0} & 0 & \rr{0} & 0 & 0 & 0 & 0 & 4 & \\
      \rr{0} & \rr{0} & 0 & \rr{0} & 0 & 0 & 0 & 0 & 0 & \\\cline{7-10}
      \rr{} & \rr{} & \rr{} & \rr{} & \rr{} & \rr{} & \rr{} & \rr{} & & \ddots
    \end{array}
  \right)
  $
  \end{small}}
\end{equation}

and 
\begin{equation}
\mbox{
  \begin{small}
  $\displaystyle
A_X = 
  \left(
    \begin{array}{c|c|cc|cc|ccc|c}
      0 & \rr{1} & 0 & \rr{0} & 0 & \rr{0} & 0 & 0 & \rr{0} & \\\cline{1-2}
      0 & 0 & 1 & \rr{\sqrt{2}} & 0 & \rr{0} & 0 & 0 & \rr{0} & \\\cline{2-4}
      \rr{0} & 0 & 0 & 0 & \sqrt{2} & \rr{0} & 0 & 0 & \rr{0} & \\
      \rr{0} & 0 & 0 & 0 & 1 & \rr{\sqrt{3}} & 0 & 0 & \rr{0} & \\\cline{3-6}
      \rr{0} & \rr{0} & 0 & 0 & 0 & 0 & \sqrt{2} & \sqrt{3} & \rr{0} & \\
      \rr{0} & \rr{0} & 0 & 0 & 0 & 0 & 0 & 1 & \rr{2} & \\\cline{5-9}
      \rr{0} & \rr{0} & 0 & \rr{0} & 0 & 0 & 0 & 0 & 0 & \\
      \rr{0} & \rr{0} & 0 & \rr{0} & 0 & 0 & 0 & 0 & 0 & \\
      \rr{0} & \rr{0} & 0 & \rr{0} & 0 & 0 & 0 & 0 & 0 & \\\cline{7-10}
      \rr{} & \rr{} & \rr{} & \rr{} & \rr{} & \rr{} & \rr{} & \rr{} & & \ddots
    \end{array}
  \right)
  $
  \end{small}}
\end{equation}

The constraints (\ref{2site-norm})--(\ref{2site-var}) can then be written as
\begin{equation}
\phi_1 = \mathbf{C}^T S \mathbf{C} - 1 = 0
\end{equation}
\begin{equation}
\phi_2 = \half \mathbf{C}^T \, S \, D \, \mathbf{C} - \nbar = 0
\end{equation}
\begin{equation}
\phi_3 = \quarter \mathbf{C}^T \, S \, D^{\prime} \, \mathbf{C} - \overline{n^2} = 0
\end{equation}
where $S$, $D$ and $D^{\prime}$ are diagonal matrices with elements given by
\begin{equation}
S_{NN^{\prime}pp^{\prime}} = \delta_{NN^{\prime}} \, \delta_{pp^{\prime}} \,(2-\delta_{p0})
\end{equation}
\begin{equation}
D_{NN^{\prime}pp^{\prime}} = \delta_{NN^{\prime}} \, \delta_{pp^{\prime}} \, N
\end{equation}
\begin{equation}
D^{\prime}_{NN^{\prime}pp^{\prime}} = \delta_{NN^{\prime}} \, \delta_{pp^{\prime}} \, (N^2\!+\!p^2)
\end{equation}
with the usual Kronecker delta function. Using these definitions, equation (\ref{2site-lm}) can then be written in matrix form as  
\begin{eqnarray}\Label{2site-eig}
(S^{-1/2} A S^{-1/2} + S^{1/2}B S^{-1/2}) \mathbf{u} = \lambda \mathbf{u}
\end{eqnarray}
where
\begin{equation}
\mathbf{u} = S^{1/2} \mathbf{C}
\end{equation}
\begin{equation}\Label{2site-eigA}
A = A_V + A_V^T + (2d\!-\!1) \langle X \rangle (A_X + A_X^T)
\end{equation}
\begin{equation}
B = \mu \, D + \half \, \nu \, D^{\prime}
\end{equation}

\begin{figure}[t]
    \begin{center}
		\mbox{
		\subfigure[Commensurate]{\includegraphics[width=0.50\textwidth]{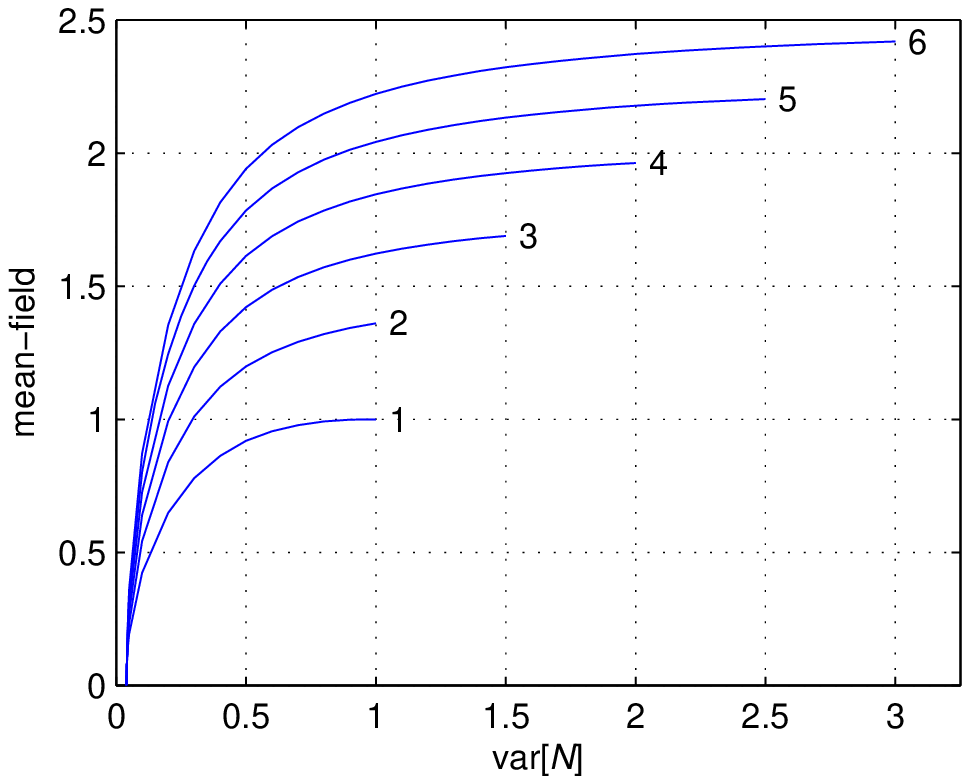}}
		\hspace{3mm}
		\subfigure[Incommensurate]{\includegraphics[width=0.50\textwidth]{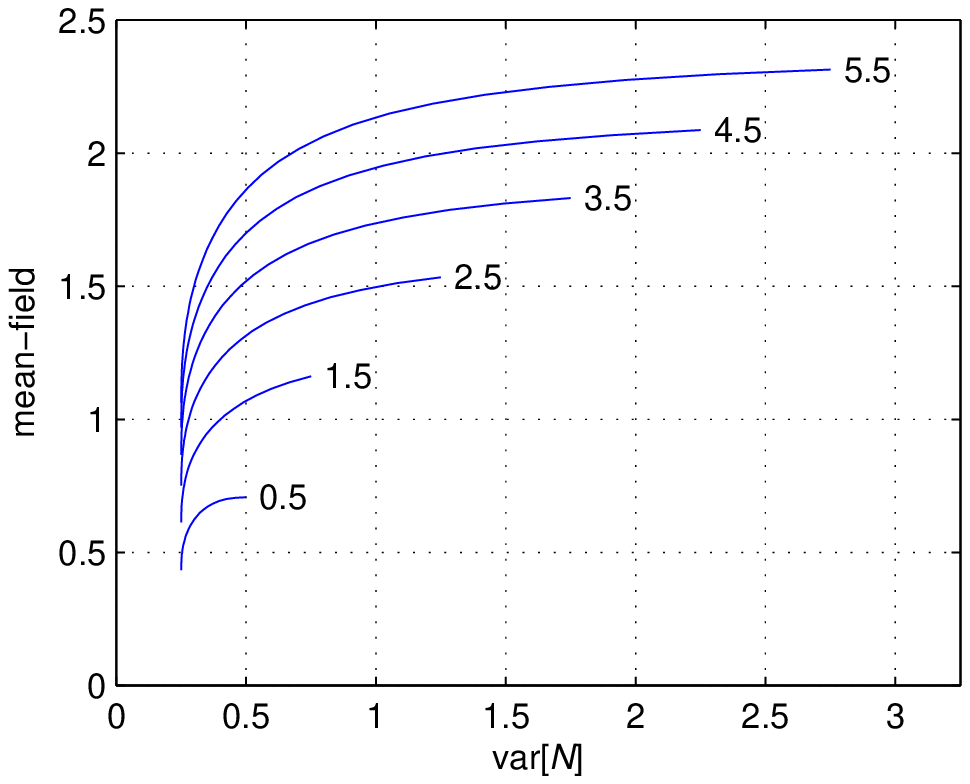}}
		}
		\caption{Mean-field as a function of number variance for commensurate (a) and incommensurate (b) occupations. In contrast to the one site approximation, in the commensurate case the Mott insulator phase with vanishing mean-field can occur for a non-zero number variance.}
		\Label{fig:2site-mfvsvar}
		\end{center}
\end{figure}

\begin{figure}[!htp]
    \begin{center}
		\mbox{
		\subfigure[Mean-field, d=1]{\includegraphics[width=0.50\textwidth]{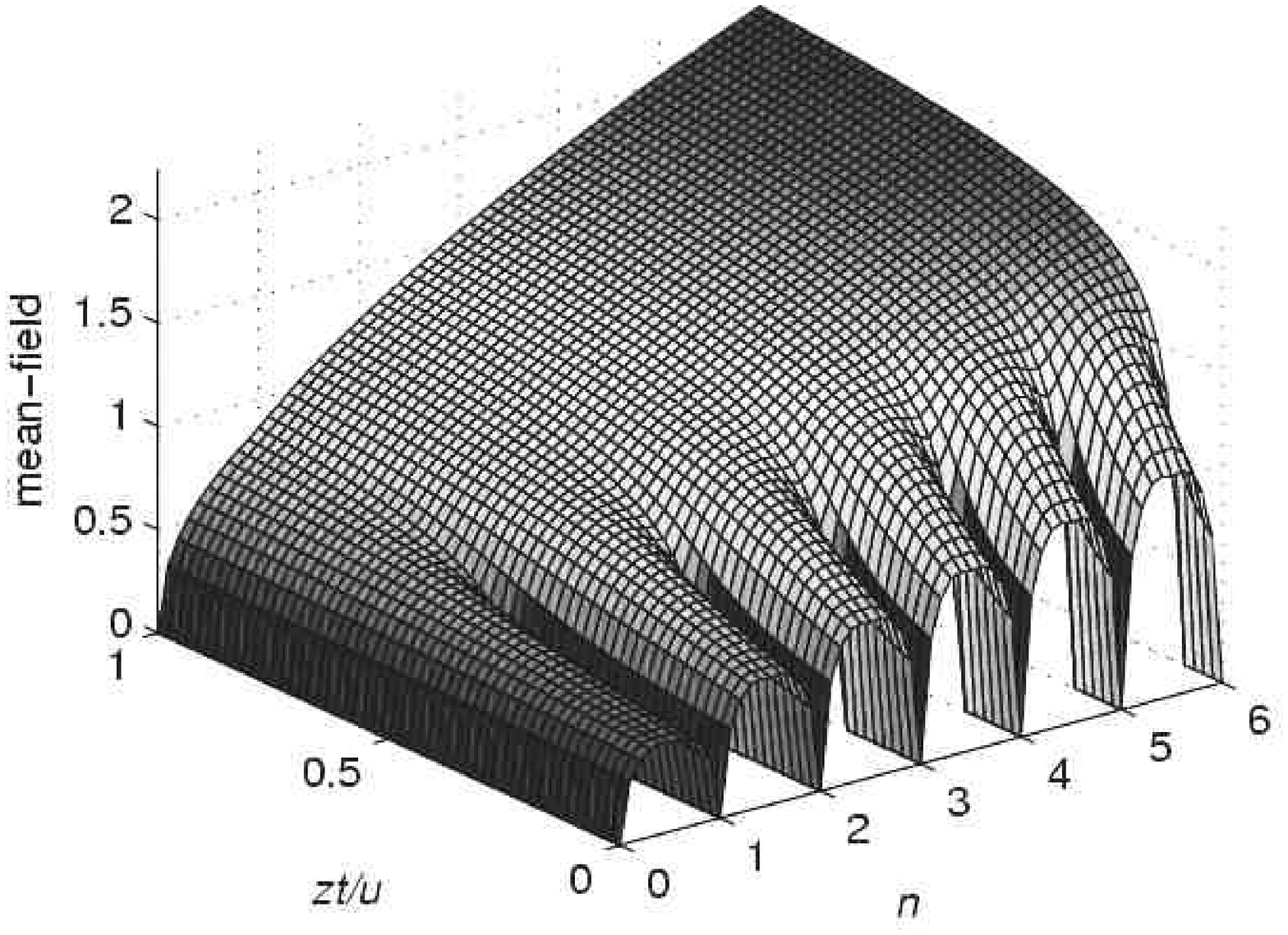}}
		\hspace{3mm}
		\subfigure[Intersite correlation, d=1]{\includegraphics[width=0.50\textwidth]{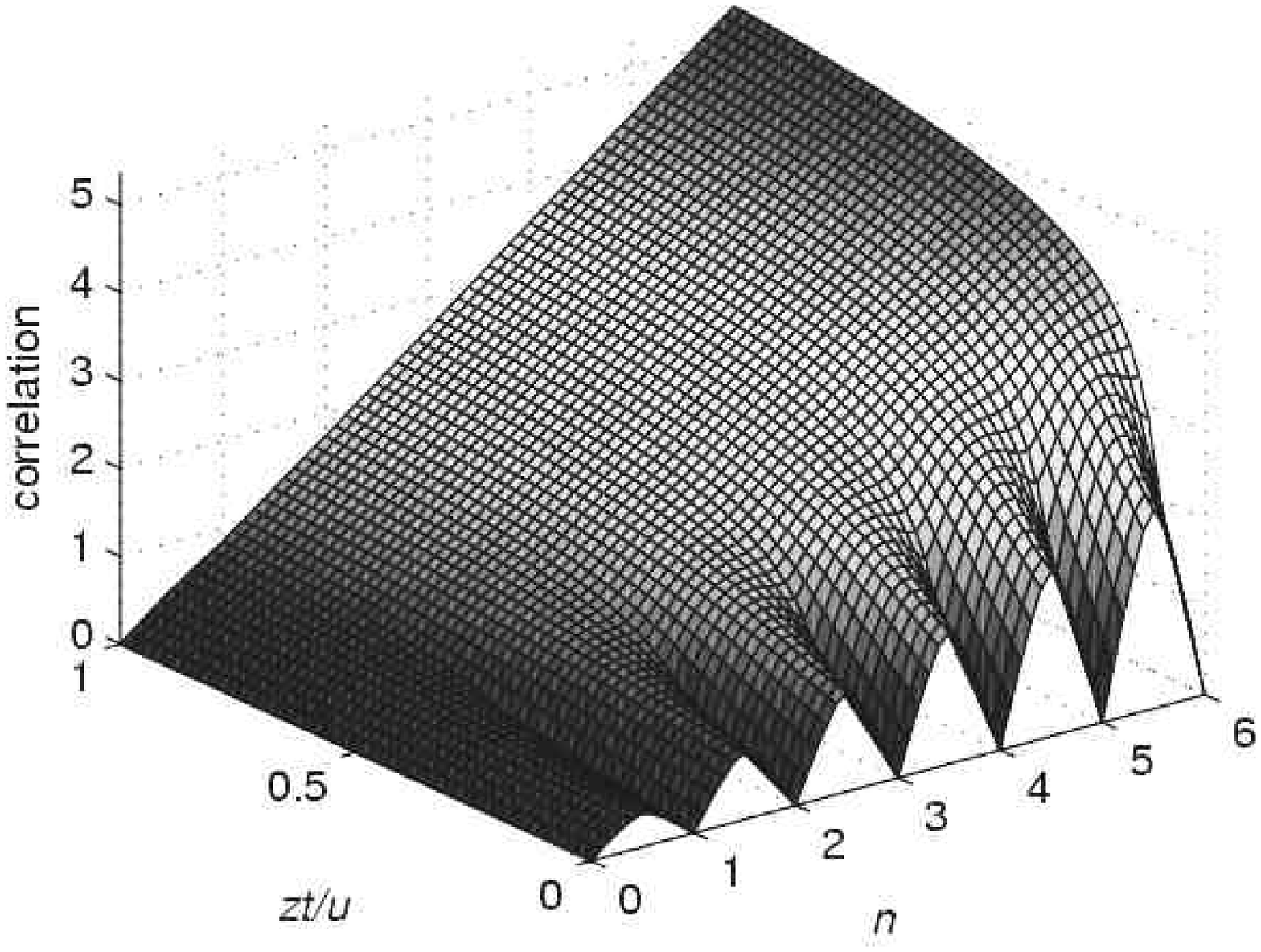}}
		}
		\mbox{
		\subfigure[Mean-field, d=2]{\includegraphics[width=0.50\textwidth]{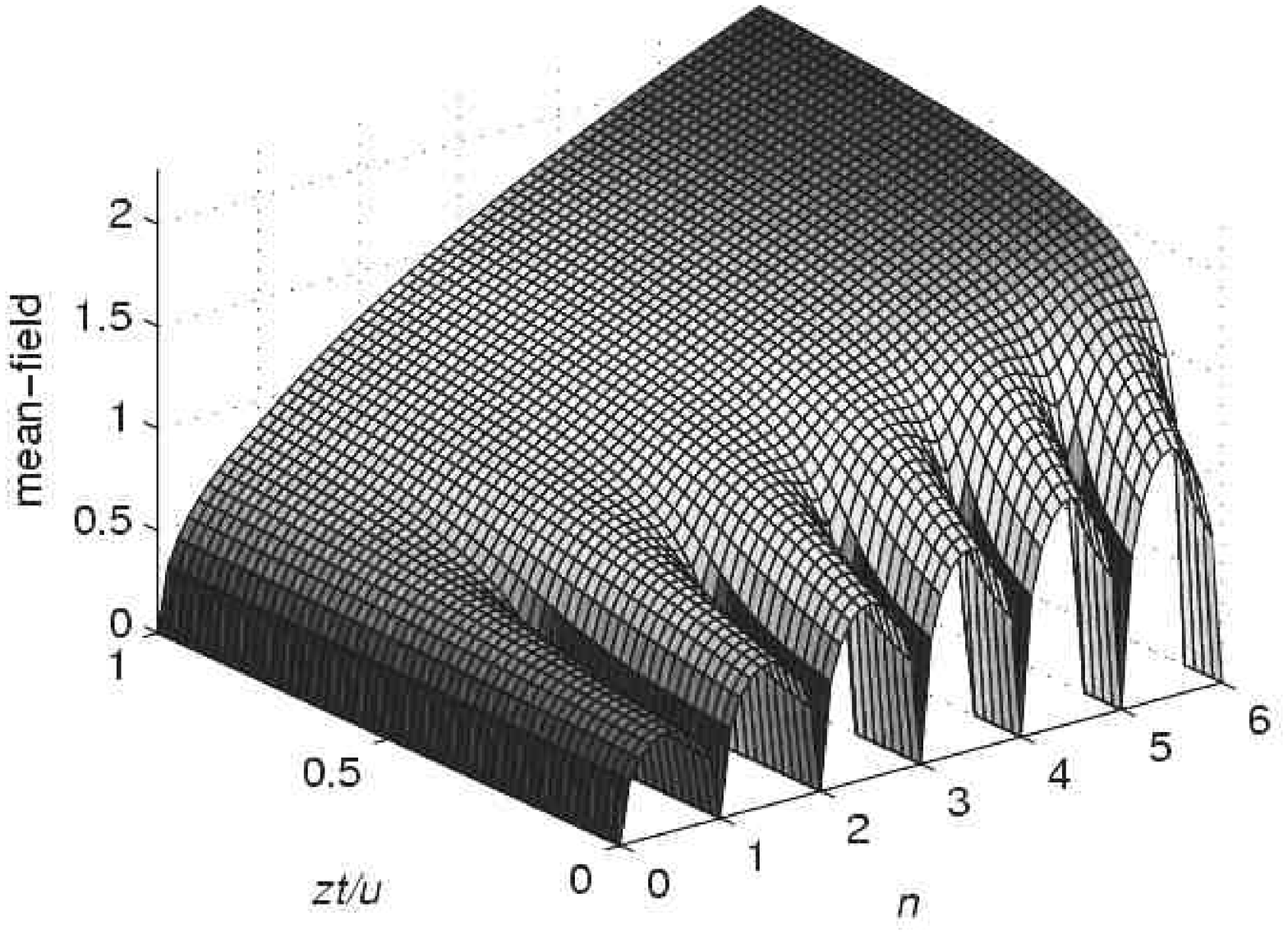}}
		\hspace{3mm}
		\subfigure[Intersite correlation, d=2]{\includegraphics[width=0.50\textwidth]{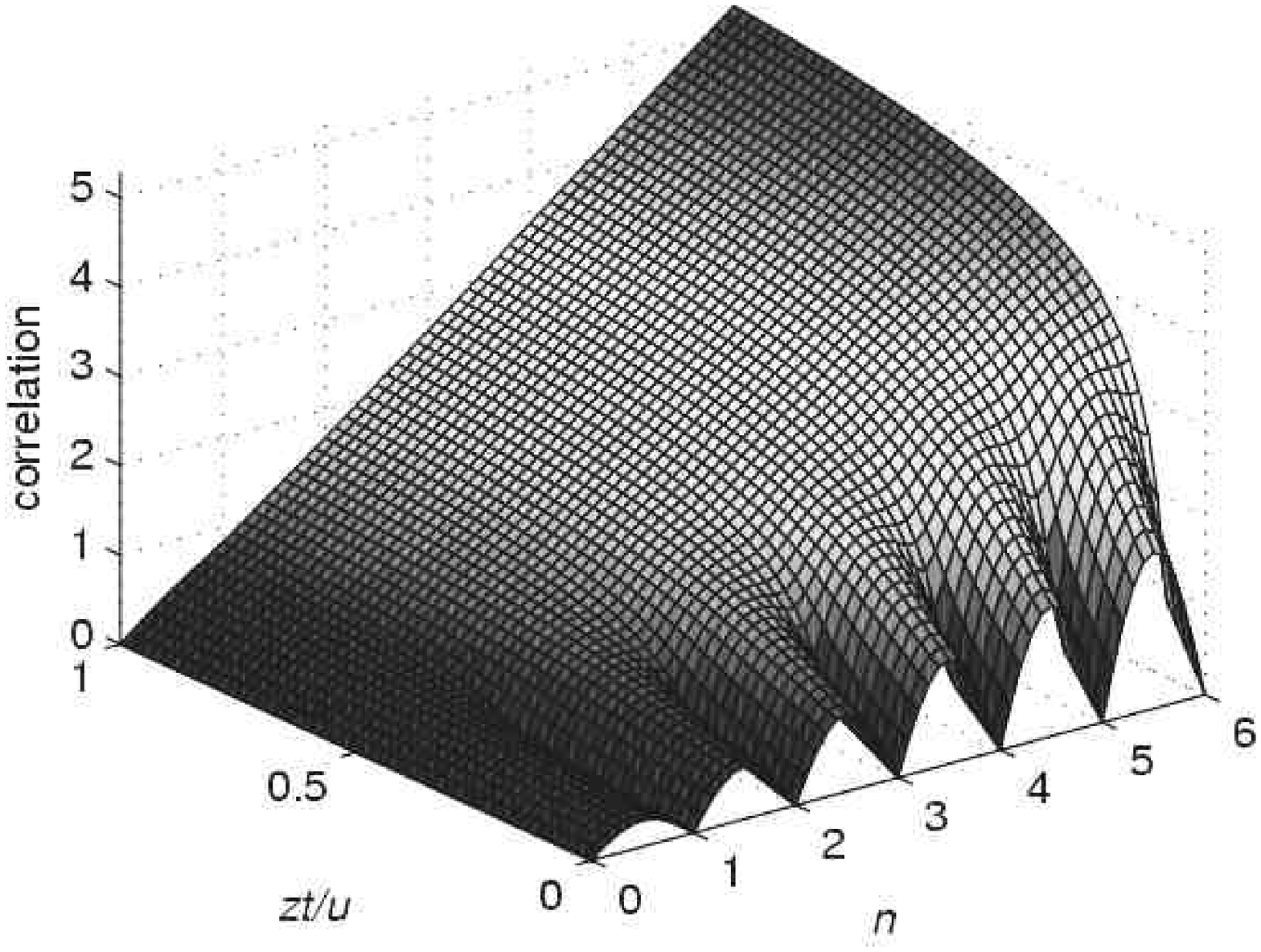}}
		}
		\mbox{
		\subfigure[Mean-field, d=3]{\includegraphics[width=0.50\textwidth]{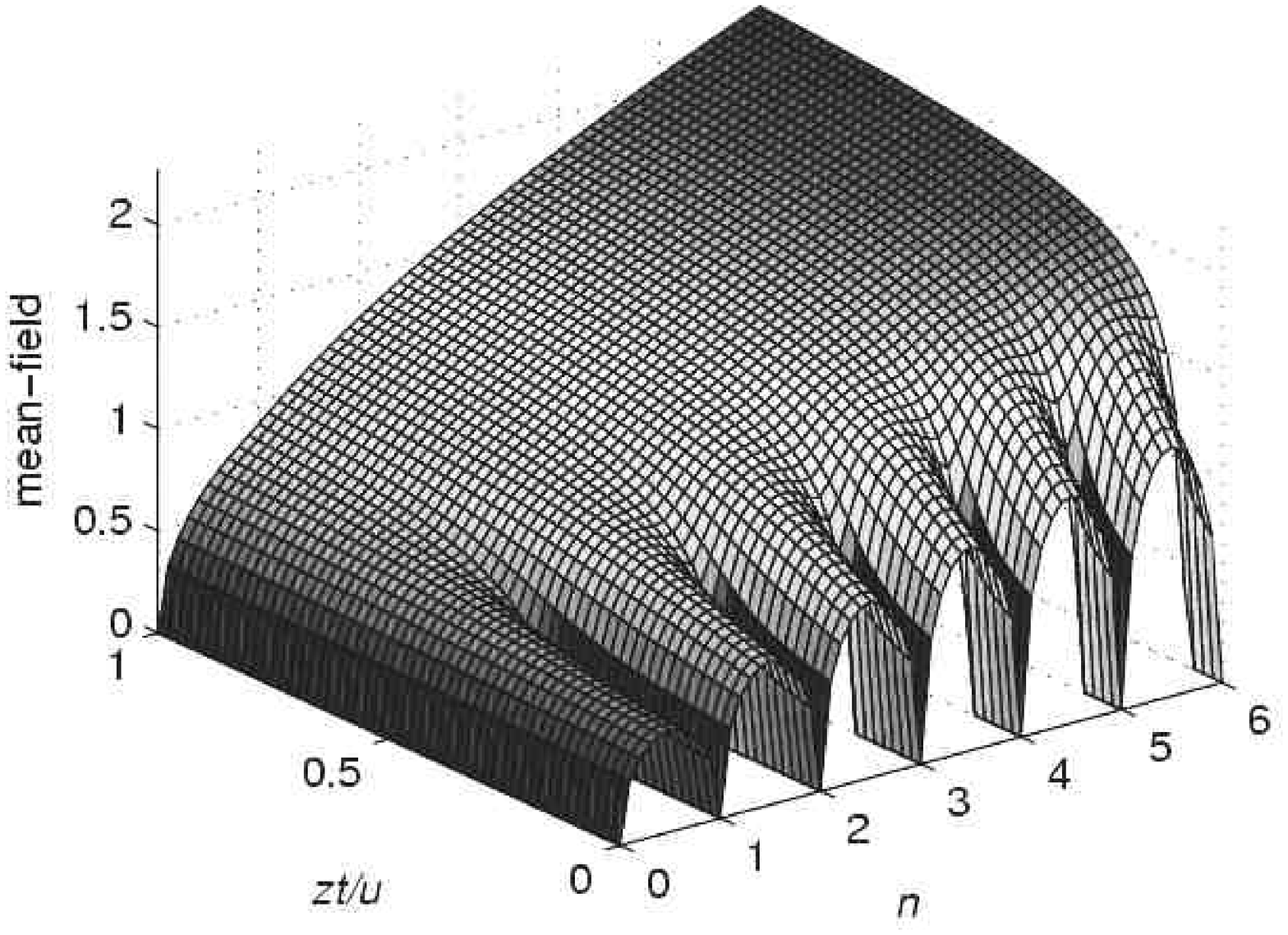}}
		\hspace{3mm}
		\subfigure[Intersite correlation, d=3]{\includegraphics[width=0.50\textwidth]{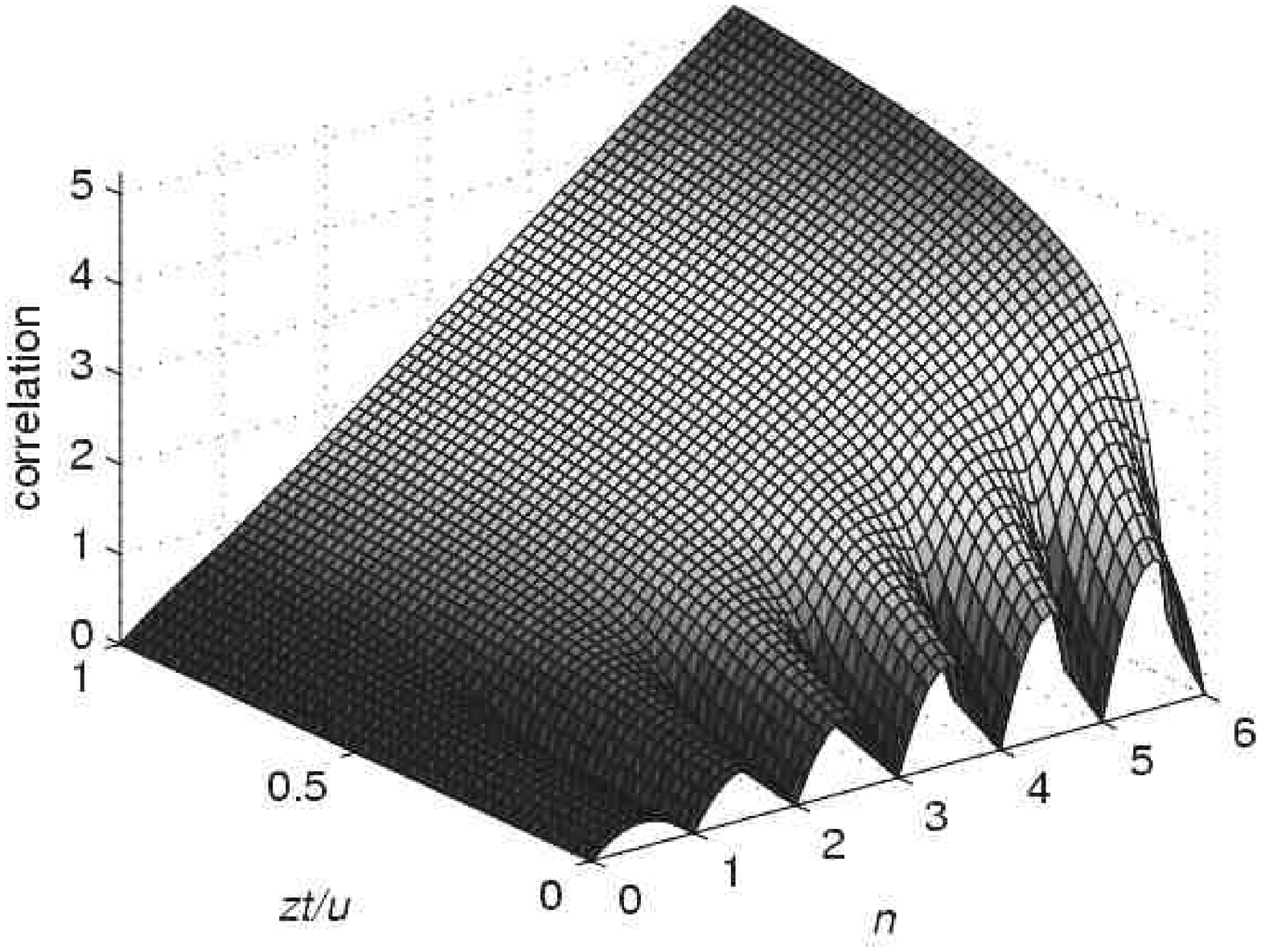}}
		}
		\caption{Ground state results for the Bose-Hubbard model in the two site mean-field approximation. The mean-field and intersite correlations are shown as a function of the mean site occupation $\nbar$ and the relative interaction strength $zt/u$. A non-zero mean-field represents the superfluid regime. The mean-field is zero below a critical value $(zt/u)_c$ for commensurate occupations, indicating the onset of the Mott insulator phase. The intersite correlations are non-zero everywhere except for commensurate occupations in the limit of no tunnelling where $zt/u = 0$. The results are qualitatively similar for 1, 2, and 3 dimensional lattices. However, a feature of the $d=1$ phase diagram is the weak suppression of the mean-field at half-integer occupations, which is seen more seen more clearly in Fig.\ref{fig:2site-lowvar}. This may be an artefact of the two site approximation and requires further study.} 
		\Label{fig:2site-phasediagram}
		\end{center}
\end{figure}

\begin{figure}[!htp]
    \begin{center}
    \mbox{
		\subfigure[Mean-field, $d=1$]{\includegraphics[width=0.31\textwidth]{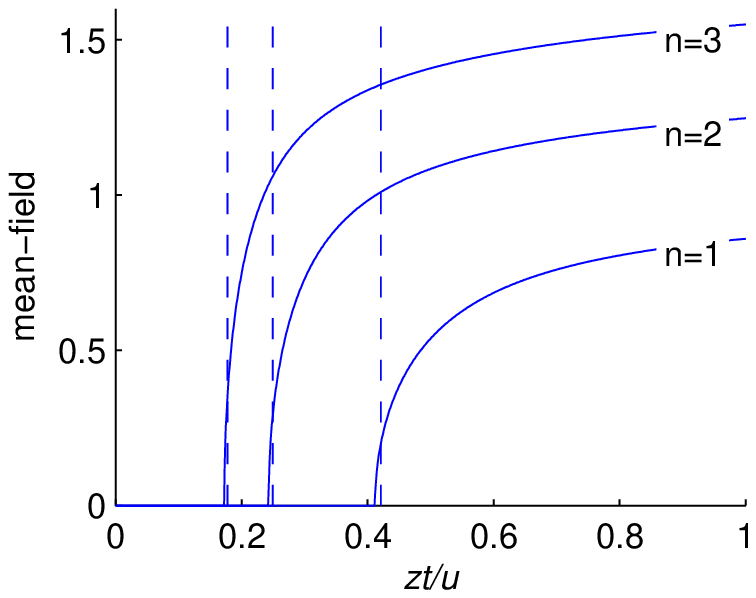}}
		\hspace{1mm}
		\subfigure[Mean-field, $d=2$]{\includegraphics[width=0.31\textwidth]{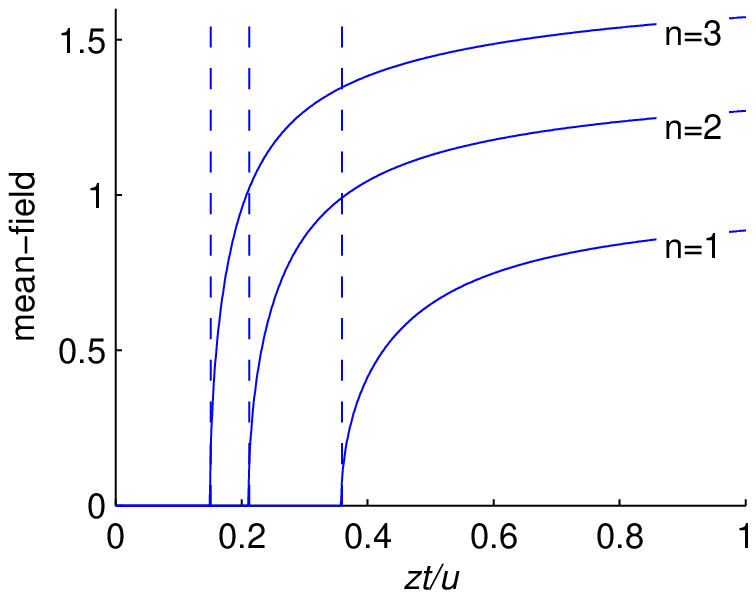}}
		\hspace{1mm}
		\subfigure[Mean-field, $d=3$]{\includegraphics[width=0.31\textwidth]{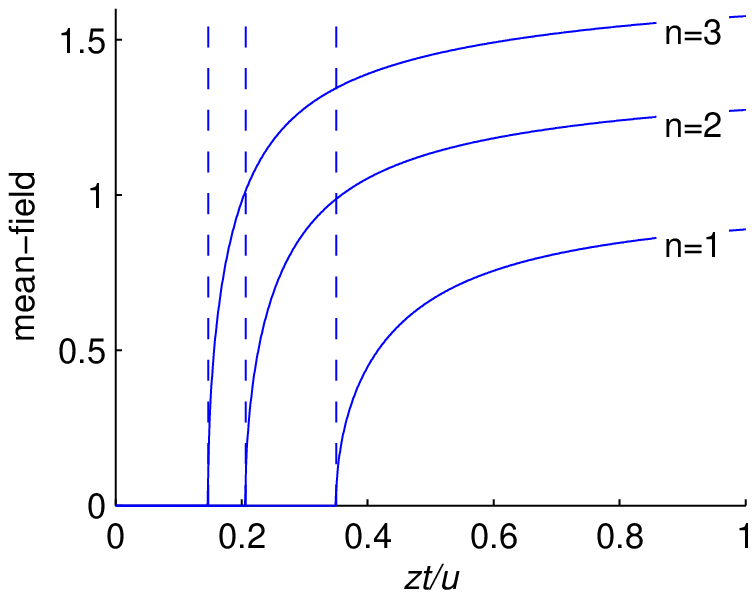}}
		}
		\mbox{
		\subfigure[Correlation, $d=1$]{\includegraphics[width=0.31\textwidth]{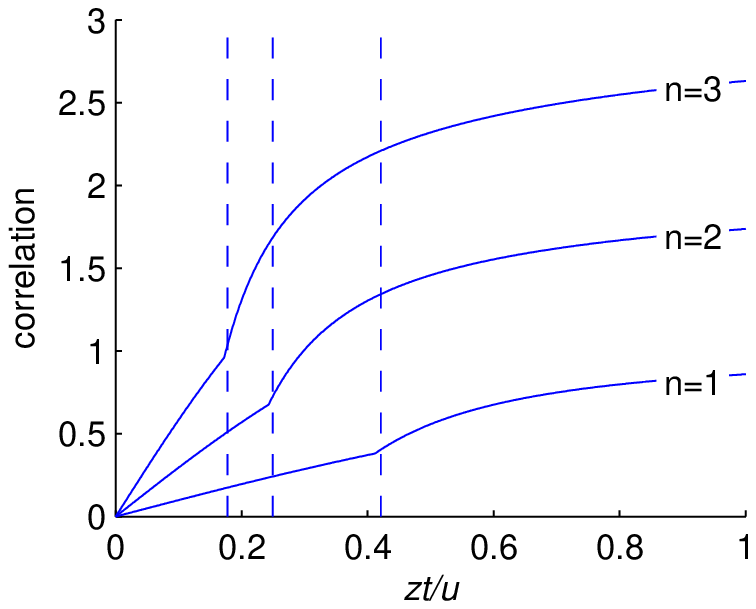}}
		\hspace{1mm}
		\subfigure[Correlation, $d=2$]{\includegraphics[width=0.31\textwidth]{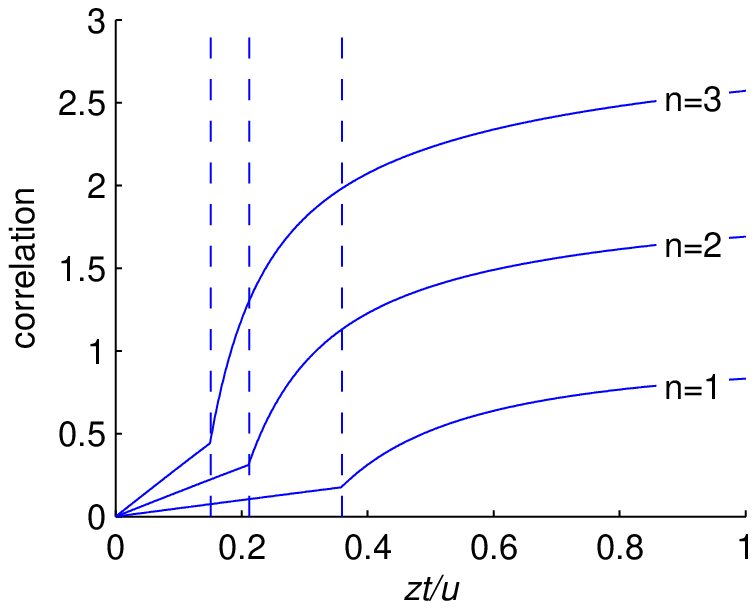}}
		\hspace{1mm}
		\subfigure[Correlation, $d=3$]{\includegraphics[width=0.31\textwidth]{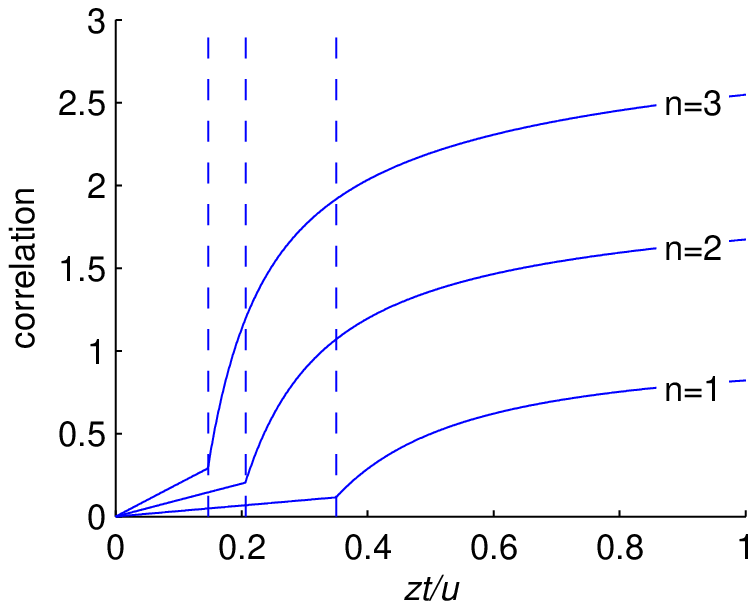}}
		}
		\mbox{
		\subfigure[Variance, $d=1$]{\includegraphics[width=0.31\textwidth]{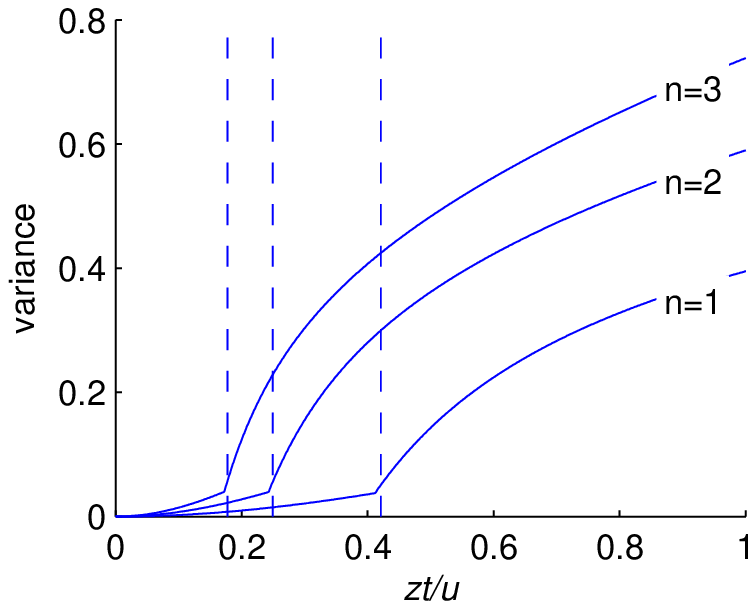}}
		\hspace{1mm}
		\subfigure[Variance, $d=2$]{\includegraphics[width=0.31\textwidth]{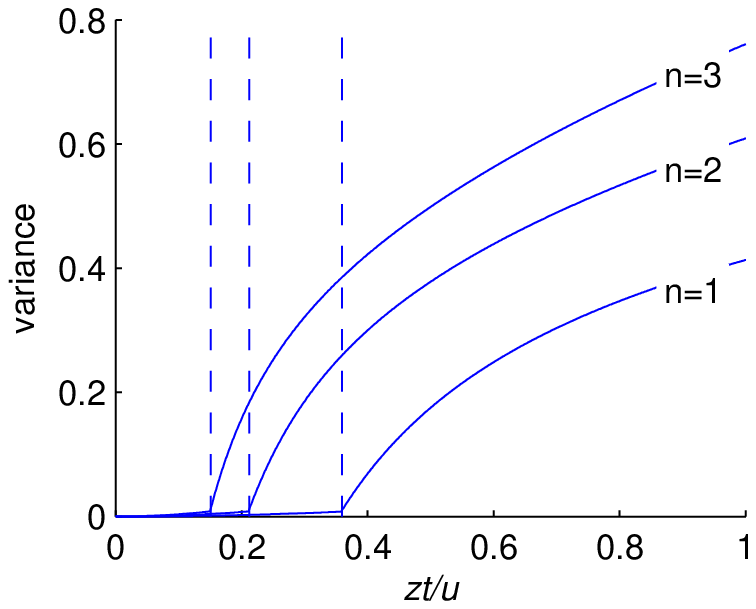}}
		\hspace{1mm}
		\subfigure[Variance, $d=3$]{\includegraphics[width=0.31\textwidth]{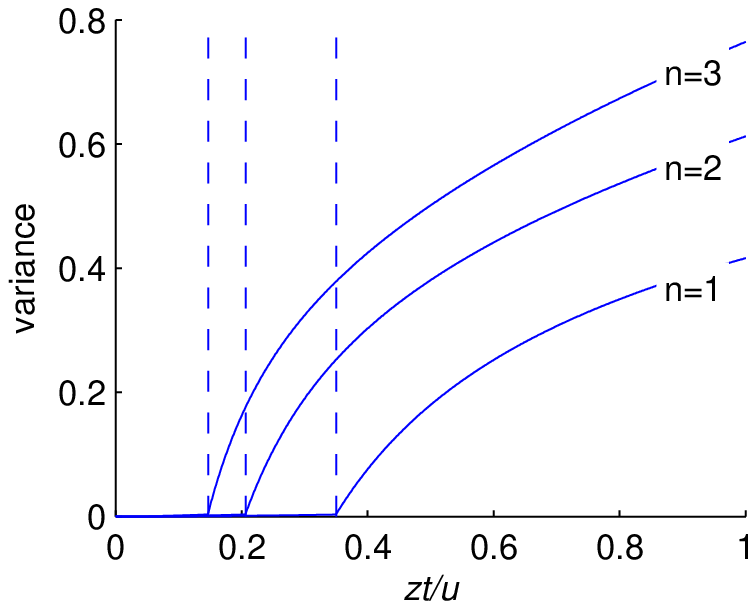}}
		}
		\end{center}
		\caption{$d=1$, $2$, $3$ results: mean-field, intersite correlation and variance for commensurate occupations of $\nbar=1$, $2$, $3$. The dashed vertical lines indicate the corresponding positions (from left to right) of the phase boundary between the superfluid and Mott insulator phases as calculated using the pertubation theory outlined in section \ref{sec-phasebound2s}} 
		\Label{fig:2site-commensurate}
\end{figure}

This can be made linear in $\mathbf{u}$ (and therefore $\mathbf{C}$) when the mean-field term $\langle X \rangle$ in (\ref{2site-eigA}) is treated as a free parameter $x$; the resulting eigenvalue equation is solved iteratively in $x$ so that it is self-consistent with the calculated mean-field $\langle X \rangle = \mathbf{C}^T A_X \mathbf{C}$. The  eigenvector solution $\mathbf{u}$ which maximizes $F$ at each iteration is nonnegative; in fact the nonnegative eigenvector, the maximizing solution, belongs to the maximum eigenvalue as we now show.

\subsubsection{Determining the maximising solution}

To see why this is the case we need to invoke the Perron-Frobenius theorem for nonnegative matrices \cite{Maccluer2000}. We first note that the $r \times r$ matrix  $G = S^{-1/2} A S^{-1/2} + S^{1/2}B S^{-1/2}$ in equation (\ref{2site-eig}) is nonnegative everywhere except along the leading diagonal where negative values are possible when either $\mu$ or $\nu$ (or both) are negative. It is then always possible to construct a matrix $G^{\prime} = G + \gamma I_r$ that is nonnegative everywhere for a sufficiently large $\gamma$, $I_r$ being the $r \times r$ identity matrix. 

The eigenvectors of $G^{\prime}$ and $G$ are clearly the same with the eigenvalues of $G^{\prime}$ given by $\lambda^{\prime}_i = \lambda_i + \gamma$ where $\lambda_i$ is the $i$th eigenvalue of $G$ corresponding to the $i$th eigenvector. The Perron-Frobenius theorem states that the spectral radius $\rho$ of a nonnegative matrix is an eigenvalue corresponding to a nonnegative eigenvector\footnote{The spectral radius for an $n \times n$ matrix $A$ with eigenvalues $\lambda_i$  ($1 \geq i \geq n$) is defined as $\rho(A) = \textrm{max}|\lambda_i|$.}. That is, $\rho(G^{\prime})$ is the eigenvalue with the desired nonnegative solution. We can relate this to the maximum eigenvalue of $G$ when $\gamma$ simultaneously satisfies the following two conditions:
\begin{enumerate}
	\item $\gamma  \geq | \textrm{min}(G) |$ so that $G^{\prime}$ is nonnegative
	\item $\gamma \geq | \textrm{min}(\lambda_i) |$ so that all $\lambda^{\prime}_i \geq 0$ and $\textrm{max}(\lambda_i) + \gamma$ is necessarily the spectral radius of $G^{\prime}$
\end{enumerate}
For any bounded matrix $G$ it is always possible to select such a value of $\gamma$; we can therefore associate the correct solution for $C_{N, p}$ with the maximum eigenvalue of (\ref{2site-eig}) which returns a self-consistent mean-field.

\subsection{The ground state solution}

The general procedure for finding the ground state solution follows the method used in the one site formulation (see section \ref{sec-exact1s}). In particular, at a given mean occupation $\nbar$, (\ref{2site-eig}) determines a locus of solutions which can be expressed as a function of number variance. Therefore, at a given $t/u$ and $n$, the two site energy can be reparameterized in terms of number variance; finding the ground state phase diagram then only requires the minimization of the energy with respect to variance. In practice to perform this procedure efficiently, it is useful to first calculate a set of solution states covering a suitable variance range, which act as initial conditions in the final minimisation step. 

In the results presented here, this was achieved by fixing $\nbar$ and initially choosing a large value for the variance corresponding to the superfluid regime (specifically $\var = \nbar$ for $\nbar \leq 1$ and $\var = \nbar/2$ for $\nbar \geq 1$). The corresponding state was found by solving equation (\ref{2site-eig}) using the initial values $x_0 = 5$, $\mu_{0} = 2$ and $\nu_{0} = -1$, which was found to be sufficient for the range of parameters considered here. A lower variance was then selected and its corresponding state was calculated, by using the previous solution to determine the initial conditions. By iteration it was then possible to find solutions efficiently over a broad variance range. 

However, although this general procedure works well for most of the parameter space, it fails to converge (to a sufficient accuracy) in two limiting cases: for Mott-insulator states where the mean-field is zero and there is degeneracy; and in the limit of the lowest possible variance, $\var = \delta \nbar (1 - \delta \nbar)$, corresponding to the localized region where the Lagrange multipliers diverge at the solution. It is therefore necessary to consider these cases separately.

\subsubsection{Calculating Mott states}\label{sec-mott2s}
For commensurate occupations and below a critical value for the variance, the system is in the Mott insulator state corresponding to a vanishing mean-field. In this case, the above approach is numerically unstable for two reasons. Firstly, the algorithm fails to correctly converge to a zero mean-field since the search direction cannot be determined. Secondly, the system states become highly degenerate in $\mu$ in the Mott insulator state (corresponding to a incompressible phase) which further prevents convergence (with respect to $\mu$). 

These problems are circumvented respectively by the following adjustments to the general procedure. Firstly, explicitly setting the free-parameter $x$ to zero in the equation (\ref{2site-eig}), ensures that any solution is self-consistent in the mean-field; this can be seen to be case by considering how the mean-field term appears in the eigenvalue equation (see (\ref{2site-quadform-mf}) and (\ref{2site-eigA}) in particular). Secondly, setting $\mu = -r \, \nu$ for a suitably chosen quantity $r$, restricts the parameter space to the degenerate region, so that $\nbar$ remains fixed while $\var \rightarrow 0$ in the limit $\nu \rightarrow \infty$. These trajectories are illustrated by the dashed curves in Fig.\ref{fig:2site-contours}. By enforcing this relationship between $\mu$ and $\nu$, we can therefore probe the Mott insulator states accurately.

\subsubsection{Solutions in the limit of zero tunneling}\Label{sec-lowvar2s}
Considering (\ref{2site-E}) in the limit of zero tunneling where $t/u \rightarrow 0$, the ground state is clearly determined by states of the lowest possible variance. In this case, we found the method of Lagrange multipliers leads to poor convergence as the Lagrange multipliers diverge at the solution. We therefore consider a more direct method whereby we minimize the two site energy in terms of a reduced parameter space, which correspond to the lowest allowed variance ($\var = \delta n (1 - \delta n)$) for a given mean occupation $\nbar$.

Recall the lowest variance state is given by equation (\ref{2site-lowvarstate}); in this basis, the hopping energy (\ref{2site-hoppingenergy}) is given by
\begin{equation}\Label{2site-lowvar-F}
F = \beta/2(\nbarint+1)(1 + (2d-1)(\sqrt{\alpha}+\sqrt{\gamma})^2)
\end{equation}
The constraints (normalization, fixed $\nbar$ and fixed variance) can be given as a set of linear equations
\begin{eqnarray}
\alpha + \beta + \gamma &=& 1 \nonumber \\
\nbarint \alpha + ( \nbarint \!+\!\half) \beta + ( \nbarint \!+\!1) \gamma &=& \nbar \nonumber \\
{\nbarint}^2 \alpha + ({\nbarint}^2\!+\! \nbarint \!+\!\half) \beta + ({\nbarint}^2 \! + \! 2\nbarint\!+\!1) \gamma 
&=& \var + \nbar^2
\end{eqnarray}
which when written as $A x = b$, admits solutions of the form $x = x_0 + \kappa \, \textrm{nullspace} [A]$ where we have taken
\begin{equation}
x_0 = 
\begin{array}{|c|} \alpha \\ \beta \\ \gamma \end{array}
= \begin{array}{|c|} [\nbar] + 1 - \nbar \\ 0 \\ \nbar - [\nbar] \end{array}
\end{equation}
as a solution of the homogenous system $A x = 0$. There is an additional requirement on $\kappa$ that the coefficients are bounded with $0 \leq \alpha \leq 1$, $0 \leq \beta \leq 2$ and $0 \leq \gamma \leq 1$.
For a given occupation $\nbar$ and dimension $d$, we can then maximise the hopping energy (\ref{2site-lowvar-F}) with respect to the remaining free parameter $\kappa$.  The ground state phase diagram is then calculated easily for the case where $zt/u=0$; the resulting mean-field and intersite correlation are shown in figures \ref{fig:2site-lowvar}(a) and \ref{fig:2site-lowvar}(b) respectively.

\begin{figure}[t]
    \begin{center}
    \mbox{
		\subfigure[]{\includegraphics[width=0.5\textwidth]{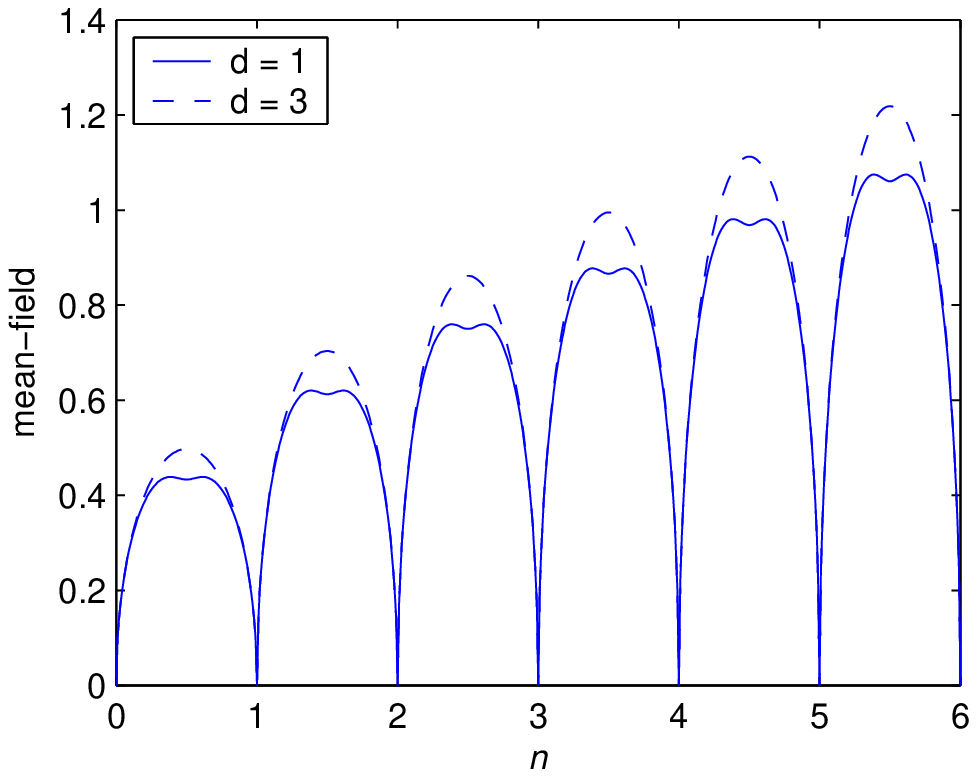}}
		\subfigure[]{\includegraphics[width=0.5\textwidth]{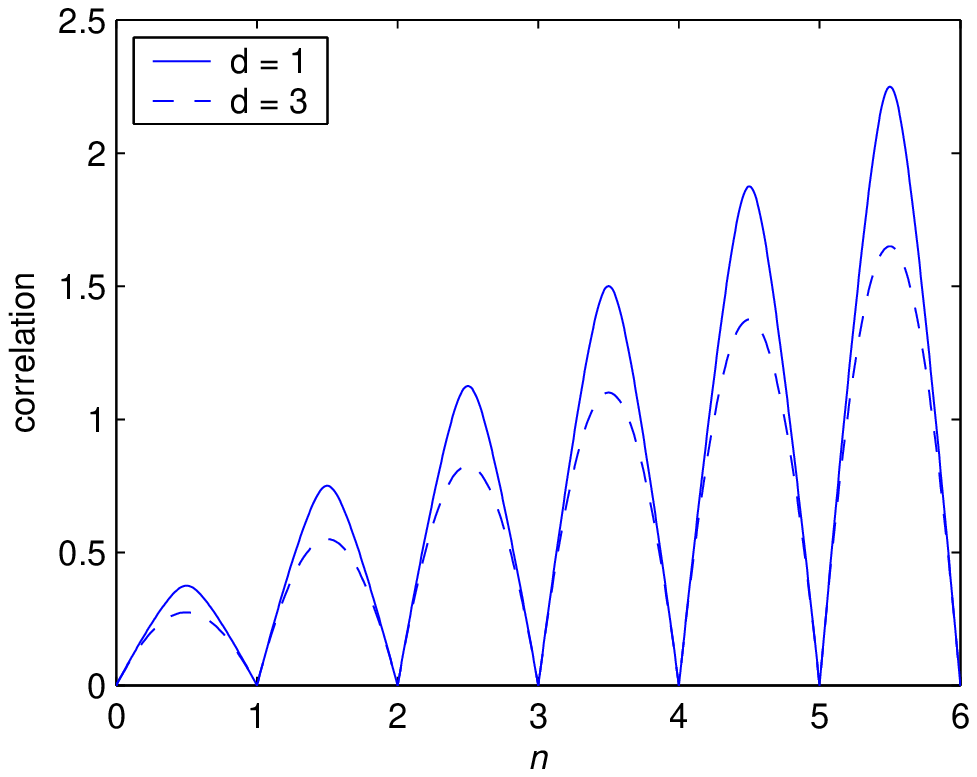}}
		}
		\caption{The mean-field (a) and intersite correlation (b) using a reduced basis of three states in the limit of zero tunelling ($zt/u = 0$) for $d=1$, $3$. The methodology is outlined in section \ref{sec-lowvar2s}. A feature of the $d=1$ case is the weak supression of the mean-field around half-integer occupations. The $d=2$ results, which been omitted for clarity, are very close to the $d=3$ case.} 
		\Label{fig:2site-lowvar}
		\end{center}
\end{figure}

\subsubsection{Numerical Solutions}

Numerical calculations were performed using a state space truncated with $N \leq 20$, corresponding to 121 coefficients $C_{N,\,p}$. The MATLAB function {\em fsolve} was used to find the self-consistent solution, for a given $\nbar$ and $\var$, by supplying the function with suitable initial values $x_0$, $\mu_{0}$ and $\nu_{0}$ for the mean-field and Lagrange multipliers. 

\subsection{Results}

Using the calculated states with maximum tunneling energy, the two site energy (\ref{2site-E}) was then minimized with respect to $\var$ for a given $\nbar$ and $Z/u$. The results of this calculation are shown in Figs. \ref{fig:2site-phasediagram}(a)-(f) for arbitrary mean occupations. To investigate the superfluid to Mott-insulator transition more clearly, these results have also been reproduced for commensurate occupations in Figures \ref{fig:2site-commensurate}(a)--(i).

This ground state phase diagram is qualitatively similar to the results for the one site model (Fig.\ref{fig:exact}); the most important feature of both models is the vanishing of the mean-field for commensurate occupations above the critical transition point $\overline{u}_c$, corresponding to the Mott insulator phase. Note that the Mott insulator to superfluid transition is clearly reproduced for commensurate mean occupations, an improvement over the corresponding two site Q-function approximation (section \ref{sec-qfunc2s}) which can be attributed to the exact treatment of the intersite correlations in the present treatment. We now consider a simple perturbative expansion from which the position of the phase transition boundary can be easily calculated for the two site formulation.

\subsection{Calculating the phase transition boundary}\label{sec-phasebound2s}

Following the one-site case, we consider the following states near the transition with commensurate occupation $\nbar$:
\begin{eqnarray}\Label{2site-pert-state}
|s\rangle = \sqrt{1-\alpha-2\beta} \, |\nbar, \nbar\rangle + \sqrt{\frac{\alpha}{2}} (|\nbar-1,\nbar+1\rangle + |\nbar+1, \nbar-1\rangle) 
\nonumber \\  + \sqrt{\frac{\beta}{2}}\bigg(|\nbar,\nbar-1\rangle + |\nbar-1, \nbar\rangle + |\nbar,\nbar+1\rangle + |\nbar+1, \nbar\rangle\bigg)
\end{eqnarray}
This state is a superposition of the minimum set of number states that give a non-zero mean-field and intersite correlation. 
The coefficients have been chosen to satisfy normalization and a fixed commensurate occupation $\nbar$. The corresponding variance on one site is given by $\var = \alpha + \beta$. Note that $|s\rangle$ represents a perturbation in $\alpha$ and $\beta$ from the Mott insulator phase.
With this state the intersite correlation (\ref{two_site_corr}) is given by 
\begin{equation}
\langle V \rangle = \sqrt{2(1-\alpha-2\beta)\alpha} \, \sqrt{\nbar(\nbar+1)} + \frac{\beta}{2}(2\nbar+1)
\end{equation}
and the mean-field (\ref{2site-mf}) is
\begin{equation}
\langle X \rangle = \sqrt{\frac{\beta}{2}}\left(\sqrt{1-\alpha-2\beta} + \sqrt{\frac{\alpha}{2}}\right)(\sqrt{\nbar}+\sqrt{\nbar+1})
\end{equation}
In the ground state, the phase transition occurs for the minimum two site energy (\ref{2site-E}); clearly this occurs with respect to the state (\ref{2site-pert-state}) when $\partial{E_{12}}/\partial{\alpha}=0$ and $\partial{E_{12}}/\partial{\beta}=0$ are both satisfied. Moreover the phase transition occurs for $\beta=0$ where the mean-field is zero. That is,
\begin{equation}
0 = \frac{\partial{E}}{\partial{\alpha}}\bigg|_{\beta = 0} = \left(\frac{u}{t}\right) - \frac{\sqrt{\nbar(1+\nbar)}(1-2\alpha)}{\sqrt{2\alpha(1-\alpha)}}
\end{equation}
\begin{eqnarray}\Label{2site-dE-dbeta}
0 = \frac{\partial{E}}{\partial{\beta}}\bigg|_{\beta = 0} &=& \left(\frac{u}{t}\right) - \half(1+2\nbar) 
\nonumber \\ && - \half(2d-1)(\sqrt{\nbar}+\sqrt{\nbar+1})^2(\sqrt{1-\alpha}+\sqrt{\alpha/2})^2
\nonumber \\ && + \frac{\sqrt{2\alpha\nbar(\nbar+1)}}{\sqrt{1-\alpha}}
\end{eqnarray}
Numerically solving these equations leads to the results shown in table \ref{tab:2sitepb}. Note that in the non-physical limit of $d\! \rightarrow\!\infty$ the variance at the transtion is $\var=\alpha=0$. Considering equation (\ref{2site-dE-dbeta}) the transition then occurs at $\Uc = 2(u/2dt)_c = (\sqrt{\nbar}+\sqrt{\nbar+1})^2$, the same value as in the one-site model. This is not unexpected as the intersite correlation $\langle V \rangle$ becomes negligible compared with the mean-field contribution in the Hamiltonian (\ref{2site-bh}).
\\
\begin{table}
\begin{center}
\begin{tabular}{|l|ccc|} \hline
$d$ & $\nbar=1$ & $\nbar=2$ & $\nbar=3$ \\ \hline
$1$ & $4.75$ & $8.02$ & $11.26$ \\
$2$ & $5.56$ & $9.43$ & $13.27$ \\ 
$3$ & $5.71$ & $9.69$ & $13.63$ \\
$10$ & $5.82$ & $9.88$ & $13.90$ \\
$10^6$ & $5.83$ & $9.90$ & $13.93$ \\ \hline
\end{tabular}
\caption{Critical value $\overline{U}_c$ calculated for two site mean-field approximation using a perturbation expansion around the Mott Insulator state}
\label{tab:2sitepb}
\end{center}
\end{table}

There are some notable differences between the one and two site mean-field approximations, which we discuss here. In particular, the inclusion of intersite correlations in the two site Hamiltonian, results in a non-zero number variance even in the Mott insulator phase where the mean-field vanishes; this can be seen in Figs. \ref{fig:2site-mfvsvar} and \ref{fig:2site-commensurate}. This contrasts with the one site results (see Fig. \ref{fig:X-max-compare}) where the mean-field and number variance are identically zero. Moreover, for the two site model the calculated transition points $\overline{U}_c$, as shown in table \ref{tab:2sitepb}, occur at lower values than for the one site model. In particular, for $\nbar = 1$ and $d=1$, the two site model yields $\overline{U}_c = 4.75$, whereas the one site model gives $\overline{U}_c = 5.83$.

These differences are most pronounced for the $d=1$ case where the mean-field approximation is no longer valid, as has also been noted elsewhere \cite{Zwerger2003}. At higher dimensions ($d=2$, $3$), where the mean-field approximation is more accurate, the number variance is close to zero in the Mott phase and the calculated transition points for the one and two site models are close. It is expected that in the (non-physical) limit of infinite lattice dimensionality, the one and two site results will converge. 

\subsection{Comparison to other work}

The results for the two site Hamiltonian given in the previous section clearly demonstrate a partial shift to lower values $\Uc$ (ie. a weaker lattice) when compared to the one site results. This shift can be attributed to the inclusion of intersite correlations in the model. 
To compare with our results with those reported elsewhere it is useful to refer to table \ref{tab:lit}. 

For the $d=1$ case, numerically exact schemes, with Density Matrix Renormalization Group (DMRG) and Quantum Monte-Carlo (QMC) simulations, predict a much lower value for the phase transition point $\Uc$ in the Bose-Hubbard model. Using the DMRG technique with the infinite-system algorithm, the transition point for $\nbar=1$ and $d=1$ has been reported as $\overline{U}_c = 1.68$ \cite{Pai96} and $\overline{U}_c = 1.81$ \cite{Kuhner98}. Using the more accurate finite-system algorithm, the same transition has been reported as $\overline{U}_c = 1.92$ \cite{Rapsch99} and $\overline{U}_c = 1.68$ \cite{Kuhner2000}. Similarly, for QMC simulations without lattice disorder, the transition has been reported as $\overline{U}_c = 2.33$ \cite{Batrouni92}. Also, pertubative expansions using defect states have yielded similar results to the that of the QMC technique in one dimension. In particular, the above transition has been reported as $\overline{U}_c = 2.33$ \cite{Freericks94} and $\overline{U}_c = 2.04$ \cite{Freericks96}.

\begin{table}
\begin{tabular}{p{1.5in}|p{2in}|l|l} \hline
Method & Position of phase transition boundary for $d=1$, $n=1$ (unless otherwise stated) & Authors & Reference \\ \hline

Mean-field approximation using second order perturbation expansion & $\Uc = 2n+1 + \sqrt{(2n+1)^2-1}$,  $\nbar = 1$: $\Uc = 5.83$ & Oosten \etal & \cite{Oosten2001} \\ \hline

Mean-field theory using variational approach &  $\Uc = 2.0$ & Amico and Penna & \cite{Amico2000} \\ \hline

Mean-field theory using variational approach &  $\nbar = 1$: $\Uc = 5.83$ & Sheshadri \etal & \cite{Sheshadri93} \\ \hline

Perturbative expansion using defect (particle/hole) states &  $d = 1$, $\nbar = 1$: $\Uc = 2.33$;  $d = 2$, $\nbar = 1$: $\Uc = 3.68$ & Freericks and Monien & \cite{Freericks94} \\ \hline

Perturbative expansion using defect (particle/hole) states &  $\Uc = 2.04$ (see table in reference for further calculated values at different $d$ and $n$) & Freericks and Monien & \cite{Freericks96} \\ \hline

Quantum Monte-Carlo &  $\Uc = 2.33$ (without disorder) & Batrouni and Scalettar & \cite{Batrouni92} \\ \hline

DMRG with infinite-system algorithm &  $\Uc = 1.68$ & Pai \etal & \cite{Pai96} \\ \hline

DMRG with infinite-system algorithm &  $\Uc = 1.81$ (without nearest-neighbour interactions) & Kuhner and Monien & \cite{Kuhner98} \\ \hline

DMRG with finite-system algorithm &  $\Uc = 1.68$ (without nearest-neighbour interactions) & Kuhner \etal & \cite{Kuhner2000} \\ \hline

DMRG with finite-system algorithm &  $\Uc = 1.92$ (for zero disorder) & Rapsch \etal & \cite{Rapsch99} \\ \hline

\end{tabular}
\caption{The calculated boundary of the Superfluid to Mott-insulator phase boundary as determined from the Bose-Hubbard model as solved using various methods in the literature.}
\label{tab:lit}
\end{table}

\newpage

\section{Conclusions}

\pj{We have presented two model Hamiltonians for the Bose-Hubbard model using two different mean-field approximations.} The simpler one site model treats all intersite correlations using a mean-field approximation that decouples the problem into the sum of one site Hamiltonians. The second model is a two site extension where intersite correlations between two adjacent sites are explicitly included while treating interactions with neighbouring sites using the mean-field approximation. Each model has been tackled using two methodologies: a treatment in terms of a Q-function representation; and a numerically exact method using either the one or two site states.

In the case of the one site Hamiltonian, we find the Q-function representation agrees well with the numerically exact treatment, but can be solved at a fraction of the computational cost. For the two site Hamiltonian, the Q-function gives a good qualitative description but \pj{does not give a clear Mott insulator to superfluid transition due to limitations of the parameterization on two sites. It is encouraging, however, that the Q-function approach gives the intersite correlations accurately when compared with the two site exact solution.} In contrast \pj{to the Q-function approach}, the two site exact solution yields a well-defined phase transition for commensurate mean occupations. In this case, the critical relative interaction strength $\Uc$ is smaller (corresponding to a weaker lattice strength) for all commensurate values of $\nbar$ than for the one site model; this shift is most pronounced for a one dimensional lattice.

\pj{
What emerges from the Q-function approach is that the quantum mechanical constraints play a primary role in determining the ground state results for the Bose-Hubbard model. This is most clearly illustrated by the one site formulation where the results are determined by the uncertainty relation in two forms: a restriction on the minimum variance permitted by the quantum state; and a relation between the number and phase fluctuations. In particular, the onset of the Mott insulator phase is characterized by supression of number fluctuations at the lower bound of the variance constraint.
}

Exact numerical results with QMC or DMRG simulations on finite one dimensional lattices by other authors indicate the actual position of the \pj{superfluid to Mott insulator transition occurs for an even lower value of $\Uc$ than is calculated by our two site model}. It is expected that if we were to extend our formalism to explicitly include the interactions between three or more sites, the predicted value $\Uc$ would shift to lower values in line with these other treatments. However, this remains a computationally difficult problem due to the dramatic increase in the size of the Hilbert space of the problem when more sites are included.


\section{Acknowlegments}

The authors would like to thank D. Jaksch and P. Zoller (of Institut f\"ur theoretische Physik, Universtit\"at Innsbruck, Austria) for useful discussions.



\newpage

\section*{References}
\bibliographystyle{unsrt}
\bibliography{BoseHub}
\appendix
\section{Values of parameters in particlar cases.}\label{appendix-a}
To fit the kinds of distribution in Fig.\ref{Qfuns} we determine the parameters
$ v$, {$ \sigma$} and $ c $ by fitting the moments $ \langle {a} \rangle$,
$ \langle {a}{a}^\dagger\rangle$ and 
$ \langle {a} {a}{a}^\dagger{a}^\dagger\rangle $.
\subsection*{Parameters for a number state}\Label{Params}
The exact values for the quantities (\ref{par3}--\ref{par7}) in the case of a 
number state $ |m\rangle$ are
\begin{eqnarray}\Label{par3n}
\langle {a} \rangle &=& 0
\\ \Label{par4n}
\langle {a}{a}^\dagger\rangle &=& m+1
\\ \Label{par5n}
\langle {a}{a} \rangle &=& 0
\\ \Label{par6n}
\langle {a}{a}{a}^\dagger\rangle &=& 0
\\ \Label{par7n}
\langle {a} {a}{a}^\dagger{a}^\dagger\rangle 
&=&  (m+1)(m+2)
\end{eqnarray}
leading to the choice
\begin{eqnarray}\Label{par8n}
v^4 &=& m^2 + {3\over 2}m + {1\over 2}
\\ \Label{par9n}
\sigma &=&  m+1 -\sqrt{m^2 + {3\over 2}m + {1\over 2}}
\\ \Label{par10n}&\approx& {1/4} \quad\mbox{ as $ m\to \infty$}
\\ \Label{par11n}
c&=& 0.
\end{eqnarray}
The two quantities (\ref{par5n},\ref{par6n}) are given exactly by this choice.

\subsection*{Parameters for a superposition of two number states}
For an equal superposition of the kind (\ref{bh3}) that
\begin{eqnarray}\Label{par8}
v^4 &=& n^2 +{n\over 2} -{1\over 8}
\\ \Label{par9}
\sigma &=& n+{1\over2}-\sqrt{ n^2 +{n\over 2} -{1\over 8}}
\\ \Label{par10}
c &=& {{\sqrt n}\over 2 v}.
\end{eqnarray}
The exact computation of the quantities (\ref{par3}--\ref{par7}) 
yields in this case
\begin{eqnarray}\Label{par11}
\langle {a} \rangle &=& {1\over 2}\sqrt{n}
\\ \Label{par12}
\langle {a}{a}^\dagger\rangle &=& n+{1\over 2}
\\ \Label{par13}
\langle {a}{a} \rangle &=& 0
\\ \Label{par14}
\langle {a}{a}{a}^\dagger\rangle &=& {1\over 2}(n+1)\sqrt{n}
\\ \Label{par15}
\langle {a} {a}{a}^\dagger{a}^\dagger\rangle 
&=& (n+1)^2
\end{eqnarray}
The first, second and last of these are fitted exactly, leading to the 
approximate values 
\begin{eqnarray}
 \Label{par16}
\langle {a}{a} \rangle &\approx& (v^2+\sigma)c^4 
\nonumber\\&=& \left(n+{1\over 2}\right)
{n^2\over 16(n^2+n/2-1/8}
\\ \Label{par17}
\langle {a}{a}{a}^\dagger\rangle &\approx& (v^3+3\sigma v)c
\nonumber\\
&=&{1\over 2}\sqrt{n}\left(3\left(n+{1\over2}\right)-
2\sqrt{n^2+n/2-1/8}\right)
\nonumber \\
\end{eqnarray}
Even for $ n=1$ these are very tolerable approximations; (\ref{par16}) is 
diminished by comparison with $ \langle {a}{a}^\dagger\rangle$
by a factor of $ c^4\approx 0.045$, compared to the exact value of 0. For large 
$ n$ (\ref{par17}) becomes equal to the true value (\ref{par14}), and even for 
$ n=1$ the true and the model results differ by less than 8\%.

\subsection*{Parameters for a coherent state}
For a coherent state $ |\beta\rangle$, where $ \beta\equiv m$ is taken to be 
real,
the quantities take the form
\begin{eqnarray}\Label{par11c}
\langle {a} \rangle &=& \beta
\\ \Label{par12c}
\langle {a}{a}^\dagger\rangle &=& \beta^2 +1 
\\ \Label{par13c}
\langle {a}{a} \rangle &=& \beta^2 
\\ \Label{par14c}
\langle {a}{a}{a}^\dagger\rangle &=& \beta(\beta^2+2) 
\\ \Label{par15c}
\langle {a} {a}{a}^\dagger{a}^\dagger\rangle 
&=& \beta^4 + 4\beta^2 + 2
\end{eqnarray}
from which we can compute
\begin{eqnarray}\Label{16c}
v^4 &=&  \beta^4 + \beta^2 +{1\over2}
\\ \Label{17c}
\sigma &=& \beta^2 = 1 - \sqrt{\beta^4 + \beta^2 +{1\over2}}
\\ \Label{17d}
c &=&  {\beta\over v}  \approx 1-{1\over2\beta^2}
 \quad \mbox{ as $ \beta\to \infty$}
\end{eqnarray}
It is easy to check that for large $ \beta$ the quantities 
(\ref{par13c},\ref{par14c}) are correctly given, and they are of course 
correct 
for $ \beta=0$.  For $ \beta=1$, we find that 
$ \langle{a}{a}\rangle=0.8 $
instead of the exact value of 1, and 
 $ \langle{a}{a}{a}^\dagger \rangle=2.84 $, instead of the exact value of 3.

\section{A rigorous lower bound on the energy}
As a spinoff from our methodology \pj{of section \ref{sec-qfunc1s}}, we find that we can also develop a rigorous 
\emph{lower bound} for the energy as a function of the mean-field, and we can 
evaluate the predictions given by assuming the lower bound is equal to the 
ground-state energy.  

\pj{To see this we note the mean energy, which is given by (\ref{1site-E}) in the mean-field 
approximation, is parametrized by number variance $ {\rm var}[N] $ and the mean-field 
$\langle X \rangle$.}
Then we know that $ {\rm var}[N] $ satisfies the two inequalities
(\ref{con4}) and (\ref{con9}), which we can combine together as
\begin{eqnarray}\Label{rig2}
{\rm var}[N] &\ge & \max\left({\langle X\rangle^2\over 4\langle Y^2\rangle},
\delta n(1-\delta n)\right).
\end{eqnarray}
We can develop a lower bound on $ \langle Y^2\rangle $ by noting that
\begin{eqnarray}\Label{rig3}
\langle Y^2\rangle &=& n-{1\over 2}- \langle X^2\rangle
\le n-{1\over 2}- \langle X\rangle^2,
\end{eqnarray}
so that we can deduce from (\ref{rig2}) that
\begin{eqnarray}\Label{rig4}
{\rm var}[N] &\ge & 
\max\left({\langle X\rangle^2\over 4( n-{1\over 2}- \langle X\rangle^2)},
\delta n(1-\delta n)\right),
\end{eqnarray}
and that the mean energy (\ref{1site-E}) satisfies the lower bound
\begin{eqnarray}\fl\Label{rig5}
\langle E\rangle &\ge& -Z\langle X\rangle^2 +
u\left\{2 + n^2 + 3 n +\max\left({\langle X\rangle^2\over 4( n-{1\over 2}- 
\langle X\rangle^2)},
\delta n(1-\delta n)\right)\right\}.
\end{eqnarray}
This bound is parametrized by the mean-field $ \langle X\rangle$ only, so the 
rigorous lower bound is given by the value, $  X_{\rm min}(u,n)$, of 
$ \langle X\rangle$ which gives the minimum value, 
$ E_{\rm min}(u,n)$, the right hand side of (\ref{rig5}), \pj{and subject to the 
restriction that $\langle X \rangle \leq \sqrt{n-1}$ (the equality holding for
a coherent state)}.  
These are easy to evaluate, and resulting chemical potential and mean-field are 
plotted in Fig.\ref{lb}. The results are less accurate than the phase space 
method (see Fig.\ref{1site-qgs}), but are still surprisingly good.

\begin{figure}[h]
\hbox{\includegraphics[width=6.5cm]{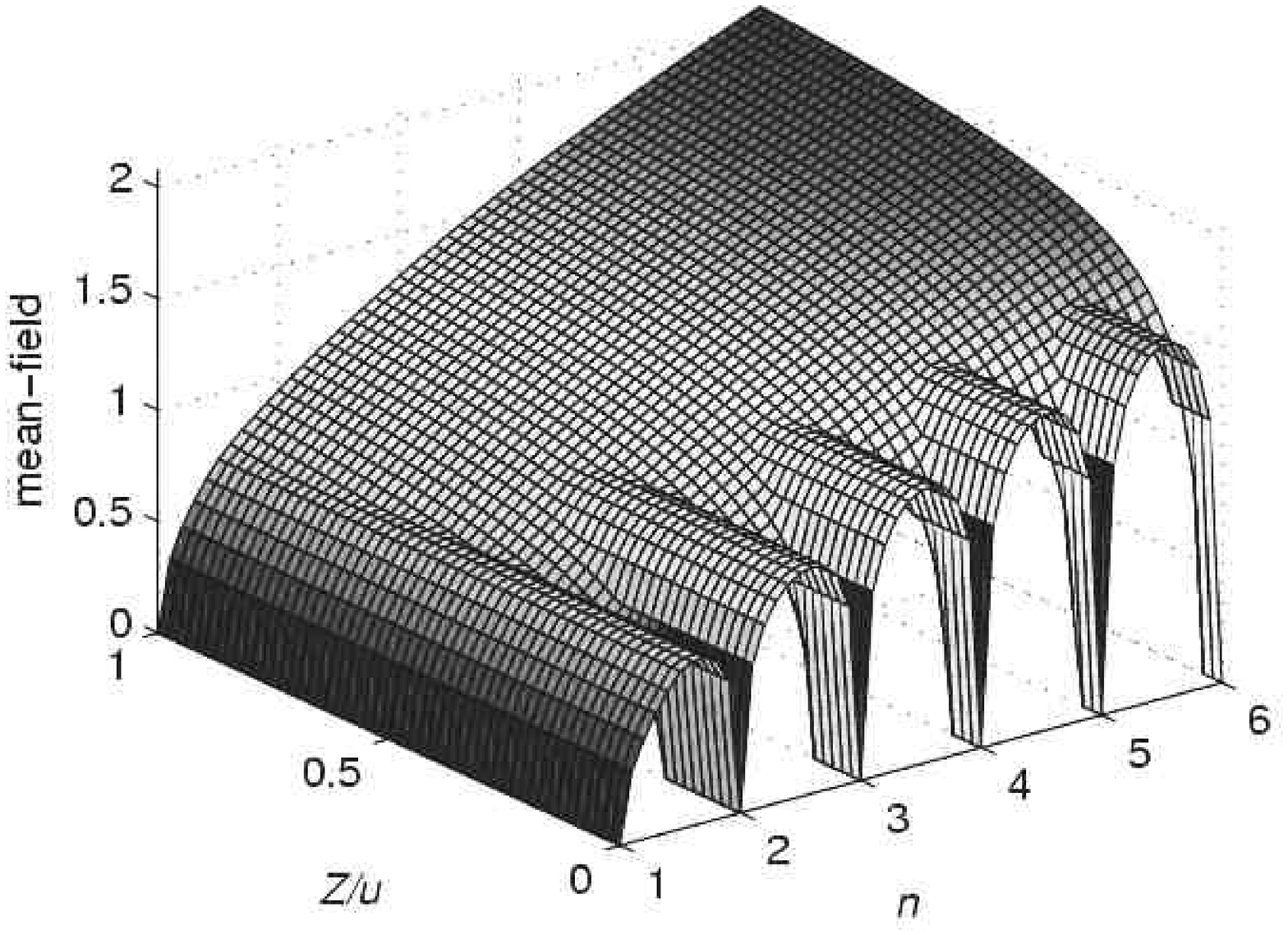}%
\includegraphics[width=6.5cm]{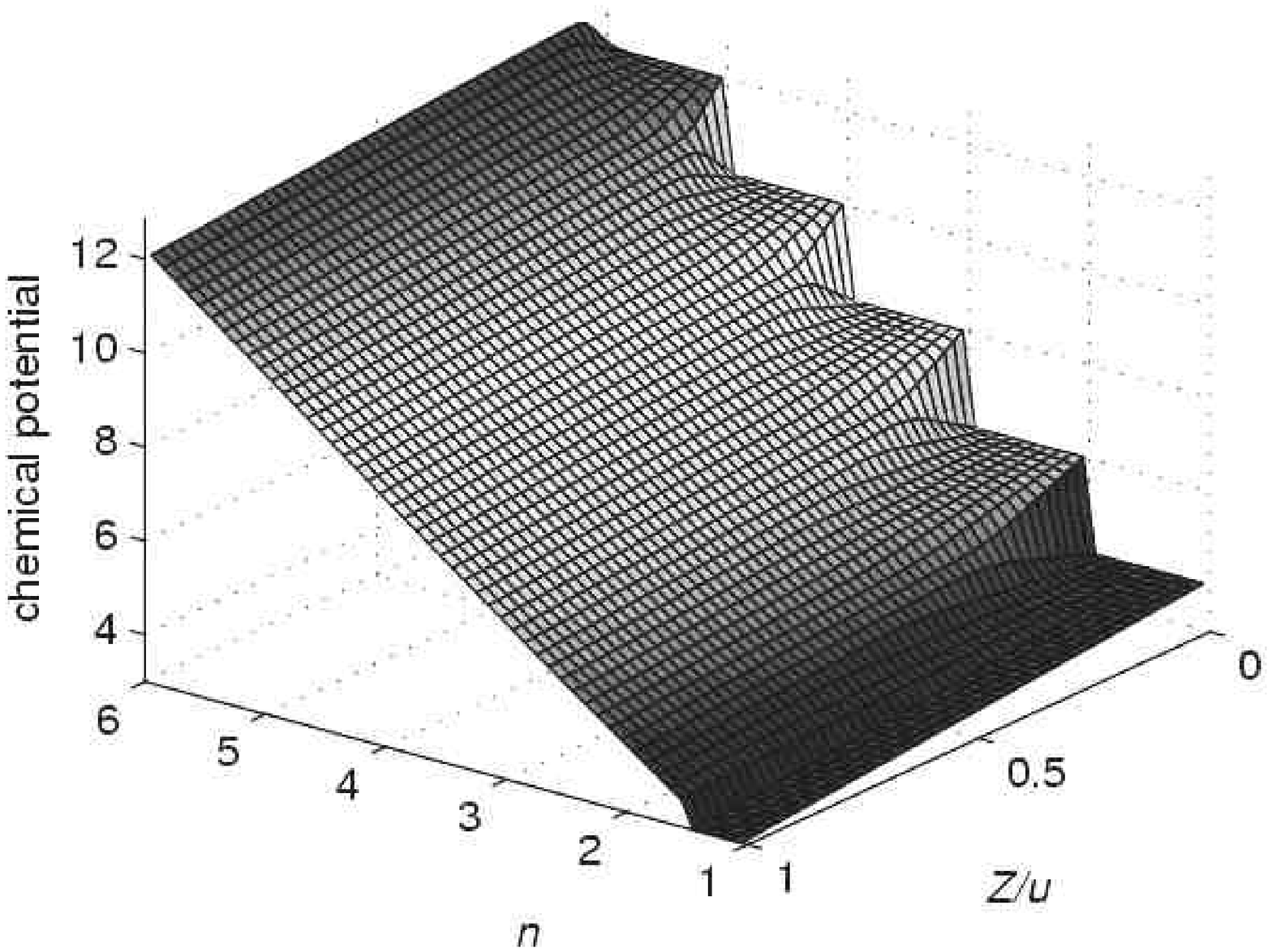}%
}
\caption{\Label{lb}Ground state phase diagram calculated using rigorous lower bound on energy (\ref{rig5}) in Q-function representation}
\end{figure}
\end{document}